\documentclass[aps,prl,twocolumn,groupedaddress,showpacs,showkeys,10pt]{revtex4-1}
\usepackage{amsmath,amsfonts,bm,mathtools,amssymb}
\usepackage[colorlinks=true,linkcolor=blue,citecolor=red,bookmarksnumbered=true]{hyperref}
\usepackage{times,mathpazo,charter,graphicx,bm,enumitem}
\usepackage{color}

\makeatletter
\let\eref=\eqref
\def\section{%
  \@startsection
    {section}%
    {1}%
    {\z@}%
    {0.7cm \@plus1ex \@minus .2ex}%
    {0.3cm}%
    {\normalfont\small\bfseries}%
}%
\def\subsection{%
  \@startsection
    {subsection}%
    {2}%
    {\z@}%
    {.5cm \@plus1ex \@minus .2ex}%
    {.15cm}%
    {\normalfont\small\bfseries}%
}%

\let\tru@int=\int
\def\int{\mathop{\textstyle\tru@int}\limits}
\def\overl@ss#1#2{\vcenter{\offinterlineskip
        \ialign{$\m@th#1\hfil##\hfil$\crcr#2\crcr<\crcr } }}
\def\overgr@at#1#2{\vcenter{\offinterlineskip
        \ialign{$\m@th#1\hfil##\hfil$\crcr#2\crcr>\crcr } }}
\def\overl@ss#1#2{\vcenter{\offinterlineskip
        \ialign{$\m@th#1\hfil##\hfil$\crcr#2\crcr<\crcr } }}
\def\overgr@at#1#2{\vcenter{\offinterlineskip
        \ialign{$\m@th#1\hfil##\hfil$\crcr#2\crcr>\crcr } }}
\def\gl{\mathrel{\mathpalette\overl@ss>}}
\def\lg{\mathrel{\mathpalette\overgr@at<}}
\def\d{\mathrm{d}}

\def\Real{\mathbb{R}}
\def\Complex{\mathbb{C}}
\def\Re{\mathop{\rm Re}\nolimits}
\def\Im{\mathop{\rm Im}\nolimits}
\def\arg{\mathop{\rm arg}\nolimits}
\def\pvint{\int\kern-0.94em-\kern0.2em}
\let\^=\hat
\let\==\bar
\let\@=\mathbf
\def\diag{\mathop{\rm diag}\nolimits}
\def\d{\mathrm{d}}
\def\e{\mathrm{e}}
\def\P{\mathrm{p}}

\def\sech{\mathrm{sech}}
\def\tw{\mathrm{tw}}

\def\tr{\mathrm{tr}}
\def\bsigma{\boldsymbol{\sigma}}

\def\~#1{\tilde{\mathbf{#1}}}
\newcommand\partialderiv[3][]{\frac{\partial^{#1}#2}{\partial {#3}^{#1}}}
\makeatother

\def\[{\begin{equation}}
\def\]{\end{equation}}

\def\be{\begin{equation}}
\def\ee{\end{equation}}
\def\bse{\begin{subequations}}
\def\ese{\end{subequations}}

\advance\abovedisplayskip -10pt
\advance\abovedisplayshortskip -10pt
\advance\belowdisplayskip 2pt
\advance\belowdisplayshortskip 2pt
\advance\parskip 1pt
\allowdisplaybreaks

\begin{document}

\title{Solitons and rogue waves in spinor Bose-Einstein condensates}
\author{Sitai Li$^1$}
\author{Barbara Prinari$^2$}
\author{Gino Biondini$^{1,3}$}
\affiliation{$^1$Department of Mathematics, State University of New York at Buffalo, Buffalo, NY 14260, USA\\
    $^2$Department of Mathematics, University of Colorado Colorado Springs, Colorado Springs, CO 80918, USA\\
    $^3$Department of Physics, State University of New York at Buffalo, Buffalo, NY 14260, USA}
\date{\today}

\begin{abstract}
    We present a general classification of one-soliton solutions
    as well as novel families of rogue-wave solutions for
     $F=1$ spinor Bose-Einstein condensates (BECs).
    These solutions are obtained from the inverse scattering transform for a focusing matrix nonlinear Schr\"{o}dinger equation
    which models condensates in the case of attractive mean field interactions and ferromagnetic spin-exchange interactions.
    In particular, we show that, when no background is present,
    all one-soliton solutions are reducible via unitary transformations
    to a combination of oppositely-polarized solitonic solutions of single-component BECs.
    On the other hand, we show that, when a non-zero background is present, not all
    matrix one-soliton solutions are reducible to a simple combination of scalar solutions.
    Finally, by taking suitable limits of all the solutions on a non-zero background we also obtain
    three families of rogue-wave (i.e., rational) solutions, two of which are novel to the best of our knowledge.
\end{abstract}
\maketitle

\section{I.~ Introduction}

\hspace*{-0.3em}
Bose-Einstein condensates (BECs) have received extensive attention since their first experimental realization~\cite{aemwc1995,ketterle1995}.
One of the mathematical models proposed to describe the time evolution of the condensate wave function in a mean field approximation is the famous
Gross-Pitaevskii (GP) equation~\cite{lsy2005,esy2007},
which in one space dimension and in the absence of external trapping potentials is known to be completely integrable.
The resulting equation is the so-called nonlinear Schr\"{o}dinger (NLS) equation,
and it describes the dynamics in single-component BECs.
Multi-component BECs have also been observed experimentally~\cite{mbgcw1997,mmrri2002}.
They can be created by overlapping two single-component BECs with atoms in two hyperfine states, or mixtures of two different atomic species.
Mathematically, these situations can be modeled by coupled NLS equations with external potentials~\cite{pb1998,ps2003,kfc2008}.

Spinor BEC models have also been proposed~\cite{hs1996,om1998,imo1999},
which correspond to multi-component BECs, with atoms in a single hyperfine state but having internal
spin degrees of freedom.
When these spinor BECs were first experimentally created,
they were shown to exhibit a much richer phenomenology than single-component BECs.
For example, the spin degrees of freedom are liberated under an optical trap,
which opens up the possibility to study spin waves in a Bose-condensed gas~\cite{sacimsk1998}.
Other interesting phenomena that can only be observed in multi-component BECs
include dark-bright soliton complexes~\cite{pra77p033612,pra84p053630,pla375p642,pra80p023613,2017arXiv170508130B}
and the formation of spin domains and spin textures \cite{lskpk2003,pre72p066604,pra76p063603}.

Subsequently, a completely integrable model for spin-one ($F=1$) BECs in one dimension
and without external magnetic fields was proposed~\cite{imw2004}.
In this model, which requires a specific ratio of the scattering lengths and hence of the coupling constants,
the internal dynamics of the condensate are described by three components $\Phi_j(x,t)$ for $j = 0,\pm1$,
representing the wave function of atoms with magnetic spin quantum number $j$.
The mean-field interaction in this model is attractive, and the spin-exchange interaction is ferromagnetic.
The time evolution of the three-component wave function is given by a matrix focusing NLS equation.
Since such matrix NLS equation is completely integrable,
several methods have been used to study the system and derive explicit solutions,
including the inverse scattering transform (IST), and some solutions were presented in Refs.~\cite{Uchiyama2007,imw2004,wt2006,uiw2006darksoliton,stir2010,stir2011,sti2012,iuw2007,kw2007,t2016}.
The model was later extended to describe BECs characterized by repulsive interatomic interactions and antiferromagnetic
spin-exchange interactions (corresponding to the opposite sign for the ratio of the coupling constants)~\cite{wt2006,uiw2006darksoliton,Uchiyama2007,t2016}.
In this case, the relevant model is a defocusing matrix NLS equation, and a non-zero background is required in order for the system
to admit soliton solutions.
The generalization to a non-zero background is particularly important for both kinds of nonlinearity (attractive/repulsive),
since the non-zero background allows for the existence of so-called domain wall solutions~\cite{pre72p066604,pra76p063603},
dark-bright soliton complexes~\cite{pra77p033612,pra84p053630,pla375p642,pra80p023613,2017arXiv170508130B},
and for the focusing scalar NLS equation it is related to the existence of ``rogue'' waves.

The term rogue wave is used to refer to waves that have unusually high amplitudes (by a factor of two or larger) compared to the background,
and that ``appear from nowhere and disappear without a trace''~\cite{aat2009}.
Besides oceans (where they are also referred to as ``freak waves'')~\cite{d1990,mgo2005},
rogue wave phenomena are also observed in the atmosphere~\cite{ss2009},
optics~\cite{srkj2007,egd2009} and plasmas~\cite{mse2011}.
Rogue waves have been studied extensively in the context of the scalar integrable focusing NLS equation,
because of its role as a model equation in deep water waves, optical fibers and BECs~\cite{as1981,a2007,ps2003,kfc2008}.
In particular,
Peregrine solitons~\cite{p1983} and higher order rational solutions~\cite{dtkm2010,Ohta1716}
have been proposed as a possible mathematical description of rogue waves in various media~\cite{sg2010},
including in single component BECs~\cite{wllszl2011}.
Rational solutions of the coupled NLS equation and of the three-wave interaction equations were also studied and used to predict the existence of matter rogue waves in BECs~\cite{bdcw2012,bcdl2013}.

The purpose of this work is twofold.
First, we present a complete classification of one-soliton solutions of the focusing spinor BEC equation on a non-zero background.
Second, we obtain novel families of rational solutions
which generalize those obtained in Ref.~\cite{qm2012}.
All these soliton solutions are formulated in the context of the IST
for this model, which was recently developed in~\cite{pdlhf}.
We also discuss explicit spin polarization transformations of all these solutions
that relate solitonic and rogue waves in spinor BECs and those in single component BECs.
We show that, in the case of zero background, all one-soliton solutions of the spinor model
are equivalent, up to unitary transformations,
either to a scalar soliton solution or to a superposition of two oppositely polarized shifted scalar solitons.
On the other hand, we show that the same statement does not apply in the presence of a non-zero background, since in this case
only some of the one-soliton solutions or rational solutions are equivalent, up to unitary similarity transformations,
to superposition of polarized scalar 
solutions.

\section{II.~ Spinor BEC model and its soliton solutions}

Atoms in $F=1$ spinor BECs can be described by the three-component
macroscopic condensate vector wave function $(\Phi_1(x,t),\Phi_0(x,t),\Phi_{-1}(x,t))^T$,
where $\Phi_j(x,t)$ describes atoms with magnetic spin quantum number $j$.
In a mean-field approximation,
$\Phi_j$ is shown to satisfy the following system of partial differential equations
\vspace*{-2ex}
\bse
\label{e:physicalequations}
\begin{multline}
i \hbar \partialderiv{\Phi_{\pm1}}t + \frac{\hbar^2}{2m}\partialderiv[2]{\Phi_{\pm1}}x
= (\bar c_o+\bar c_2)(|\Phi_{\pm1}|^2+|\Phi_0|^2)\Phi_{\pm1}\\
+ (\bar c_o-\bar c_2)|\Phi_{\mp1}|^2\Phi_{\pm1} + \bar c_2\Phi_{\mp1}^*\Phi_0^2\,,
\end{multline}
\vspace*{-3ex}
\begin{multline}
i \hbar \partialderiv{\Phi_{0}}t + \frac{\hbar^2}{2m}\partialderiv[2]{\Phi_{0}}x
= (\bar c_o+\bar c_2)(|\Phi_{1}|^2+|\Phi_{-1}|^2)\Phi_{0}\\
+ \bar c_o |\Phi_0|^2\Phi_0 + 2\bar c_2\Phi_{0}^*\Phi_1\Phi_{-1}\,,
\end{multline}
\ese
where $\bar c_j$ are the coupling constants (related to the scattering lengths),
and asterisk denotes complex conjugate~\cite{imw2004}. The above set of equations
admits special reductions which are integrable.
The case $\bar{c}_2=0$ yields the three-component NLS equation.
The case $\bar{c}_o=\bar{c}_2=\sigma $ yields the matrix NLS
equation, with $\sigma=\pm 1$ corresponding to the focusing/defocusing regimes.
In the focusing case, Eqs.~\eref{e:physicalequations}
are equivalent to the following integrable model
\bse
\label{e:spinor_zbc}
\[
i \Phi_t + \Phi_{xx} + 2 \Phi\Phi^\dagger \Phi = 0\,,
\label{e:matrixNLS}
\]
where the coordinates $x$ and $t$ have been suitably nondimensionalized,
subscripts $x$ and $t$ denote partial derivatives, the dagger denotes conjugate transpose, and
\[
\Phi(x,t) = \begin{pmatrix}
\phi_1 & \phi_0 \\ \phi_0 & \phi_{-1}
\end{pmatrix},
\label{e:symmetryconstraint}
\]
\ese
where $\phi_{j}(x,t)$ for $j=0,\pm 1$ represent the normalized wave functions.
We refer the reader to Ref.~\cite{imw2004} for a detailed derivation of Eq.~\eref{e:spinor_zbc} in the context
of BECs.
It is important to point out the difference between the spinor model corresponding to the matrix NLS equation~\eref{e:spinor_zbc} and the vector NLS models.
As it is evident from the comparison of  Eqs.~\eref{e:physicalequations} with the vector NLS
(namely: $i \Phi_t+\Phi_{xx}\pm 2\|\Phi\|^2\Phi=0$, $\Phi(x,t)$ being an N-component vector),
the corresponding equations describe different physical models,
with different kinds of nonlinear terms.
Explicitly, the nonlinearity in vector NLS equations only accounts for self-phase and cross-phase modulation,
whereas the nonlinearity in the square matrix model also includes four-wave mixing terms,
and allows to describe spin-exchange interaction.

\subsection{II.A~ Non-zero background}

The above focusing matrix NLS equation admits a Lax pair,
and thus can be studied via IST~\cite{ki1977,ki1978,iuw2007,pdlhf}.
In particular, in~\cite{pdlhf} we considered the initial value problem for Eq.~\eref{e:spinor_zbc} with the boundary conditions (BC)
\[
\Phi(x,t)\to \Phi_\pm\,,\qquad x\to\pm\infty\,.
\label{e:BC}
\]
Physically, the significance of Eq.~\eref{e:BC} is that we consider BECs whose spatial extent is much broader than that of the solution structures being studied.
We refer to $\Phi_\pm = 0$ as the case of zero background,
and to $\Phi_\pm \ne 0$ as the case of non-zero background.
We further assumed that
\[
\Phi_\pm^\dagger \Phi_\pm = \Phi_\pm \Phi_\pm^\dagger = k_o^2 I_2\,,
\label{e:constraint}
\]
where $I_2$ is the $2\times2$ identity matrix and $k_o\ge0$ is the amplitude of the background.
The two definitions~\eref{e:BC} and~\eref{e:constraint} are consistent with those in previous works~\cite{uiw2006darksoliton,Uchiyama2007,iuw2007,kw2007}.
According to above definitions,
$k_o = 0$ corresponds to a zero background and is referred to here as the case of ``zero BC'' (ZBC);
the case $k_o>0$, corresponding to a non-zero background, is referred to here as ``non-zero BC'' (NZBC).
Of course Eq.~\eqref{e:constraint} restricts the class of solutions that one can describe.
On one hand, this condition is similar to the constraint that is commonly placed when looking for solutions of
focusing and defocusing vector NLS equations \cite{PAB2006,SAPM2011,PV2013,SIMA2015,KBK15,JPA2015,JMP2015,CMP2016},
and in  those cases it is a necessary condition for the existence of pure soliton solutions.
On the other hand, we show below that, even with this restriction, the system admits a large variety of soliton solutions.

It is worth at this stage to point out the difference between the current work and previous works on vector NLS equations with NZBC.
First of all, as already mentioned above,
the corresponding equations describe different physical models,
with different kinds of nonlinear terms.
When a non-zero background is considered,
this crucial difference is reflected both in the formulation of the IST and in the behavior of the solutions.
From a spectral point of view,
in the formulation of the IST,
one can show that all eigenfunctions of the square matrix model are analytic in specific regions of the spectral plane,
whereas only two eigenfunctions of the vector model are analytic.
As a result, the scattering data for the two associated spectral problems are different,
and so are the soliton solutions.
Moreover, as we show below, the soliton solutions of the $2\times2$ matrix NLS equation are associated with matrix norming constants,
and the matrix nature of the norming constants plays a crucial role in the properties of the corresponding soliton solutions.

In general, the boundary conditions $\Phi_\pm$ must be time-dependent in order to be compatible with the time evolution.
Time-independent BC can be achieved via a simple
gauge transformation, however.
Explicitly, with the transformation $\Phi(x,t)\mapsto \Phi(x,t)\,\e^{2i k_o^2t}$,
Eq.~\eref{e:spinor_zbc} can be written as
\[
\label{e:spinor_nzbc}
i \Phi_t + \Phi_{xx} + 2 (\Phi\Phi^\dagger - k_o^2 I_2) \Phi = 0\,,
\]
so that the values $\Phi_\pm$ are independent of~$t$.

Importantly, the matrix NLS equation~\eqref{e:matrixNLS} is invariant under unitary transformations.
Namely,
if $\Phi(x,t)$ is a solution of Eq.~\eref{e:matrixNLS},
\vspace*{-0.6ex}
\[
\label{e:UphiV}
\tilde\Phi(x,t) = U \Phi(x,t) V
\]
is also a solution of Eq.~\eref{e:matrixNLS}
for arbitrary constant unitary matrices $U$ and~$V$.
Of course, in order for this invariance to also apply to the full spinor BEC system~\eqref{e:spinor_zbc},
the unitary matrices $U$ and $V$ must be chosen so that $\tilde\Phi(x,t)$ is also symmetric.
Such general unitary transformations are then associated with spin rotations in the spinor BEC.
Thanks to this invariance, one can assume without loss of generality that
\vspace*{-1ex}
\[
\label{e:NZBC}
\Phi_+ = k_o I_2\,,
\]
since an arbitrary boundary condition can be reduced to the above by an appropriate choice of $U$ and $V$
in~\eref{e:UphiV}~\cite{pdlhf}.
Therefore, in the rest of this work we discuss solitons and rogue waves on a non-zero background
with asymptotic behavior as in Eq.~\eref{e:NZBC}, since solutions with a different asymptotic state
can be reconstructed from them by means of the above mentioned unitary transformations.
Note, however, that one does not have the freedom to specify both $\Phi_+$ and $\Phi_-$.
Once $\Phi_+$ has been chosen, $\Phi_-$ is determined by the specific solution considered, and
is not necessarily diagonal, even when the constraint provided by Eq.~\eqref{e:constraint} is satisfied,
as we discuss later.

The Lax pair of the spinor model~\eref{e:spinor_nzbc} is given by
\[
\nonumber
\psi_x = (-ik \underline{\sigma}_3 + \underline{\Phi})\psi\,,\qquad
\psi_t = V \psi\,,
\]
where
\begin{gather*}
V = -2ik^2 \underline{\Phi} + 2k \underline{\Phi}
    + i \underline{\sigma}_3(\partial_x\underline{\Phi} - k_o^2 I_4 + \underline{\Phi}\,\underline{\Phi}^\dagger )\,,
\\[0.5ex]
\underline{\sigma}_3 = \begin{pmatrix} I_2 & 0 \\ 0 & -I_2\end{pmatrix}\,,\qquad
\underline{\Phi} = \begin{pmatrix} 0 & \Phi \\ -\Phi^\dagger & 0\end{pmatrix}\,,
\end{gather*}
$I_4$ is the $4\times4$ identity matrix and $k\in \Complex$ denotes the spectral parameter.
In~\cite{pdlhf}, we formulated the IST for Eq.~\eref{e:spinor_nzbc} satisfying the BC~\eref{e:constraint},
and we derived an expression for general $N$-soliton solutions.
%
From the formulation of the IST for the spinor model~\eref{e:spinor_nzbc},
$N$-soliton solutions are completely determined by $N$ discrete eigenvalues and $N$ associated norming constants.
The discrete eigenvalues are scalar complex numbers,
whereas the norming constants are $2\times2$ symmetric complex-valued matrices.
In the rest of this work we will focus on one-soliton solutions, i.e. we take $N = 1$.

\subsection{II.B~ One-soliton solutions}

The one-soliton solution corresponding to a discrete eigenvalue $\zeta$
(with $|\zeta| \ge k_o$ and $\Im\zeta>0$)
and a norming constant $K$
(which must be a $2\times2$ symmetric complex-valued matrix)
is given by
\[
\label{e:1soliton_nzbc}
\Phi(x,t) = k_o^2I_2 - i X_1\e^{-2i\theta^*}K^\dagger + i k_o^2 X_2 \e^{2i\theta}K/\zeta^2\,
\]
(see~\cite{pdlhf} for details),
where
\[
\label{e:theta}
\theta(x,t) = (\zeta^2 + k_o^2)[\zeta x + (\zeta^2 - k_o^2)t]/(2\zeta^2)\,,
\]
and $X_1,X_2$ solve the following linear system
\[
\label{e:X1X2}
X_1 D = I_2 - i k_o X_2 c/\zeta\,,\quad
X_2 D^\dagger = I_2 - i\zeta X_1 c^\dagger/k_o\,,
\]
with
\[
\nonumber
c = \frac{K}{\zeta^*-\zeta}\e^{2i\theta}\,,\quad
D = I_2 + \frac{ik_o K^\dagger}{(\zeta^*)^2 + k_o^2} \e^{-2i\theta^*}\,.
\]

In Appendix~I, we show that the behavior of the soliton solutions crucially depends on the rank of the norming constant,
i.e., on the matrix nature of $K$.
We do so by calculating the total spin of the one-soliton solutions~\eref{e:1soliton_nzbc}.
We show that:
if $\det K=0$ the BEC has non-zero total spin and thus is in a ferromagnetic state;
if $\det K\ne0$ the BEC has zero total spin and thus is in a polar state~\cite{h1998}.

Moreover, it was also shown in \cite{pdlhf} that, when $\det K\ne0$ (i.e., for a polar solution), $\Phi_-= \e^{-4i\alpha} k_o I_2$,
(with $\alpha =\arg \zeta$).
Conversely, when $\det K = 0$ (i.e., for a ferromagnetic solution), in general $\Phi_-$ is not diagonal.
Clearly, the ferromagnetic solutions ($\det K = 0$) are genuine matrix solutions,
and do not admit any analogues in other models with scalar or vector norming constants
(e.g., scalar and vector NLS equations).
Moreover, because of the different asymptotics of $\Phi_\pm$ when $\det K = 0$,
one expects the solutions to exhibit a kink-like behavior,
i.e., a domain wall may form.
Indeed, we will show later that in some cases Eq.~\eref{e:1soliton_nzbc} gives rise to topological solitons.


In the limit $k_o \to 0$,
Eq.~\eref{e:1soliton_nzbc} yields soliton solutions of the spinor model with ZBC,
i.e., with $\Phi(x,t)\to0$ as $x\to\pm\infty$.
Simple calculations from Eq.~\eqref{e:1soliton_nzbc} show that the one-soliton solution with ZBC is given by
\[
\label{e:1soliton_zbc}
\Phi(x,t) = -i\, \e^{-2i\theta^*}(I_2 + c^\dagger c)^{-1}K^\dagger\,,
\]
where in this case
\[
\theta(x,t) = (\zeta x + \zeta ^2 t)/2\,,\quad
c = K\e^{2i\theta}/(\zeta^*-\zeta)\,.
\]
Similarly to the case of NZBC, solutions are in a ferromagnetic state
when $\det K = 0$, and in a polar state when
$\det K\ne0$.
The soliton solutions~\eref{e:1soliton_zbc} with ZBC were first derived in Ref.~\cite{imw2004}.

\section{III.~ Classification of soliton solutions}

In this section we present a complete classification of the one-soliton solutions
of the spinor BEC model both with ZBC and with NZBC, given respectively by
Eqs.~\eref{e:1soliton_zbc} and~\eref{e:1soliton_nzbc}.
First, in section~III.A we discuss the classification of the soliton solutions on a zero background.
Then, in section III.B we show how similar methods can be used to classify soliton solutions with NZBC.

Recall that, if $\Phi(x,t)$ is a solution of Eqs.~\eref{e:spinor_zbc},
$\tilde\Phi(x,t)$ defined by Eq.~\eqref{e:UphiV} is also a solution
provided that $U$ and $V$ are two constant unitary matrices and $\tilde\Phi(x,t)$ is also symmetric.
We can then formulate the concept of \textit{equivalence classes} of solutions.  That is,
we say that two given solutions $\Phi(x,t)$ and $\tilde\Phi(x,t)$ of the spinor BEC model~\eref{e:spinor_zbc} are equivalent
if there exist two unitary matrices $U$ and $V$ such that Eq.~\eqref{e:UphiV} holds.

\subsection{III.A~ Soliton solutions with ZBC}

In \cite{sti2012} it was shown that up to a rotation of the quantization axes, soliton solutions with ZBC can be written as a ``superposition
of two oppositely polarized displaced solitons''
of the focusing NLS equation.
We show below how this result can be obtained using a method that can be generalized to classify soliton solutions with NZBC.

Since the norming constant $K$ is symmetric,
Takagi's factorization~\cite{t1924} ensures that there exists a unitary constant matrix $U$ such that
\[
\label{e:Kdiagonal_zbc}
U K U^T = \Gamma\,,\qquad
\Gamma = \diag(\gamma_1,\gamma_{-1})\,,
\]
where $\gamma_j\ge0$ and $\gamma_j^2$ are the eigenvalues of $K^\dagger K$.
(Notice that the matrix $K^\dagger K$ is Hermitian and positive-semidefinite,
so its eigenvalues are real and non-negative.)
Therefore, one can write any norming constant as
\[
\label{e:C1diagzbc}
K = U^\dagger\Gamma U^*\,.
\]
Substituting Eq.~\eref{e:C1diagzbc} into the solution~\eref{e:1soliton_zbc},
one has
\[
\Phi(x,t) = U^T Q(x,t) U\,,
\label{e:solutiondecomposition}
\]
where $Q(x,t)$ is a diagonal matrix given by
\[
\hspace*{-0.4em}
Q(x,t) = \diag(q_1,q_{-1})
= -i\, \e^{-2i\theta^*}(I_2 + \tilde c^\dagger \tilde c)^{-1}\Gamma^\dagger\,,
\]
and $\tilde c = \Gamma\e^{2i\theta}/(k^*-k)$.
Since $U$ and $U^T$ are unitary matrices,
and $Q$ is diagonal and hence symmetric,
we conclude that $Q$ is also a solution of the spinor model~\eref{e:spinor_zbc}
and $Q\to0$ as $x\to\pm\infty$.
Moreover, because $Q$ is in the form of Eq.~\eref{e:1soliton_zbc},
it is also a one-soliton solution with the same discrete eigenvalue $\zeta$ and a diagonal norming constant $\Gamma$.

Whenever the solution of the spinor model~\eref{e:spinor_zbc} is diagonal,
like $Q(x,t)$ above,
its diagonal components, i.e., $q_{\pm1}(x,t)$,
are decoupled, and each individually satisfies the scalar focusing NLS equation:
\[
\label{e:nls_zbc}
i q_t + q_{xx} + 2|q|^2 q = 0\,.
\]
It then follows that each
$q_j(x,t)$ with $j=\pm1$ is a one-soliton solution of Eq.~\eref{e:nls_zbc}
with discrete eigenvalue $\zeta$ and norming constant $\gamma_j$.
More precisely,
if we write the discrete eigenvalue as $\zeta = V + i A$ with $A>0$ and $V\in\Real$,
each $q_j$ will have the form of the sech-shaped one-soliton solution of the focusing NLS equation:
\[
\nonumber
q_{\sech,j} = -i A\,\sech[A(x + 2Vt-\xi_j)]\,\e^{i[-Vx+(A^2-V^2)t]}\,,
\]
where $A\xi_j = \ln[\gamma_j/(2A)]$,
$A$ is the soliton amplitude and $-2V$ is the soliton velocity.
Notice that since the norming constant $\gamma_j$ is real,
each solution $q_{\sech,j}$ depends on three free parameters instead of four.
(An overall phase for each component is absorbed by the above unitary transformation.)

Furthermore, from Eq.~\eref{e:Kdiagonal_zbc} we have
\[
\nonumber
|\det K\,| = \gamma_1\gamma_{-1}\,.
\]
So, if the solution $\Phi(x,t)$ describes a ferromagnetic state, i.e., $\det K = 0$,
one of the $\gamma_j$ must be zero.
Without loss of generality, we can take $\gamma_{-1} = 0$ and $\gamma_1>0$.
[Note that the case $\gamma_1 = \gamma_{-1} =0$ is trivial,
because Eq.~\eref{e:1soliton_zbc} implies $\Phi(x,t)\equiv0$ in this case.]
On the other hand, if the solution $\Phi(x,t)$ describes a polar state, i.e., $\det K \ne 0$,
then both $\gamma_1$ and $\gamma_{-1}$ are strictly positive.

\begin{figure}[t!]
    \centering
    \includegraphics[scale=0.22]{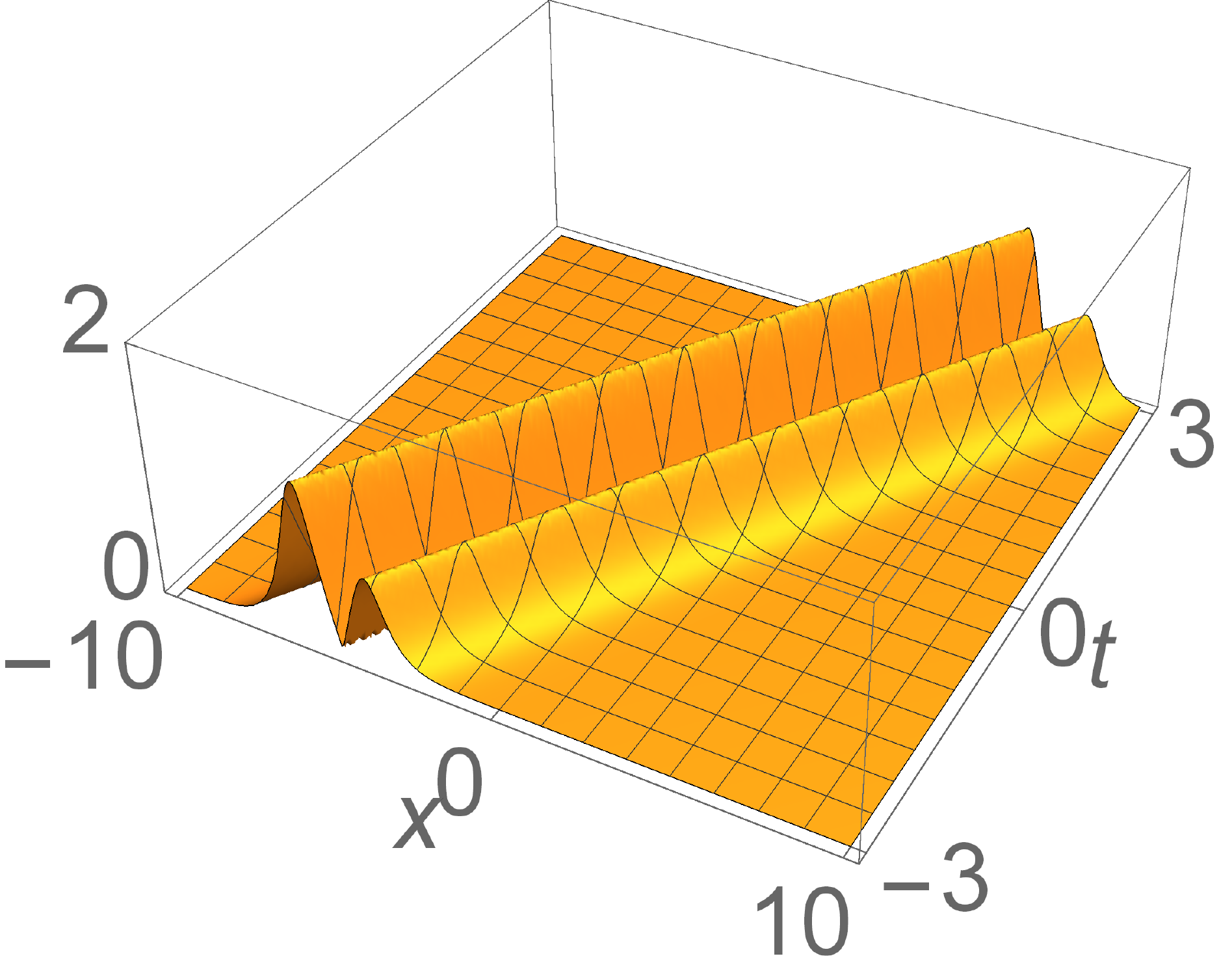}\,
    \includegraphics[scale=0.22]{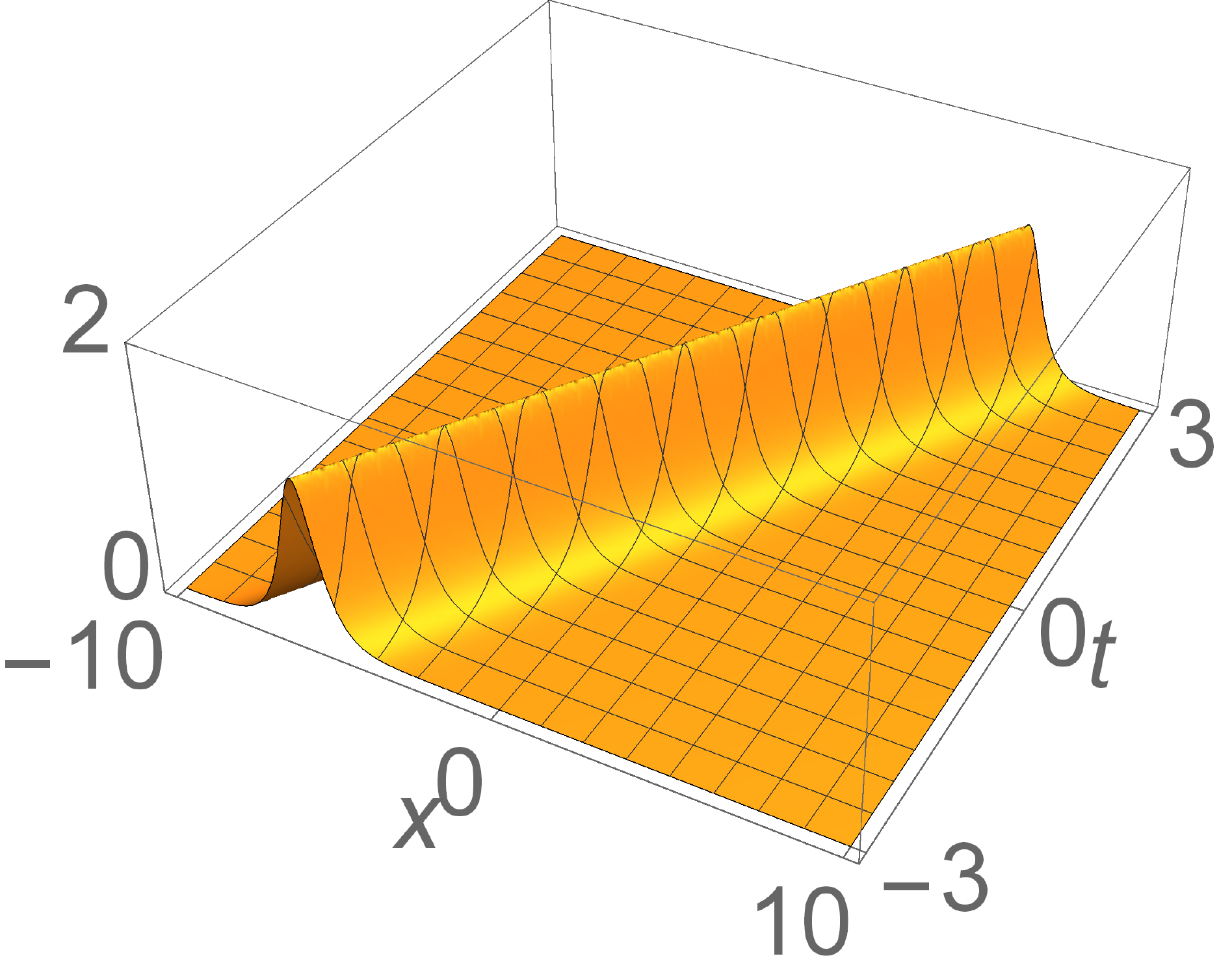}\\
    \includegraphics[scale=0.22]{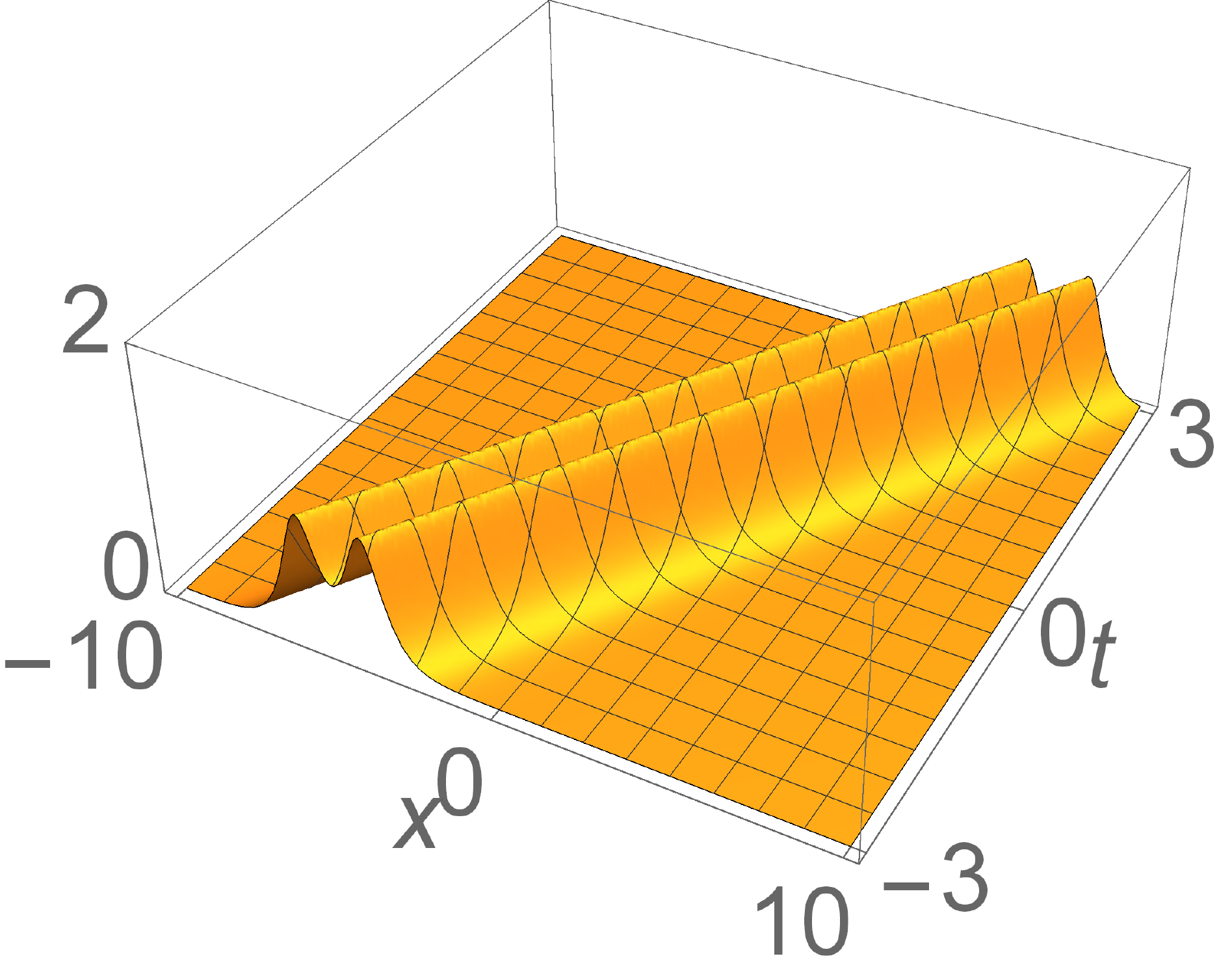}\,
    \includegraphics[scale=0.22]{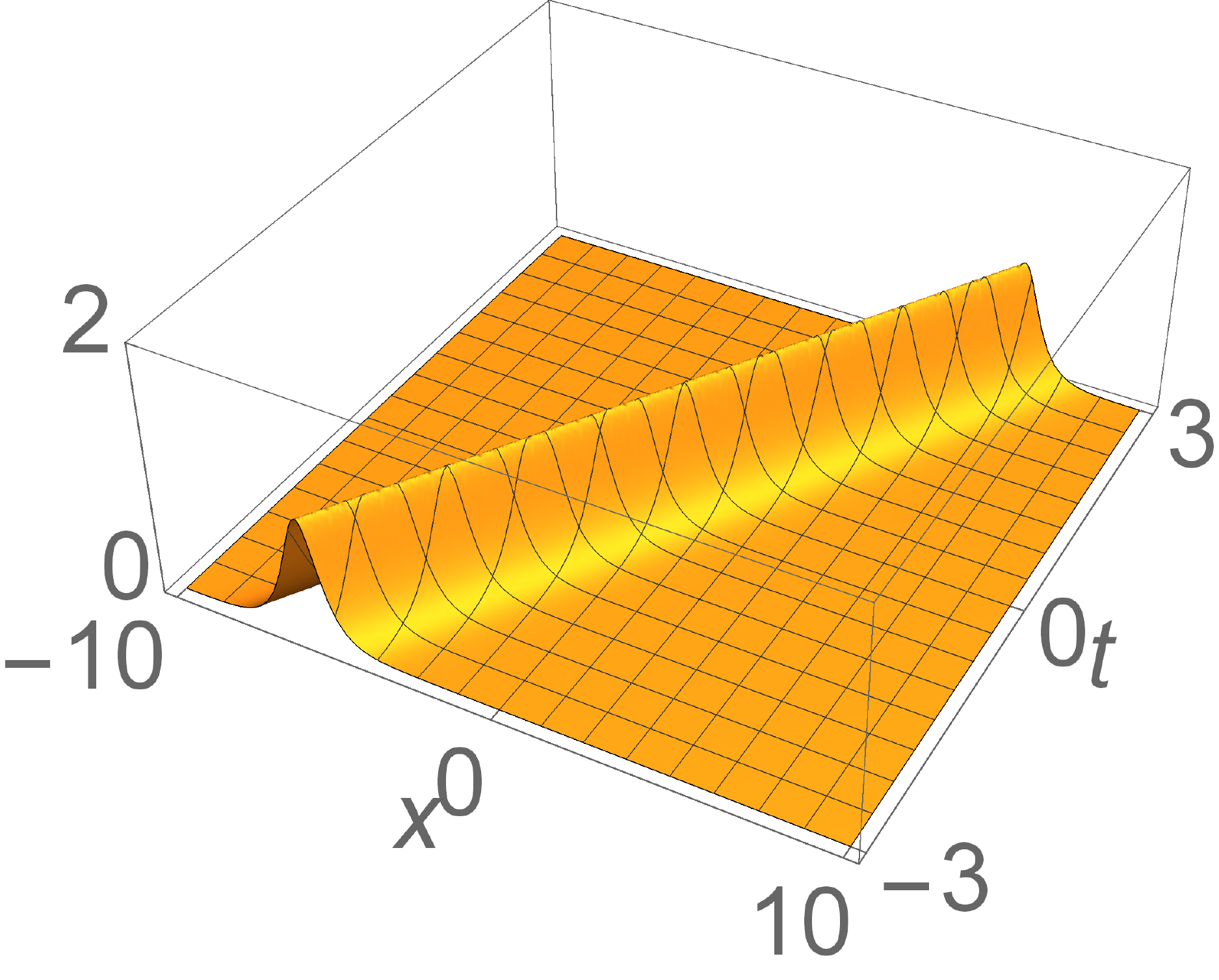}\\
    \includegraphics[scale=0.22]{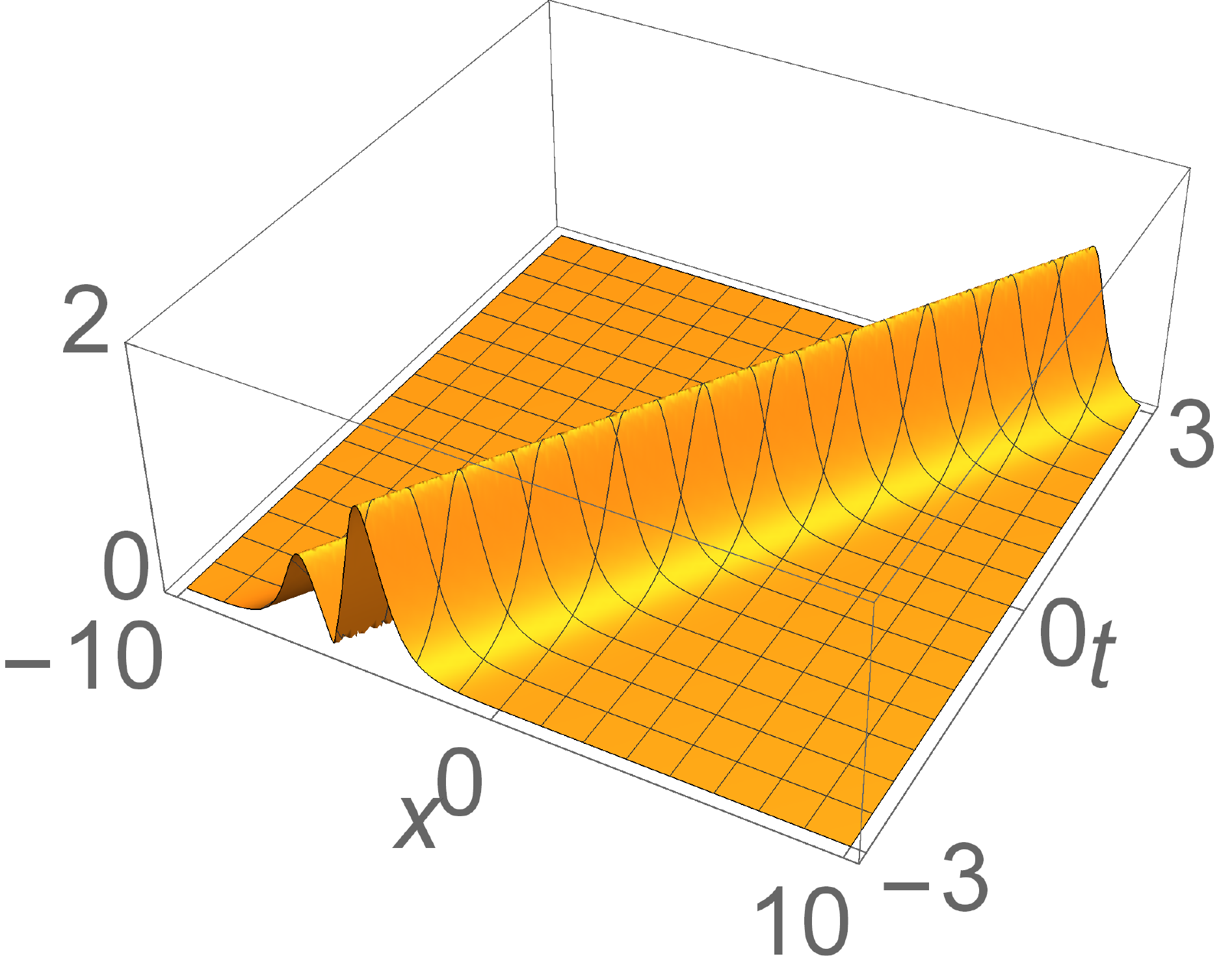}\,
    \includegraphics[scale=0.22]{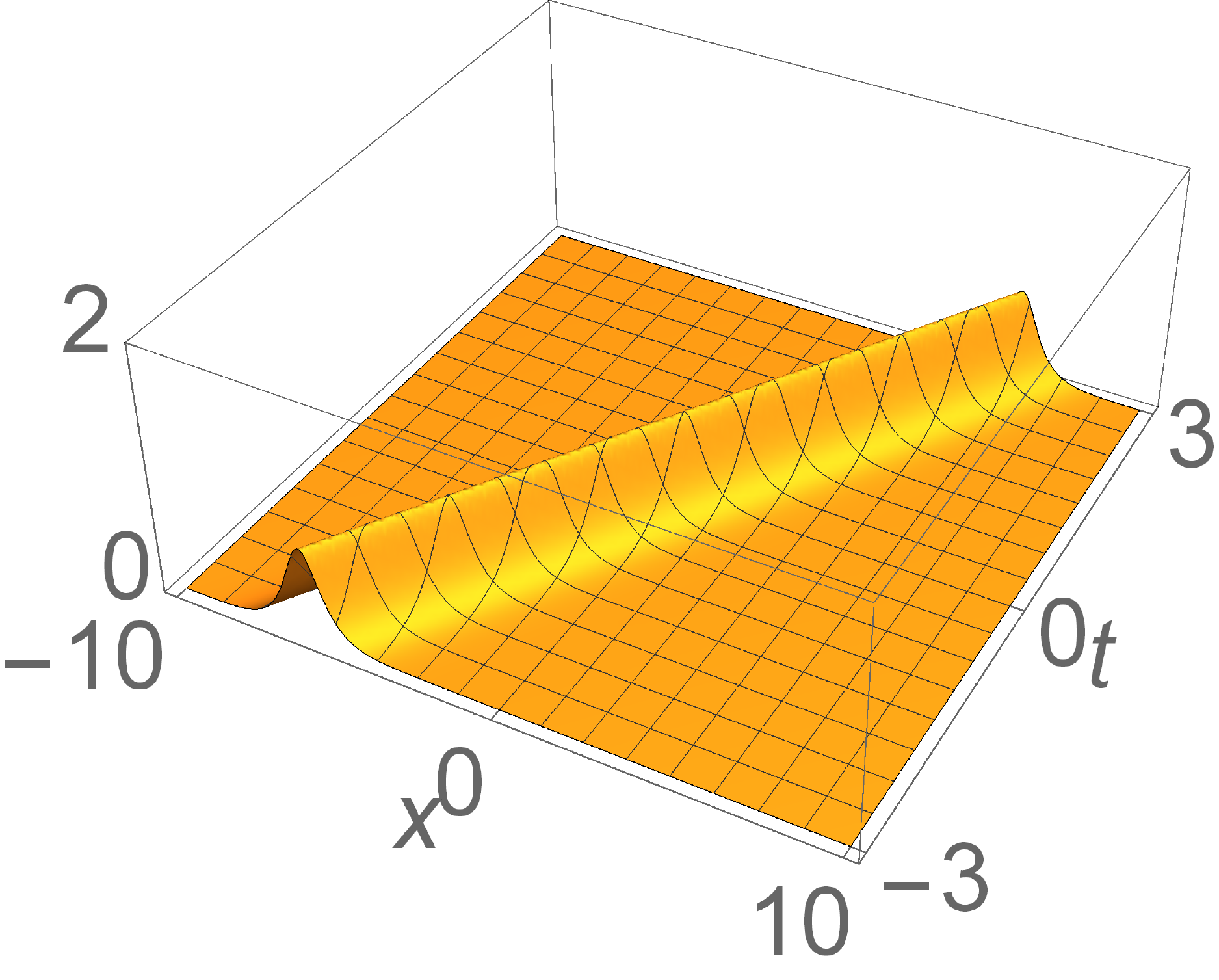}
    \caption{Amplitudes of one-soliton solutions of the spinor BEC model with ZBC, with discrete eigenvalue $\zeta = -1+2i$
        and unitary matrix $U = -\frac45 I_2 + \frac35 i \sigma_1$.
        Top: $|\phi_1(x,t)|$.
        Center: $|\phi_0(x,t)|$.
        Bottom: $|\phi_{-1}(x,t)|$.
        Left: a polar state given by Eq.~\eref{e:Phisech_zbc_p} with $\gamma_1 = 4$ and  $\gamma_{-1} = 4\e^4$.
        Right: a ferromagnetic state given by Eq.~\eref{e:Phisech_zbc_f} with $\gamma_1 = 4$.
    }
    \label{f:zbc}
\end{figure}

As an example, Fig.~\ref{f:zbc} shows two one-soliton solutions with ZBC
obtained from the same discrete eigenvalue,
but with different norming constants, one giving rise to a ferromagnetic state and the other one to a polar state.

Since eigenvalues are preserved by unitary transformations,
from the above discussion it follows that
soliton solutions with ZBC divide into two equivalence classes:

\paragraph{\bf Class~A.}
Any one-soliton solution describing a polar state (i.e., with a non-singular norming constant $K$)
can be written in the form~\eqref{e:solutiondecomposition} with
\bse
\[
\label{e:Phisech_zbc_p}
Q(x,t) = \diag(q_{\sech,1}(x,t), q_{\sech,-1}(x,t))\,,
\]
where $U$ is a constant unitary matrix
[as determined by Takagi's factorization algorithm to reduce $K$ to
its diagonal form \eref{e:C1diagzbc} with $\Gamma=\mathrm{diag}(\gamma_{1},\gamma_{-1})$],
$q_{\sech,j}(x,t)$ for $j=\pm 1$ is the classical sech-shaped soliton solution of the scalar NLS equation
with discrete eigenvalue $\zeta$ and norming constant $\gamma_j$.

\paragraph{\bf Class~B.}
Any one-soliton solution describing a ferromagnetic state
(i.e., with a rank-one norming constant $K$) can be written in the form~\eqref{e:solutiondecomposition} with
\label{e:Phisech_zbc}
\[
\label{e:Phisech_zbc_f}
Q(x,t) = \diag(q_{\sech,1}(x,t), 0)\,,
\]
\ese
where again $U$ is a constant unitary matrix
[as determined by Takagi's factorization algorithm to reduce $K$ to
its diagonal form \eref{e:C1diagzbc} with $\Gamma=\diag(\gamma_1,0)$],
and $q_{\sech,1}(x,t)$ is the classical sech-shaped soliton solution of the scalar NLS equation
with discrete eigenvalue $\zeta$ and norming constant $\gamma_1$.

\paragraph{\bf Additional remarks.}
Equations~\eref{e:Phisech_zbc} relate one-soliton solutions of the spinor model~\eref{e:spinor_zbc}
to those of the scalar focusing NLS equation~\eref{e:nls_zbc} with a zero background.
Any one-soliton solution of the spinor model~\eref{e:spinor_zbc} with ZBC is reducible,
i.e., is equivalent to either a single scalar one-soliton solution or two shifted scalar one-soliton solutions
(one in each of the two oppositely polarized states).
Conversely, given any one-soliton solutions of the focusing NLS equation,
or any two such solutions with the same discrete eigenvalue
(with either equal or different norming constants),
one can always construct a one-soliton solution of the spinor model~\eref{e:spinor_zbc} by using an arbitrary unitary matrix $U$ such
that the unitary transformation~\eref{e:solutiondecomposition} keeps the solution symmetric.

It is also worth to point out that the diagonal forms in Eq.~\eref{e:Phisech_zbc} are unique,
in the sense that the scalar solitons $q_{\sech,j}(x,t)$ are uniquely determined by the discrete eigenvalue $\zeta$
and the non-negative eigenvalues of $K^\dagger K$.
Thus, if two solutions $\Phi_1$ and $\Phi_2$ have the same diagonal scalar solitons $q_{\sech,j}$,
then they differ only by a constant unitary transformation of the form \eref{e:solutiondecomposition},
i.e., a spin rotation.

\subsection{III.B~ Soliton solutions with NZBC: Schur classes}

We next discuss one-soliton solutions in spinor BECs with a non-zero background.
In this case the phenomenology is much richer than in the case of zero background,
as it crucially depends on the location of the discrete eigenvalue in the spectral plane,
as well as the structure of the norming constant.

\begin{figure}[b!]
    \centering
    \includegraphics[scale=0.265]{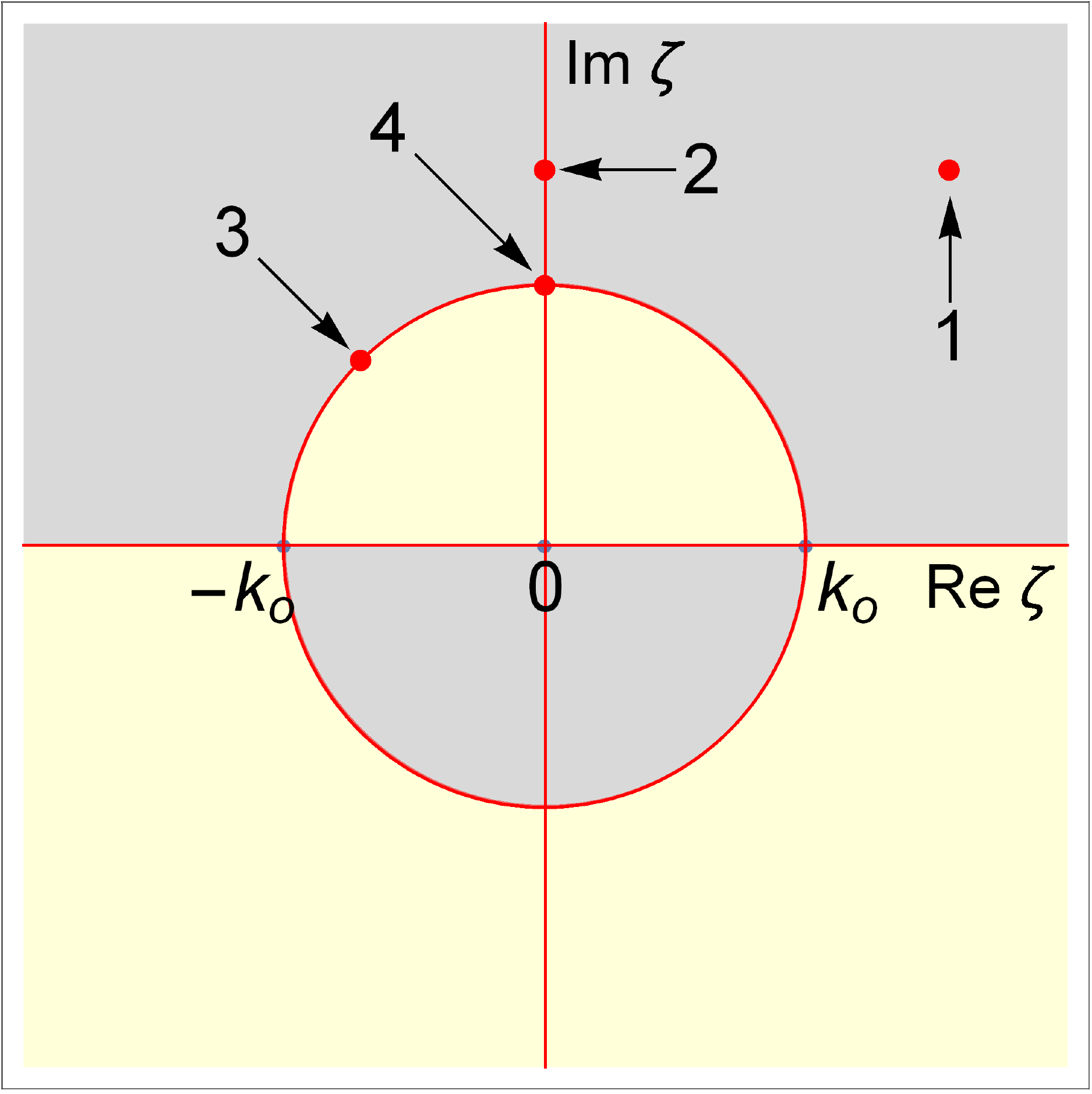}
    \caption{
        The four kinds of discrete eigenvalues in the spectral plane for the spinor BEC model.
        1. Eigenvalue in general position, corresponding to traveling solitons;
        2. Imaginary eigenvalue, corresponding to stationary solitons;
        3. Eigenvalue on the circle, corresponding to periodic solutions;
        4. Eigenvalue on the branch point, corresponding to rational solutions.
        The four kinds of solutions are the analogues of the Tajiri-Watanabe,
        Kuznetsov-Ma, Akhmediev and Peregrine solitons of the scalar focusing NLS equation, respectively.}
    \label{f:zeta}
\end{figure}

Similarly to the scalar focusing NLS equation with NZBC,
there exist four kinds of soliton solutions depending on the location of the discrete eigenvalue, namely (cf.\ Fig.~\ref{f:zeta}):
1. Traveling solitons ($|\zeta|>k_o$ with $\Re\zeta\ne0$ and $\Im\zeta>0$);
2. Stationary solitons ($\zeta = iZ$ with $Z>k_o$);
3. Periodic solutions ($\zeta= ik_o\e^{i\alpha}$ with $|\alpha|<\pi/2$); and
4. Rational solutions (corresponding to the limit of stationary solitons as $Z\to1$, or of periodic solitons as $\alpha\to0$).
For future reference we write below the general traveling soliton solution (known as Tajiri-Watanabe soliton) of the focusing NLS equation
with discrete eigenvalue $\zeta=i k_o Z\e^{i\alpha}$,
$Z>1$, $|\alpha|<\pi/2$, and norming constant $\gamma=\xi\e^{i\varphi}$,
as given in \cite{bk2014}:
\[
\label{e:qtw}
q_{\tw}(x,t) = k_o\e^{-2 i \alpha }
\frac{\cosh (\chi + 2i \alpha) + (c_{+,2} K_s - i c_{-,2} K_c)\,d}
{\cosh\chi + 2 K_sd}\,,
\]
where
\bse
\begin{gather}
K_s(x,t) = Z^2 \sin (s+2 \alpha)-\sin s\,,\\
K_c(x,t) = Z^2 \cos (s+2 \alpha)- \cos s\,,\\
\chi(x,t) = k_o c_{-,1} x\cos\alpha - k_o^2 c_{+,2} t\sin2\alpha + \log \frac{2 c_o \cos\alpha}{c_{+,1} \xi}\,,\\
s(x,t) = k_o c_{+,1} x\sin\alpha + k_o^2 c_{-,2} t\cos2\alpha -\varphi\,,\\
c_{\pm,n} = Z^n\pm 1/Z^n \qquad n=1,2\,, \\
c_o = |1-\e^{-2i\alpha}Z^2|\,,\qquad
d = \cos\alpha/(c_o c_{+,1})\,.
\end{gather}
\ese
The other three kinds of solitons are special cases of Eq.~\eqref{e:qtw} when
the discrete eigenvalue~$\zeta$ is taken as in Fig.~\ref{f:zeta}.
Note, however, that for rational solutions
a suitable limiting procedure and rescaling of the norming constant are necessary
to obtain nontrivial solutions, cf.~\cite{bk2014} for the scalar case.
The generalization of this procedure for the spinor model is discussed in Section~IV.

\begin{table}[t!]
\medskip
        \begin{tabular}{|c|c|c|c|c|c|}
\hline
            Classes& A & B & C & D & E\\
            \hline
            & & & & & \\[-2ex]
            $\Gamma$ &
            $\begin{bmatrix} \gamma_1 & 0 \\ 0 & \gamma_{-1} \end{bmatrix}$&
            $\begin{bmatrix} \gamma & 0 \\ 0 & 0 \end{bmatrix}$&
            $\begin{bmatrix} 0 & \gamma \\ 0 & 0 \end{bmatrix}$&
            $\begin{bmatrix} \gamma_1 & \gamma_0 \\ 0 & 0 \end{bmatrix}$&
\rule[-3.4ex]{0ex}{0ex}
            $\begin{bmatrix} \gamma_1 & \gamma_0 \\ 0 & \gamma_{-1} \end{bmatrix}$ \\
\hline
        \end{tabular}
    \caption{The five equivalence classes for the Schur form of the norming constant $K$.
        In Appendix~II we show that any norming constant can be reduced to one of the above classes by a transformation $K = U \Gamma U^\dagger$, where
        $U$ is a unitary matrix.
        Explicit expressions for each class of norming constant $K$ and corresponding unitary matrix $U$ are also provided in Appendix~II.
    }
    \label{t:norming}
\end{table}

Another crucial difference from the case of ZBC discussed in Section III.a is that
in general Takagi's factorization does not diagonalize the solution in the case of NZBC.
The reason is twofold.
On one hand, in general the transformation does not preserve the BC~\eqref{e:NZBC}.
That is,
the multiplication by $U$ from the left and $U^T$ from the right, which diagonalizes $K$,
changes the BC in Eq.~\eqref{e:NZBC} into $\Phi_+ = k_o UU^T$, which is not necessarily
proportional to the identity, or even diagonal.
On the other hand, the matrix $U$ in Takagi's factorization for $K$ does not diagonalize $K^\dagger$ in general.
Therefore, in general the last two terms in Eq.~\eqref{e:1soliton_nzbc} cannot be diagonalized simultaneously,
which means that the solution of the spinor model cannot always be decomposed into a simple combination of scalar solutions.

In order to study soliton solutions with NZBC
we instead find it more convenient to use the Schur decomposition~\cite{ZAMM:ZAMM19870670330} to express the norming constant in the simplest possible form.
The Schur decomposition theorem ensures that,
for any matrix $K$,
there exists a unitary matrix $U$ such that
\vspace*{-0.6ex}
\[
\label{e:Schur}
K=U \Gamma U^\dagger \,,
\]
where $\Gamma$ is an upper triangular complex-valued matrix,
called the \textit{Schur form} of the matrix $K$.
As shown in Table~\ref{t:norming} and in Appendix~II,
all (in our case, symmetric) norming constants $K$ can be divided into
five equivalence classes (here labeled classes A--E), depending on the structure of their Schur forms
(i.e., depending on whether $K$ is diagonalizable or not, and whether none, one or both of its eigenvalues are zero).
Families of norming constants and corresponding unitary transformations are shown in Appendix~II.
(Similarly to our classification of solitons with ZBC,
we ignore the trivial case in which $\Gamma$ is the zero matrix.)

Note that the Schur form $\Gamma$ of a matrix is not unique.
(For example, one can switch the two diagonal entries of $\Gamma$
and/or change the complex phase of the off-diagonal entry
via an additional unitary similarity transformation.)
Nonetheless, the \textit{structure} of $\Gamma$ \textit{is} unique.
Indeed, since both the trace and the determinant of a matrix are invariant under similarity transformations by a unitary matrix,
and since the different Schur forms can be uniquely distinguished in terms of the trace and determinant of the matrix $\Gamma$,
it follows that
norming constants belonging to different Schur classes
are not related by unitary similarity transformations.

Importantly, the above discussion implies that
\textit{soliton solutions obtained from norming constants in different Schur classes
are inequivalent}.
To see why, note that, even though the transformation~\eqref{e:UphiV} that defines equivalence classes of solutions
allows for two unrelated unitary matrices $U$ and $V$,
in order to preserve the BC~\eqref{e:NZBC} one must choose $V = U^\dagger$.
Therefore, in the case of NZBC, all solutions in the same equivalence class are related to each other
via a similarity transformation with a unitary matrix.
Thus, soliton solutions obtained from norming constants belonging to different Schur classes
belong to different equivalence classes.
The converse is also true.
That is, if two one-soliton solutions are not in the same equivalence class,
then the two norming constants are inequivalent as well.

In Section~III.C we show that solutions obtained from classes~A and~B are reducible, i.e., equivalent to a simple combination of scalar solitons,
similarly to the case of ZBC,
whereas those obtained from classes~C--E are irreducible, i.e., not representable in terms of simple combinations of scalar solitons.

Finally, recall that a soliton solution corresponds to a ferromagnetic or polar state depending on whether $\det K=0$ or $\det K\ne0$,
respectively.
Thus, if the solution describes a ferromagnetic state
(i.e., one of the eigenvalues of the norming constant is zero),
then it belongs to one of the equivalence classes B--D.
Conversely, if the solution describes a polar state
(i.e., both eigenvalues of the norming constant are non-zero),
then the it belongs to either class~A or class~E.

\subsection{III.C~ Soliton solutions with NZBC: Core components}

Substituting Eq.~\eqref{e:Schur} into Eq.~\eref{e:1soliton_nzbc},
we have
\[
\label{e:PhiQ}
\Phi(x,t) = U Q(x,t) U^\dagger\,,
\]
where
\[
\label{e:Q_NZBC}
Q(x,t) = k_o^2I_2 - i \tilde X_1\e^{-2i\theta^*}\Gamma^\dagger
+ i k_o^2 \tilde X_2 \e^{2i\theta}\Gamma/\zeta^2\,,
\]
and $\tilde X_j$ solve the following linear system
\begin{gather*}
\tilde X_1 \tilde D = I_2 - i k_o \tilde X_2 \tilde c/\zeta\,,\qquad
\tilde X_2 \tilde D^\dagger = I_2 - i\zeta \tilde X_1 \tilde c^\dagger/k_o\,,
\end{gather*}
with
\[
\nonumber
\tilde c = \frac{\Gamma}{\zeta^*-\zeta}\e^{2i\theta}\,,\quad
\tilde D = I_2 + \frac{ik_o \Gamma^\dagger}{(\zeta^*)^2 + k_o^2} \e^{-2i\theta^*}\,.
\]
Since $\Phi(x,t)$ is obtained from $Q(x,t)$ via the transformation~\eref{e:PhiQ},
we refer to $Q(x,t)$ as the \textit{core soliton component} of the solution $\Phi(x,t)$ of the spinor model.
This definition holds for all classes A--E.

Notice that for classes~A and~B the norming constant $K$ is a normal matrix
(because it is diagonalizable by a unitary similarity transformation).
In these cases, since $\Gamma$ is symmetric and diagonal,
$Q(x,t)$ is also symmetric, and thus is itself a one-soliton solution of the spinor model~\eref{e:spinor_nzbc},
with norming constant $\Gamma$.
Conversely, for classes C--E the norming constant $K$ is not a normal matrix,
(because it is not diagonalizable by a unitary similarity transformation).
In these cases, $\Gamma$ is non-diagonal
and $Q(x,t)$ is not symmetric,
and thus is \textit{not} a soliton solution of the spinor BEC model.
Nonetheless, $\Phi(x,t) = U Q(x,t) U^\dagger$ is always symmetric, and therefore \textit{is} a solution of the spinor BEC model.

Below we study the solutions obtained from each Schur class in Table~\ref{t:norming} separately.
It will be convenient to parametrize the Schur forms as follows:
\bse
\label{e:Schurparametrization}
\begin{gather}
\Gamma_{A} = \begin{pmatrix} \gamma_1 & 0 \\ 0 & \gamma_{-1}
\end{pmatrix},\quad
\Gamma_{B} = \begin{pmatrix} \gamma & 0 \\ 0 & 0
\end{pmatrix},\quad
\Gamma_{C} = \begin{pmatrix} 0 & \gamma \\ 0 & 0
\end{pmatrix},\\
\Gamma_{D} = \begin{pmatrix} \gamma & \gamma \tan(2\eta) \\ 0 & 0
\end{pmatrix},\\
\Gamma_{E} = \begin{pmatrix} \gamma + \e^{i \beta } \gamma_0  \cot (2 \eta) & \gamma_0 \\
0 & \gamma
\end{pmatrix}.
\end{gather}
\ese

\paragraph{\bf Class~A.}

\begin{figure}[t!]
    \centering
    \includegraphics[scale=0.22]{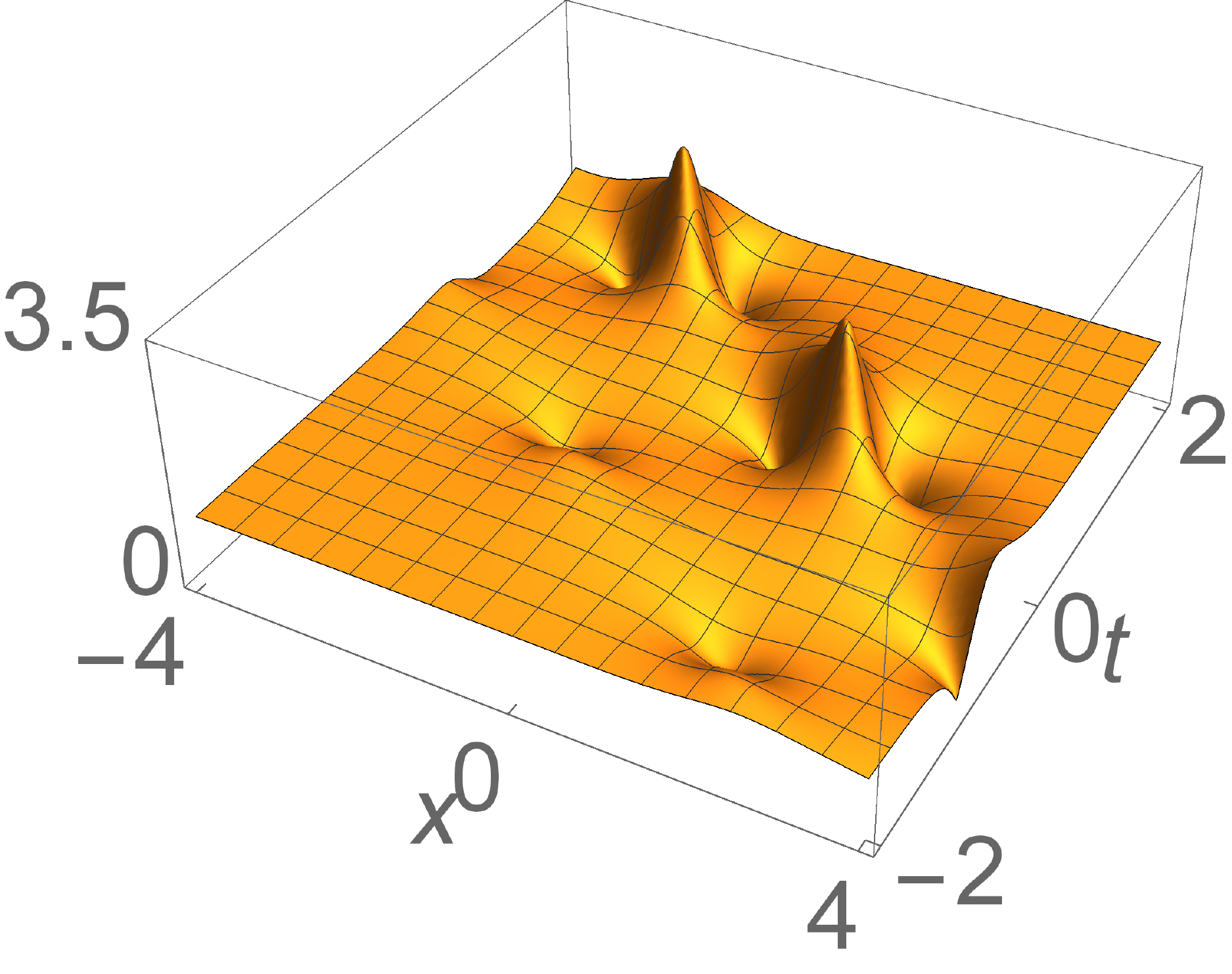}\,
    \includegraphics[scale=0.22]{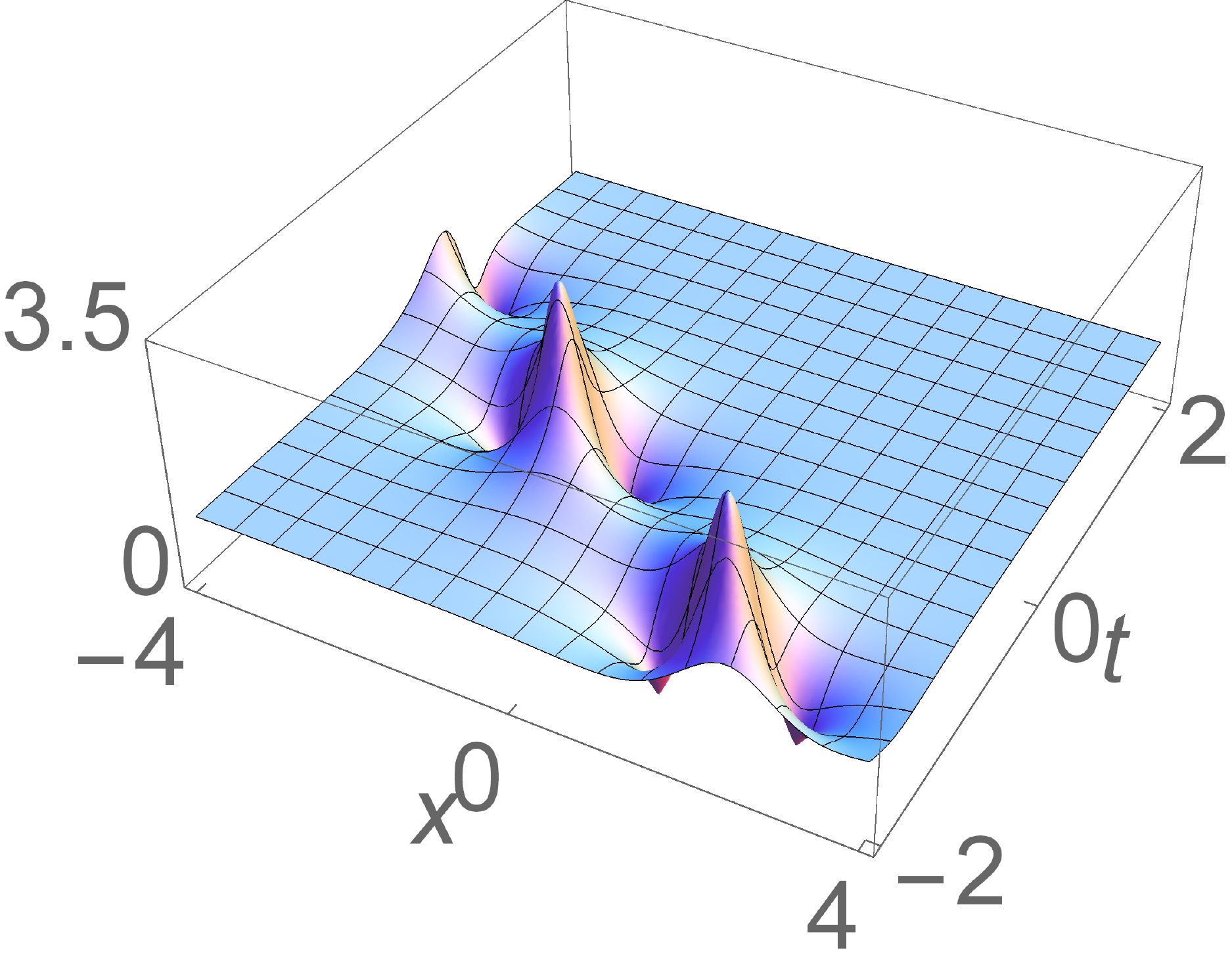}\,\\
    \includegraphics[scale=0.22]{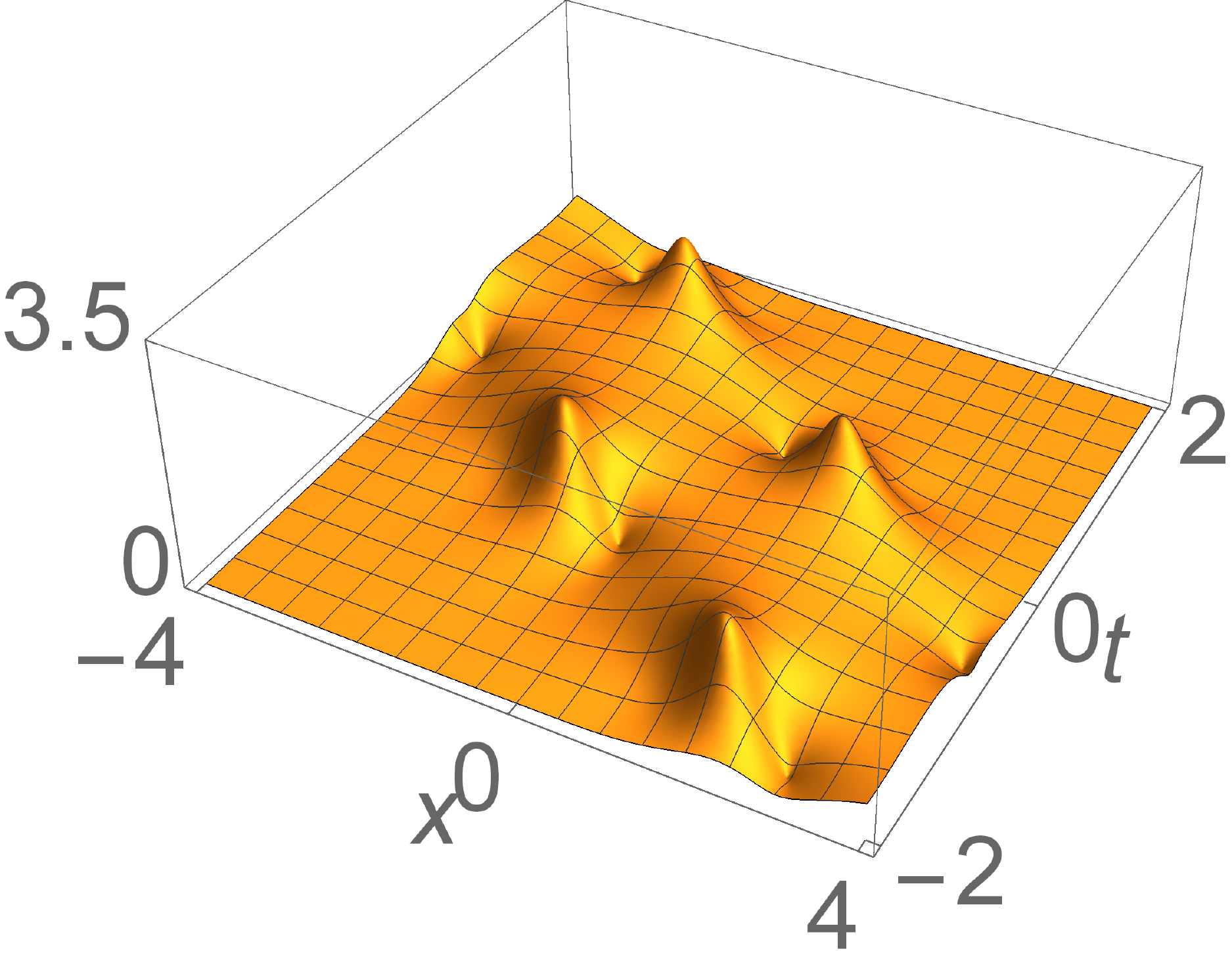}\,
    \includegraphics[scale=0.22]{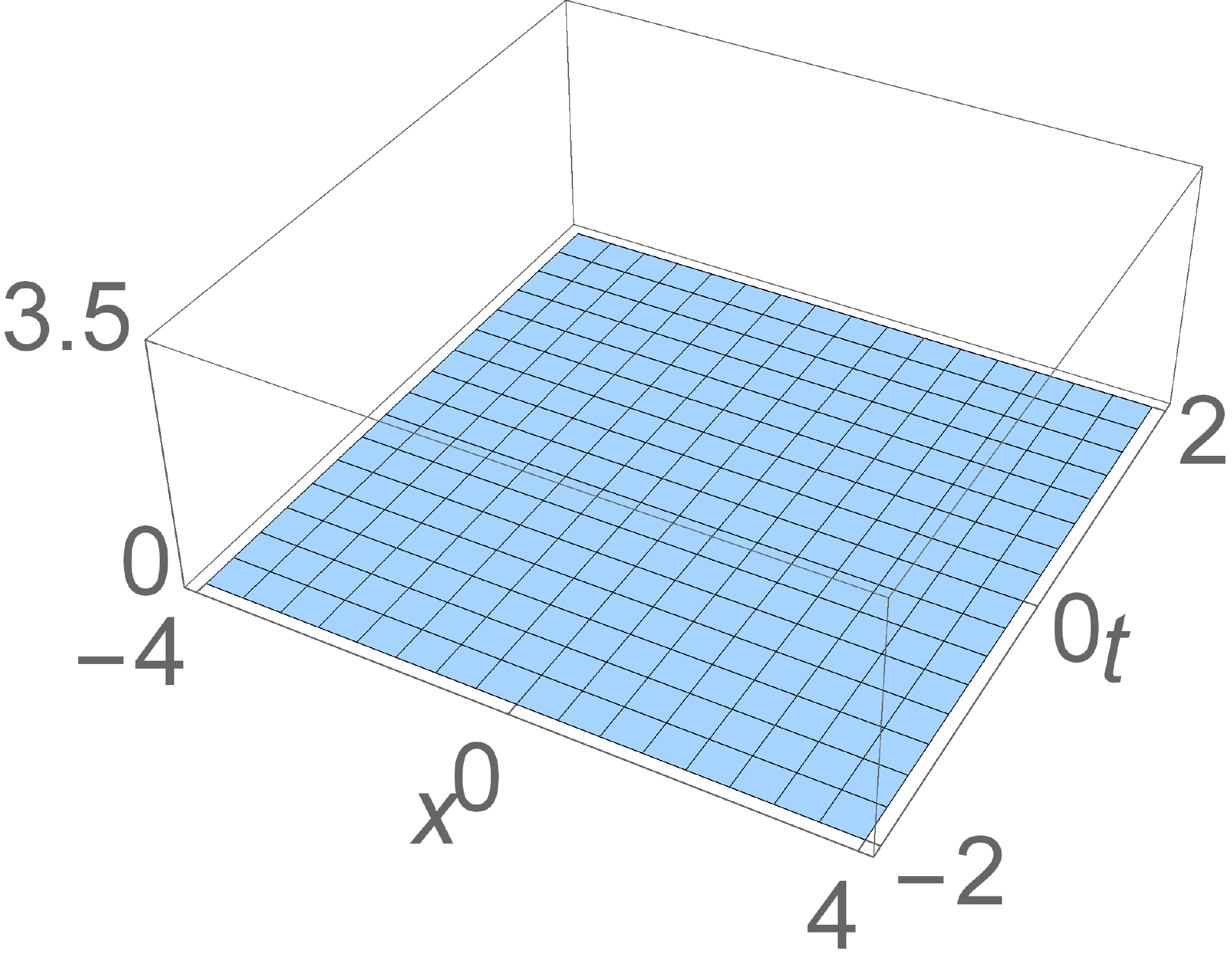}\,\\
    \includegraphics[scale=0.22]{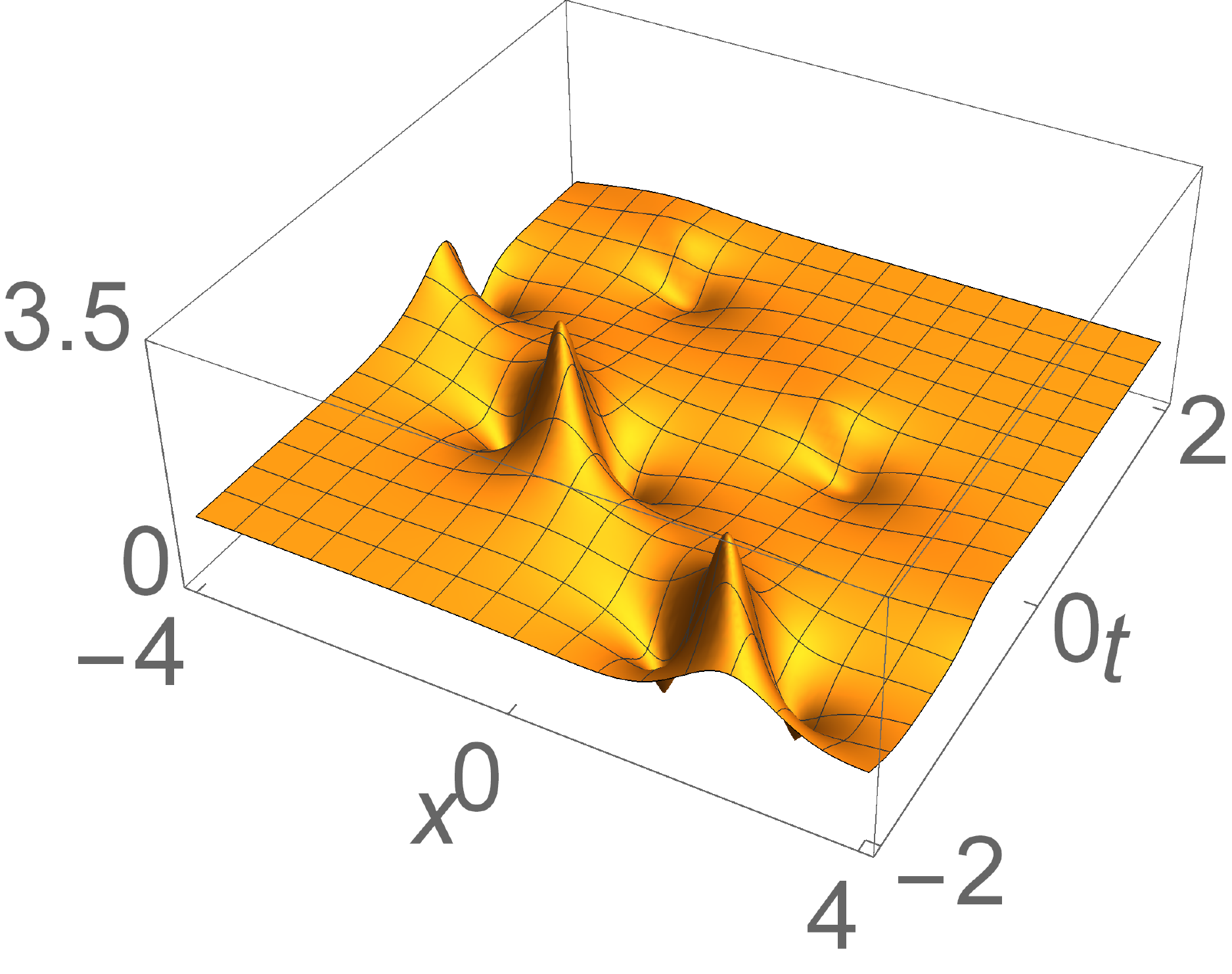}\,
    \includegraphics[scale=0.22]{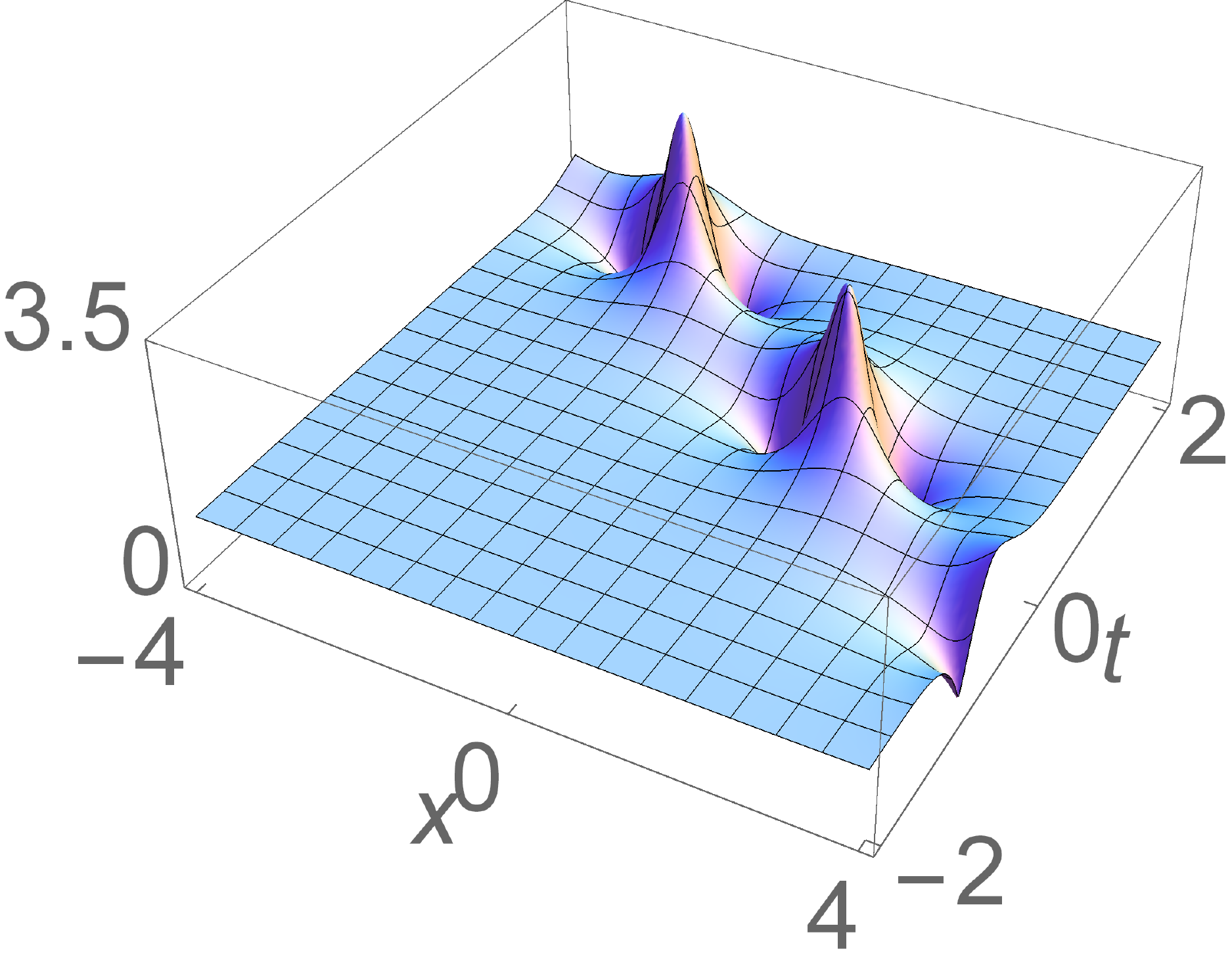}\,
    \caption{
        Amplitudes of a one-soliton solution of the spinor BEC model with NZBC
        with $k_o = 1$, discrete eigenvalue $\zeta = 2i\e^{-i\pi/6}$ and norming constant in class~A.
        Left: entries of $\Phi(x,t)$ from Eq.~\eref{e:PhiA} with $\gamma_1 = \e^{i \pi /3}/10$ and $\gamma_{-1} = 50\,\e^{-i\pi/3}$, and $\eta = \pi/8$.
        From top to bottom: $|\phi_1(x,t)$, $|\phi_0(x,t)|$ and $|\phi_{-1}(x,t)|$.
        Right: entries of the corresponding core soliton component $Q(x,t)$ from Eq.~\eref{e:QA}.
        From top to bottom: $|q_{1,1}(x,t)|$, $|q_{1,2}(x,t)| = |q_{2,1}(x,t)|$ and $|q_{2,2}(x,t)|$.
    }
    \label{f:solitonA}
\end{figure}

Combining $\Gamma_{A}$ in Eq.~\eqref{e:Schurparametrization} and Eq.~\eref{e:Q_NZBC},
simple calculations show that $Q(x,t) = \diag(q_1(x,t),q_{-1}(x,t))$, where,
similarly to the case of ZBC,
each $q_j(x,t)$ solves the scalar NLS equation
\[
\label{e:nls_nzbc}
i q_t + q_{xx} + 2(|q|^2 - k_o^2) q = 0\,,
\]
with the NZBC $|q_j(x,t)|\to k_o$ as $x\to\pm\infty$.
More precisely,
since the asymptotics as $x\to\infty$ of $Q(x,t)$ is fixed by Eq.~\eref{e:NZBC},
it follows that $q_j(x,t)\to k_o$ as $x\to\infty$.
Let $q_{\mathrm{tw},j}(x,t)$ denote the Tajiri-Watanabe (TW) soliton solution~\eqref{e:qtw}
with discrete eigenvalue $\zeta=i k_o Z\e^{i\alpha}$,
and norming constant $\gamma = \gamma_j=\xi_j\e^{i\varphi_j}$ for $j=\pm 1$.
As a consequence of the decoupling provided by the Schur decomposition of the norming constant,
$q_j(x,t) = q_{\mathrm{tw},j}(x,t)$ for $j=\pm1$ are the general one-soliton solutions of the scalar NLS equation.
Thus, the core soliton component $Q(x,t)$ in class~A is diagonal, and it is given by
\[
\label{e:QA}
Q(x,t) = \diag(q_{\mathrm{tw},1}(x,t),q_{\mathrm{tw},-1}(x,t))\,.
\]
As discussed in Appendix~II,
the most general form for the unitary matrix $U_A$ that converts the norming constant into its Schur form,
and hence the general one-soliton soliton into the core soliton component,
in class~A is given by Eq.~\eref{e:UA}.
Therefore, the general solution $\Phi(x,t)$ in class~A is a one-parameter family of transformations of two shifted scalar TW solitons:
\bse
\label{e:PhiA}
\begin{gather}
\phi_1(x,t) = q_{\tw,1}(x,t) \sin^2\eta + q_{\tw,-1}(x,t) \cos^2\eta\,,\\
\phi_0(x,t) = \frac{1}{2}(q_{\tw,1}(x,t) - q_{\tw,-1}(x,t)) \sin2\eta\,,\\
\hspace*{-0.2em}
\phi_{-1}(x,t) = q_{\tw,1}(x,t) \cos^2\eta + q_{\tw,-1}(x,t) \sin^2\eta\,,
\end{gather}
\ese
where $-\pi/2<\eta < \pi/2$.
Note that $\Phi(x,t)$ is determined by a total of eight real parameters via Eq.~\eref{e:PhiA}.
An example of a soliton solution $\Phi(x,t)$ in class~A and the corresponding core soliton component $Q(x,t)$ is shown in Fig.~\ref{f:solitonA}.
It can be seen from Fig.~\ref{f:solitonA}(left), that in general,
the solution~\eref{e:PhiA} exhibits a one-to-one correspondence between potential traps and peaks in different spin states.
In particular, the background develops holes in the two components $\phi_{\pm1}$ corresponding to potential traps.
Such potential traps in turn create peaks in the component $\phi_0$.
The existence of pairs of holes and peaks is reminiscent of the dark-bright soliton complexes considered in Refs.~\cite{pra77p033612,pra84p053630,pla375p642,pra80p023613,2017arXiv170508130B}.
In this sense, the soliton solution~\eref{e:PhiA} exhibits oscillatory dark-bright behavior among the spin states.
This behavior manifests itself only in $\Phi(x,t)$,
and not in the corresponding core component $Q(x,t)$,
which is a direct result of the spin rotation $U$ of the core solutions $Q(x,t)$ from Eq.~\eref{e:PhiQ}.
It is also clear that such oscillatory dark-bright behavior travels with the same velocity of the TW solitons $q_{\tw,\pm1}(x,t)$,
and is fully determined by the discrete eigenvalue $\zeta$.
The frequency of oscillation is also the same of the TW soliton.
Since the TW solitons are well known and has been studied extensively in the past,
the corresponding results can be simply carried over to the soliton solutions in class~A.

\paragraph{\bf Class~B.}

\begin{figure}[t!]
    \centering
    \includegraphics[scale=0.22]{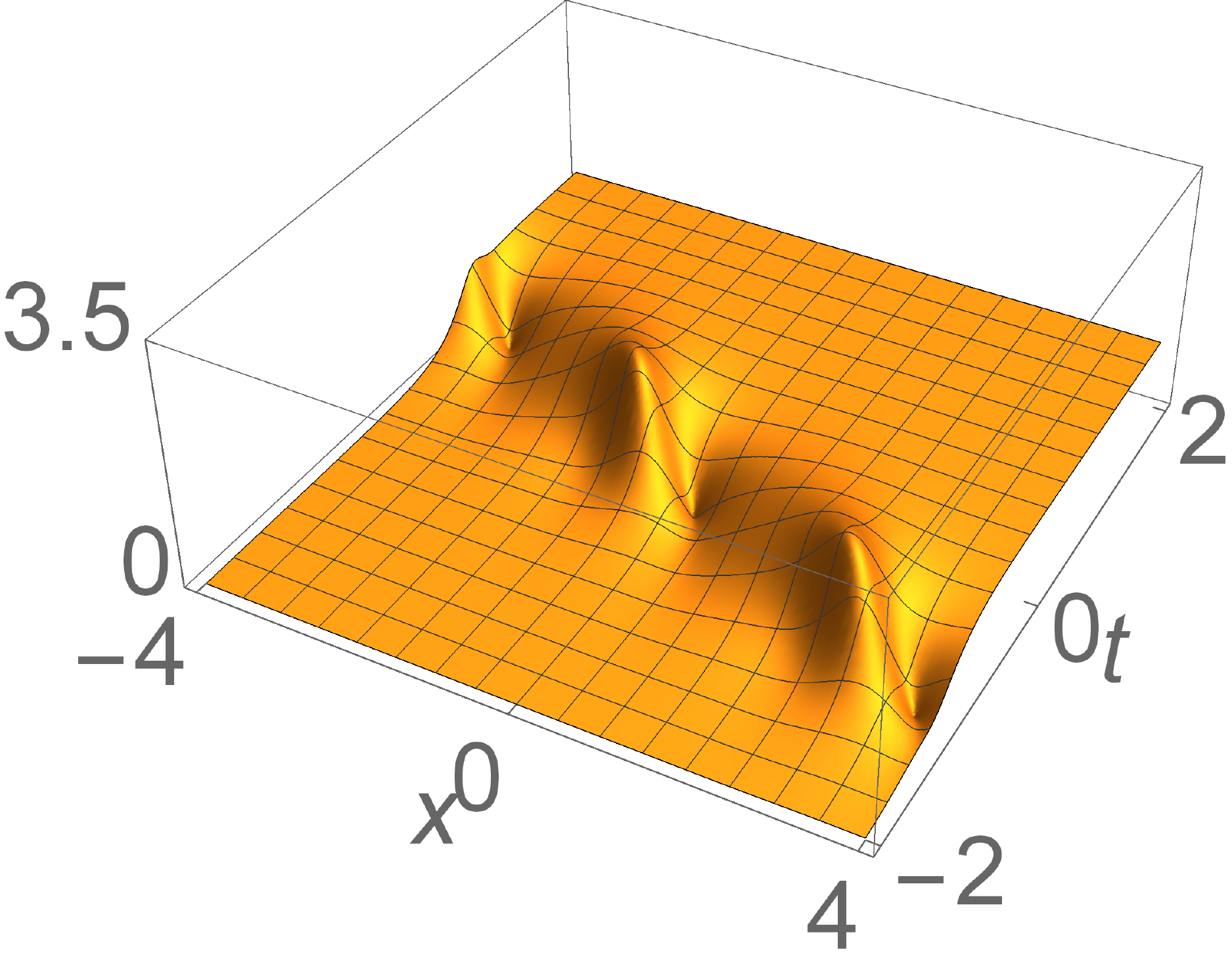}\,
    \includegraphics[scale=0.22]{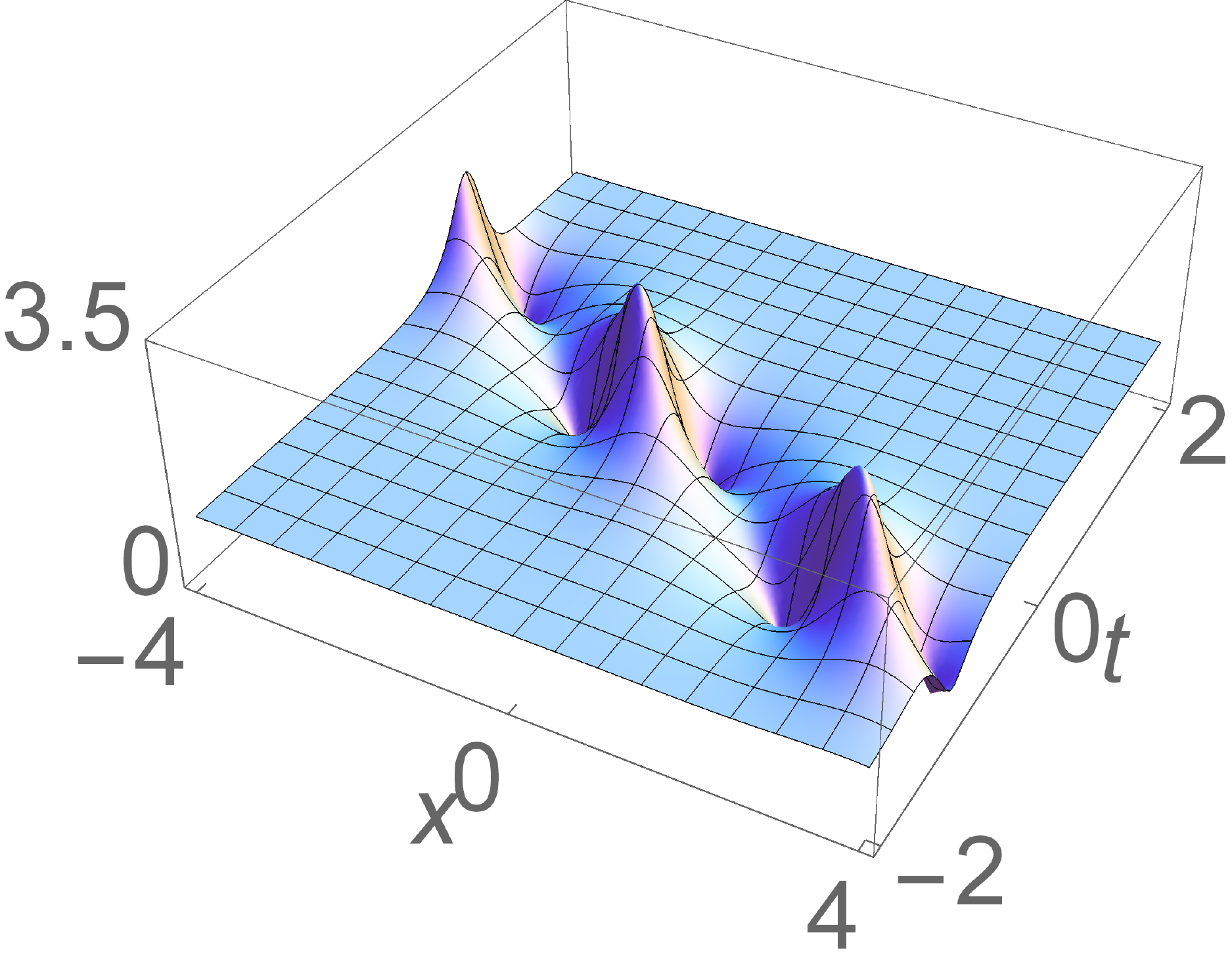}\,\\
    \includegraphics[scale=0.22]{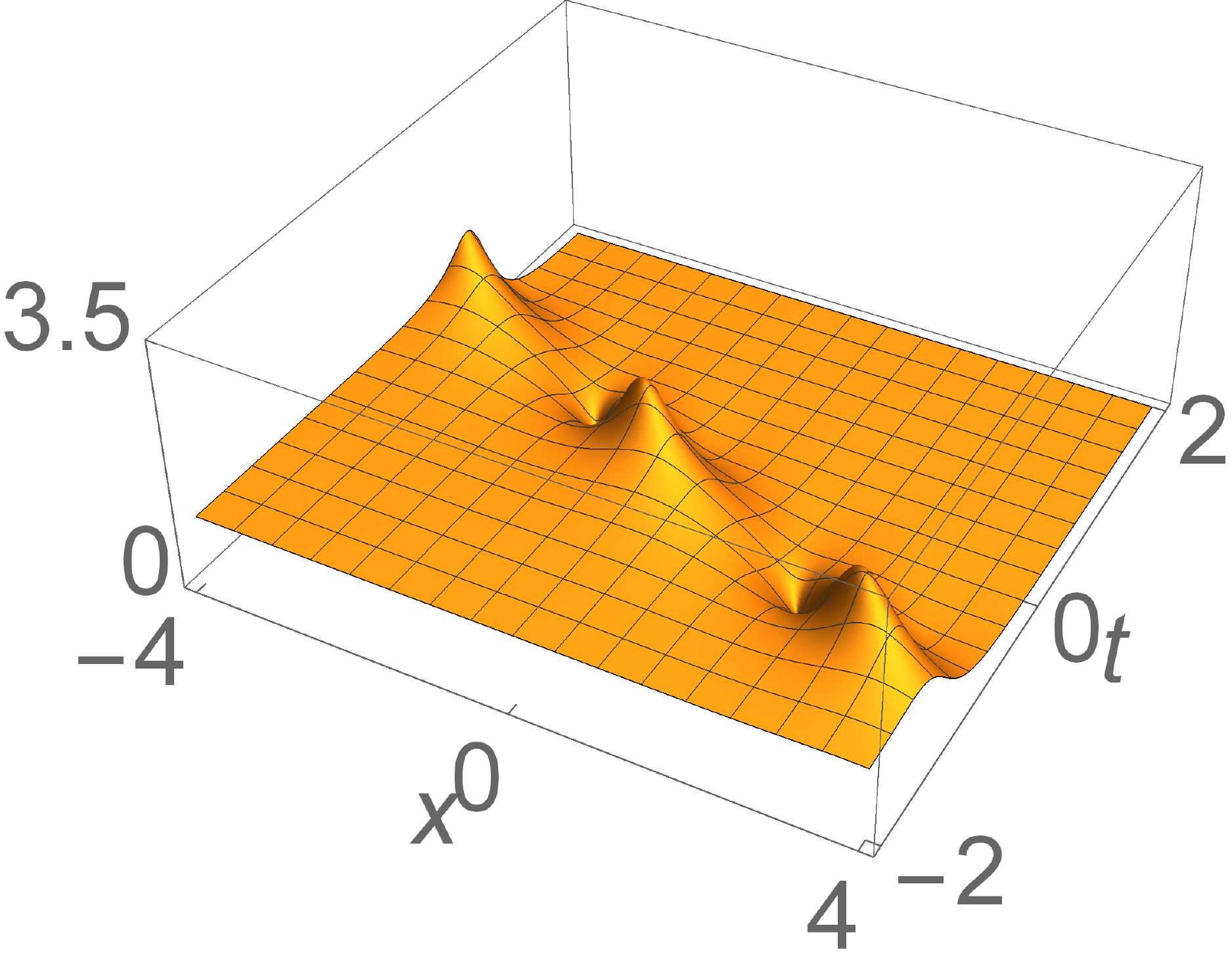}\,
    \includegraphics[scale=0.22]{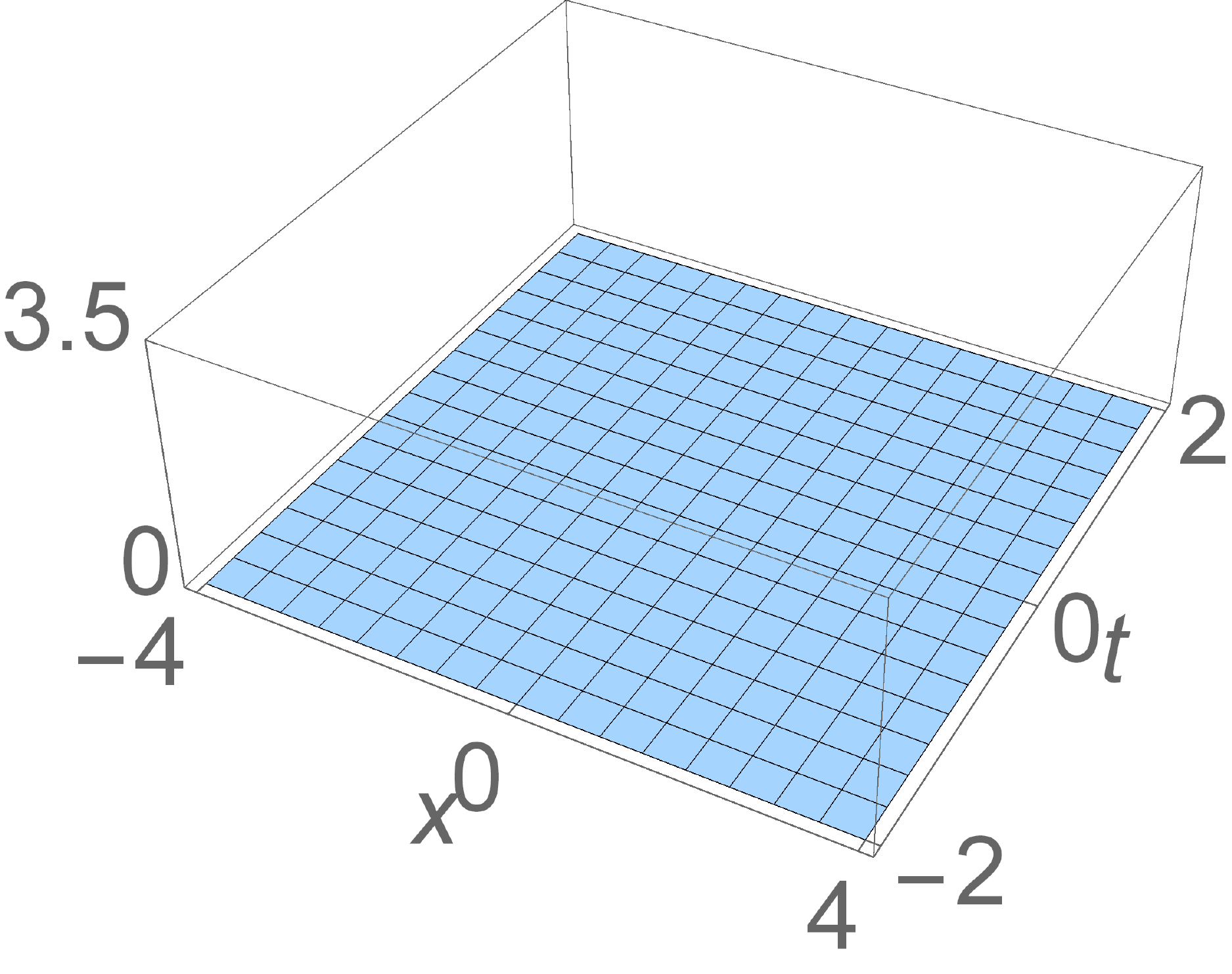}\,\\
    \includegraphics[scale=0.22]{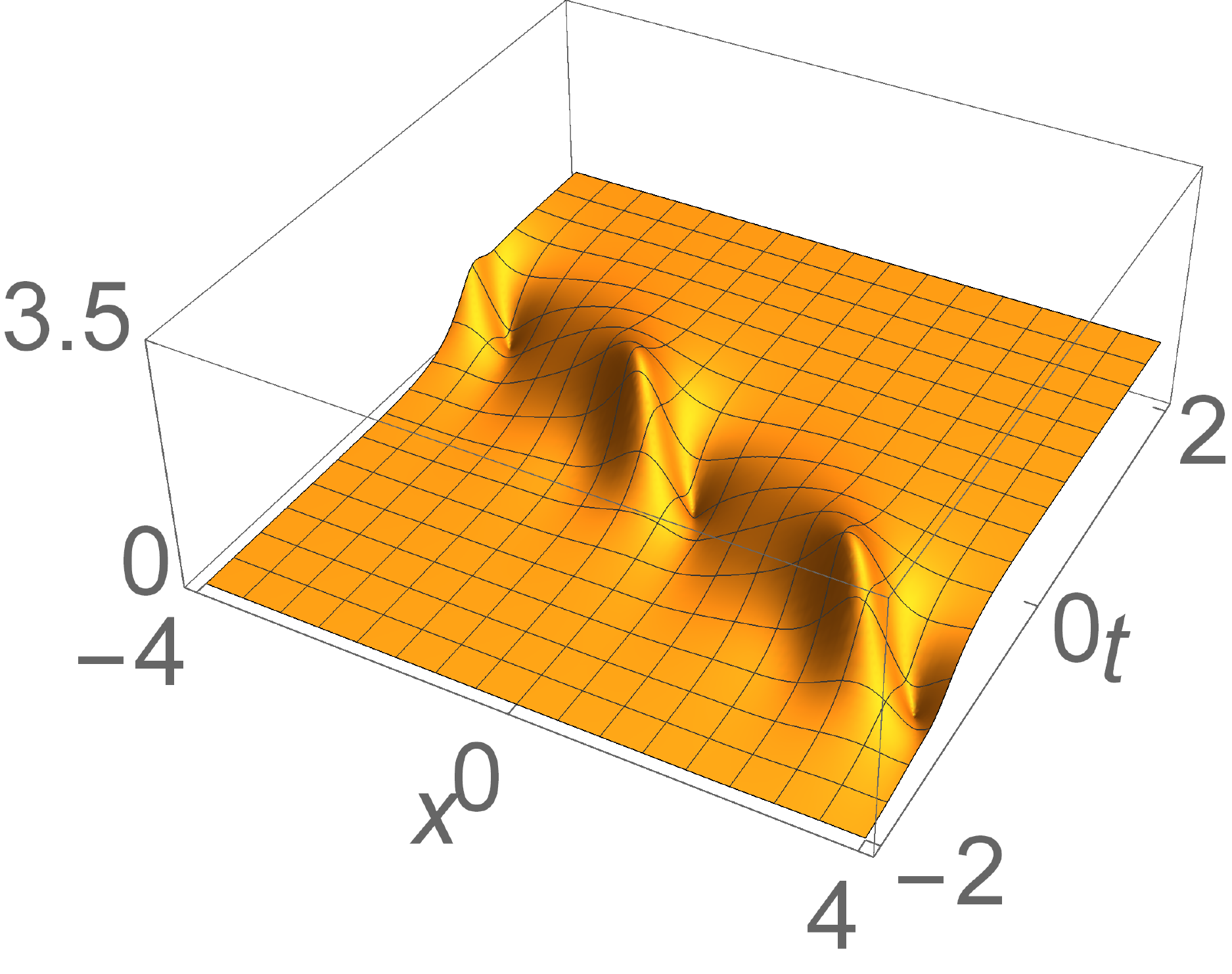}\,
    \includegraphics[scale=0.22]{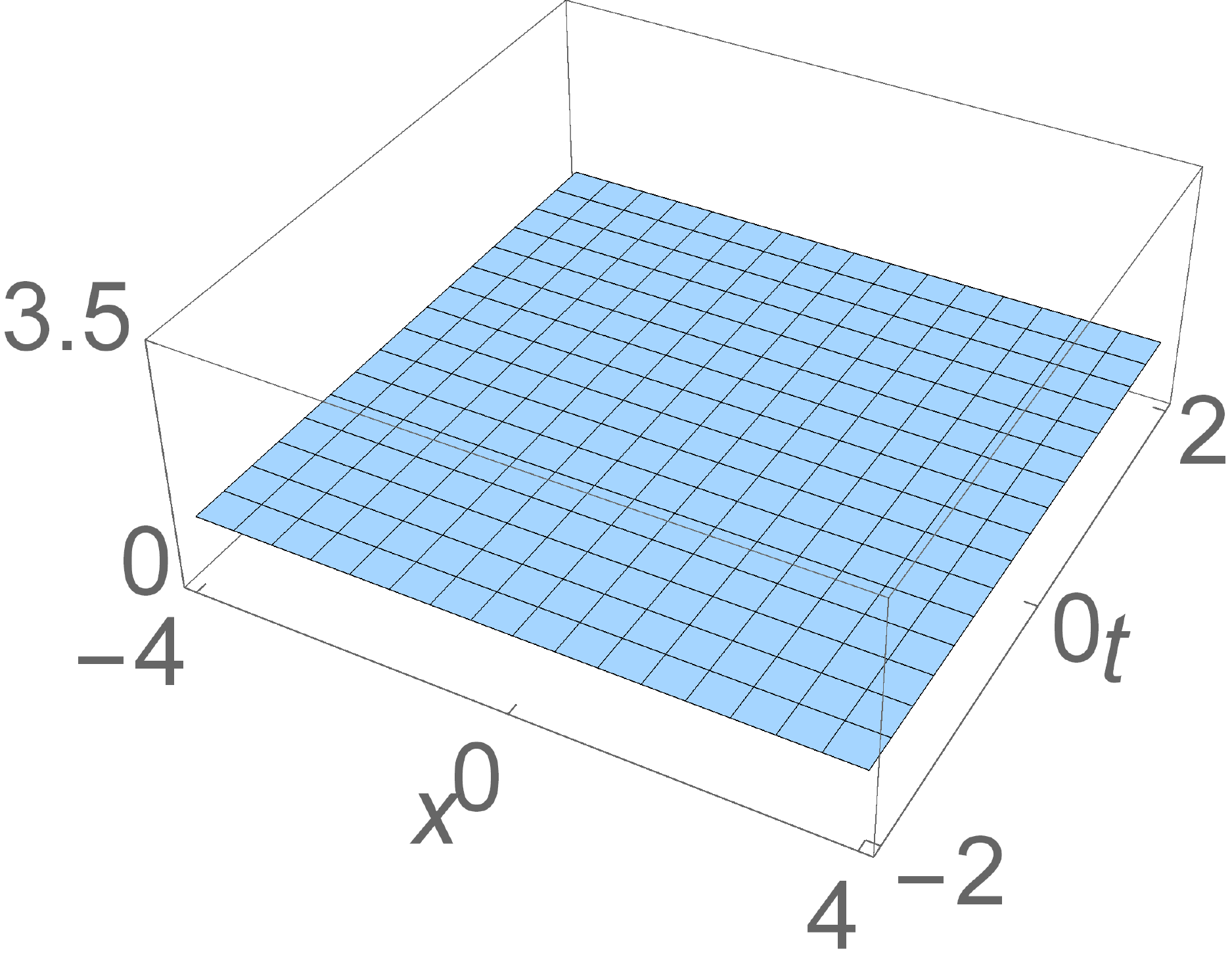}\,
    \caption{
        Same as Fig.~\ref{f:solitonA},
        but for a one-soliton solution in class~B and discrete eigenvalue $\zeta = 2i\e^{-\pi/4}$.
        Left: the soliton solution from Eq.~\eref{e:PhiB} with $\gamma = \e^{i\pi/3}$ and $\eta = \pi/4$.
        Right: the corresponding core soliton component from Eq.~\eref{e:QB}.
    }
    \label{f:solitonB}
\end{figure}

Combining $\Gamma_B$ in Eq.~\eqref{e:Schurparametrization} with Eq.~\eref{e:Q_NZBC},
simple calculations show that $Q(x,t) = \diag(q_1(x,t),k_o)$.
Similarly to class~A,
$q_1(x,t)$ satisfies the scalar NLS equation~\eref{e:nls_nzbc} with NZBC $q_1(x,t)\to k_o$ as $x\to\pm\infty$.
Thus, we have $q_1(x,t) = q_{\tw}(x,t)$ from Eq.~\eref{e:qtw}.
The core soliton component is given by
\vspace*{-0.4ex}
\[
\label{e:QB}
Q(x,t) = \diag(q_{\tw}(x,t),k_o)\,.
\]
The general one-soliton solution $\Phi(x,t)$ in class~B is defined by
a one-parameter family of transformations that couple a scalar TW soliton~\eref{e:qtw} and the non-zero background $k_o$,
namely
\bse
\label{e:PhiB}
\begin{gather}
\phi_1(x,t) = q_{\tw}(x,t) \sin^2\eta + k_o \cos^2\eta\,,\\
\phi_0(x,t) = \frac{1}{2}(q_{\tw}(x,t) - k_o) \sin2\eta\,,\\
\phi_{-1}(x,t) =  q_{\tw}(x,t) \cos^2\eta + k_o \sin^2 \eta\,,
\end{gather}
\ese
where $-\pi/2< \eta < \pi/2$.
Notice the family of solutions in Eq.~\eref{e:PhiB} depend on six real parameters.
An example of a soliton solution $\Phi(x,t)$ in class~B and the corresponding
core soliton component $Q(x,t)$ is shown in Fig.~\ref{f:solitonB}.
With a nontrivial parameter $\eta$, this solution forms a domain wall.
In particular, the location of the wall coincides with the location of the TW soliton $q_\tw$,
and the velocity of the wall coincides with the soliton velocity.
Hence the properties of the wall are encoded into $q_{\tw}(x,t)$ by means of the discrete eigenvalue $\zeta$ and the norming constant $\gamma$.

\paragraph{\bf Class~C.}

\begin{figure}[b!]
    \centering
    \includegraphics[scale=0.208]{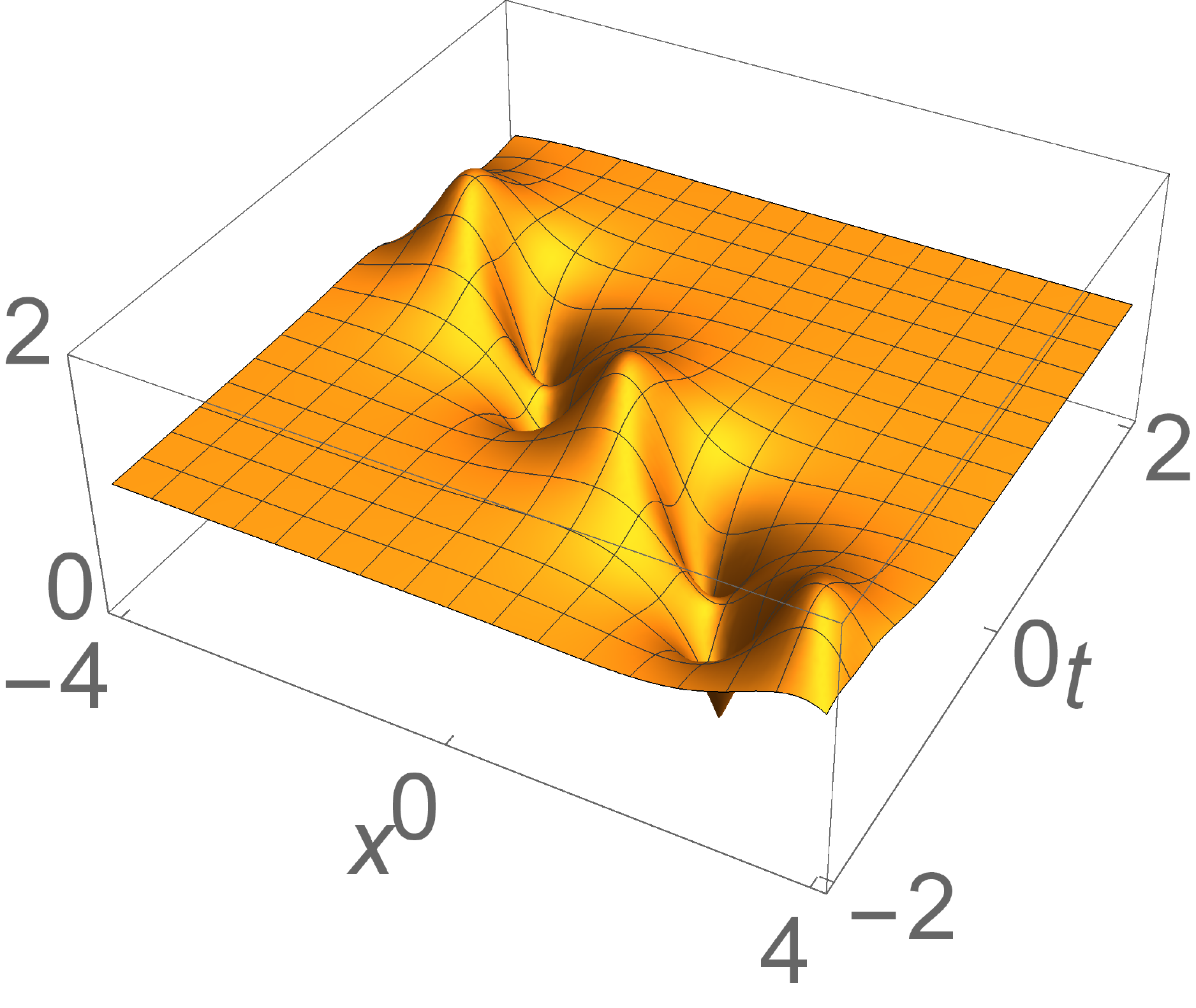}\,
    \includegraphics[scale=0.208]{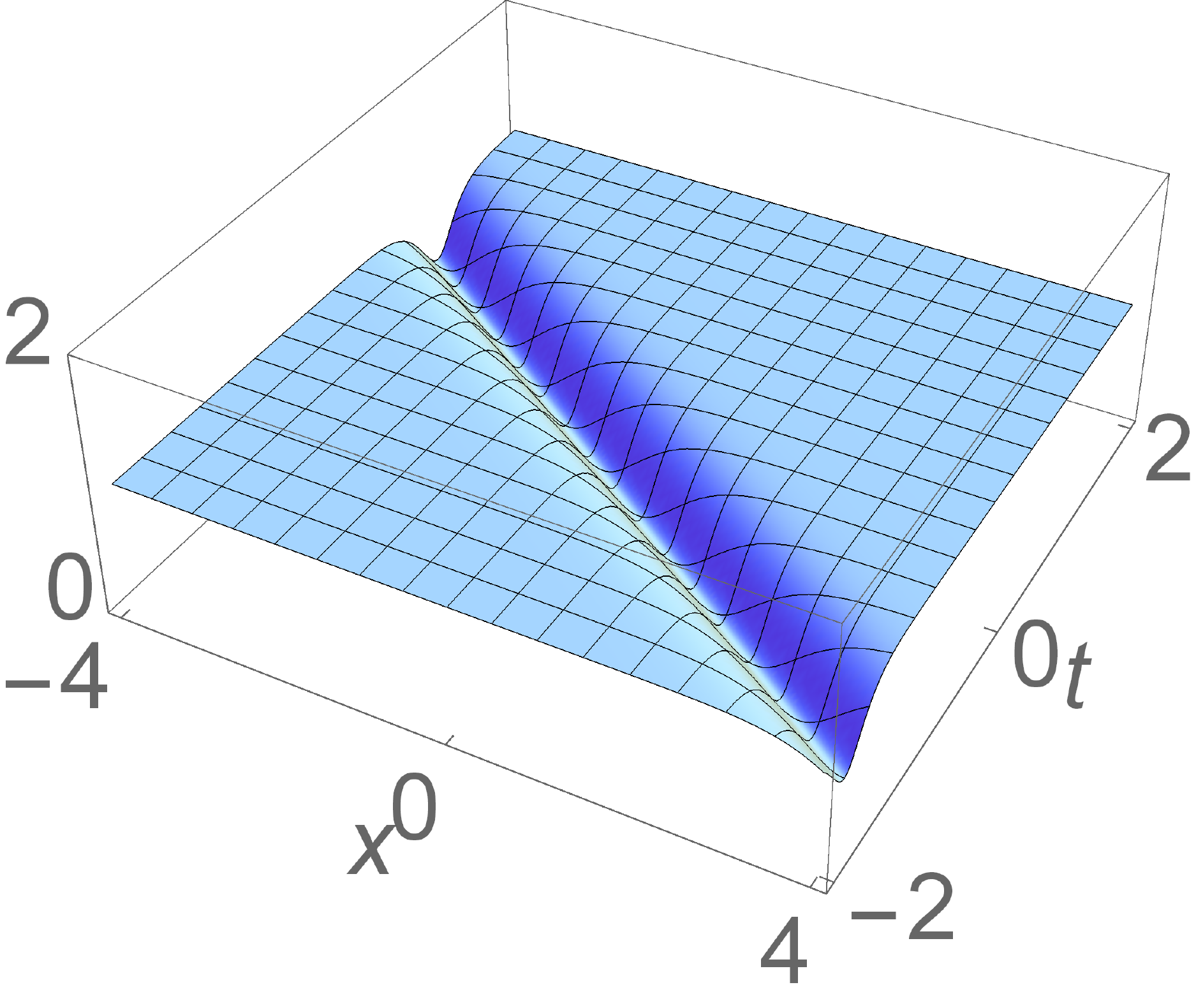}\\
    \includegraphics[scale=0.208]{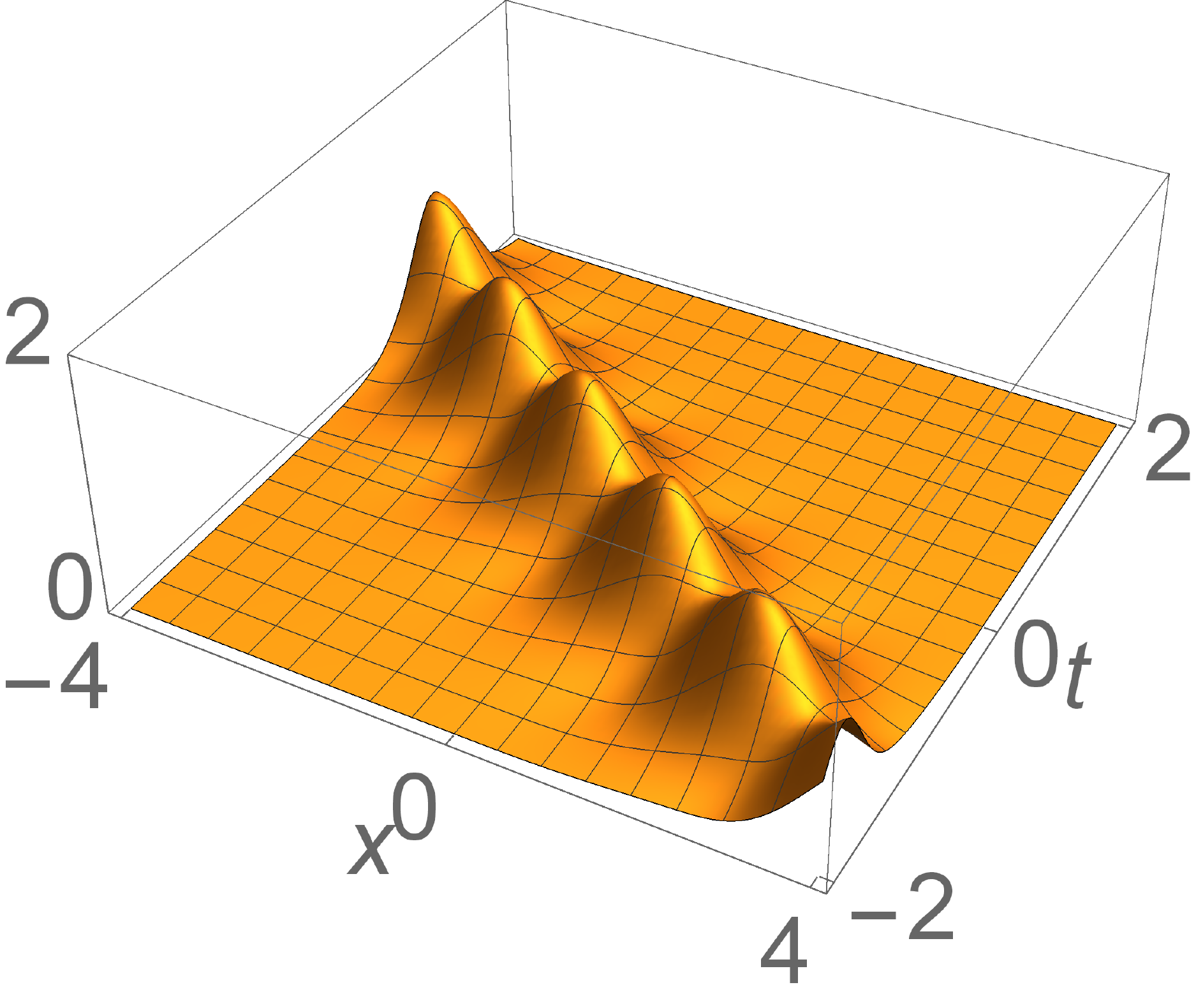}\,
    \includegraphics[scale=0.208]{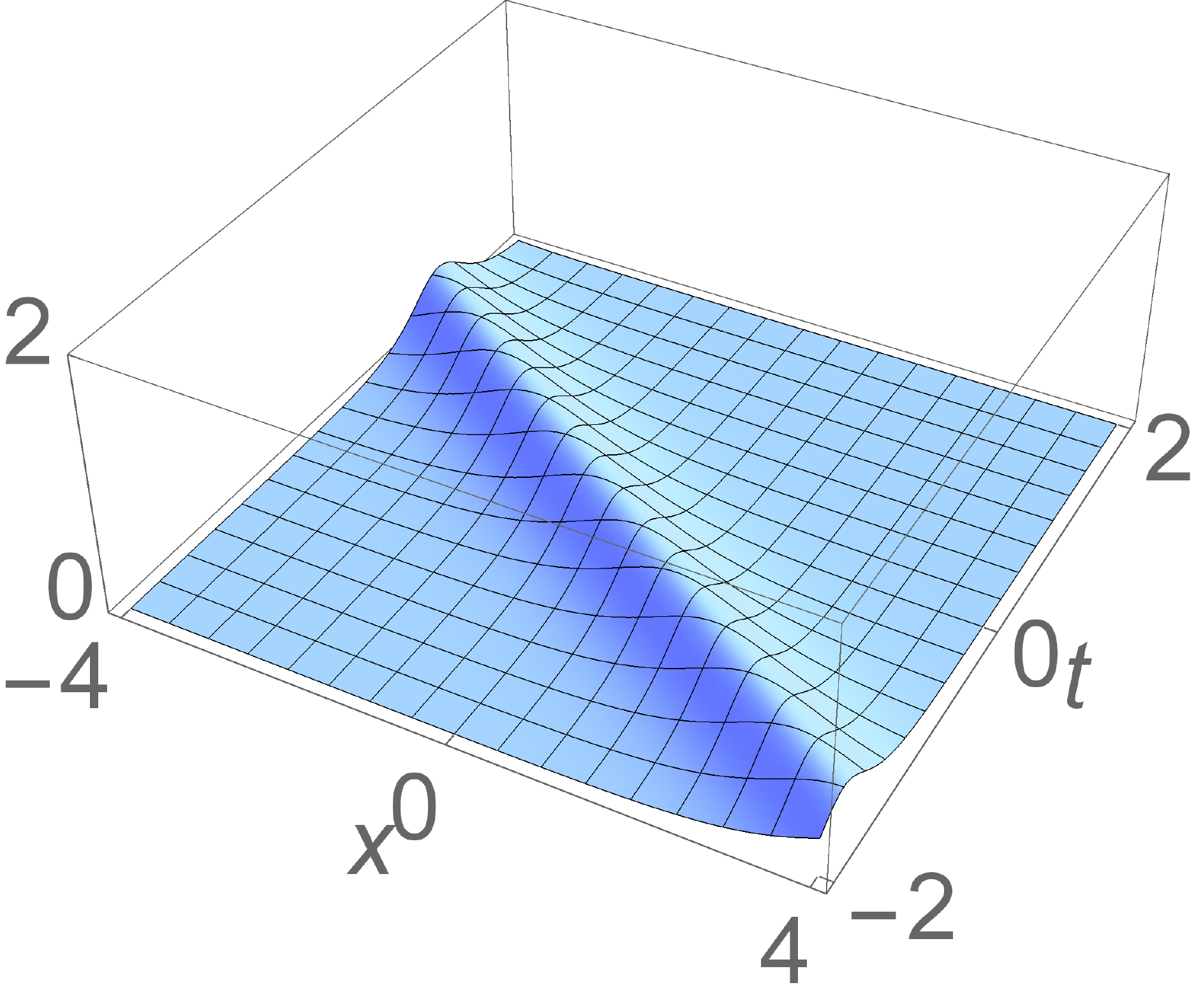}\\
    \includegraphics[scale=0.208]{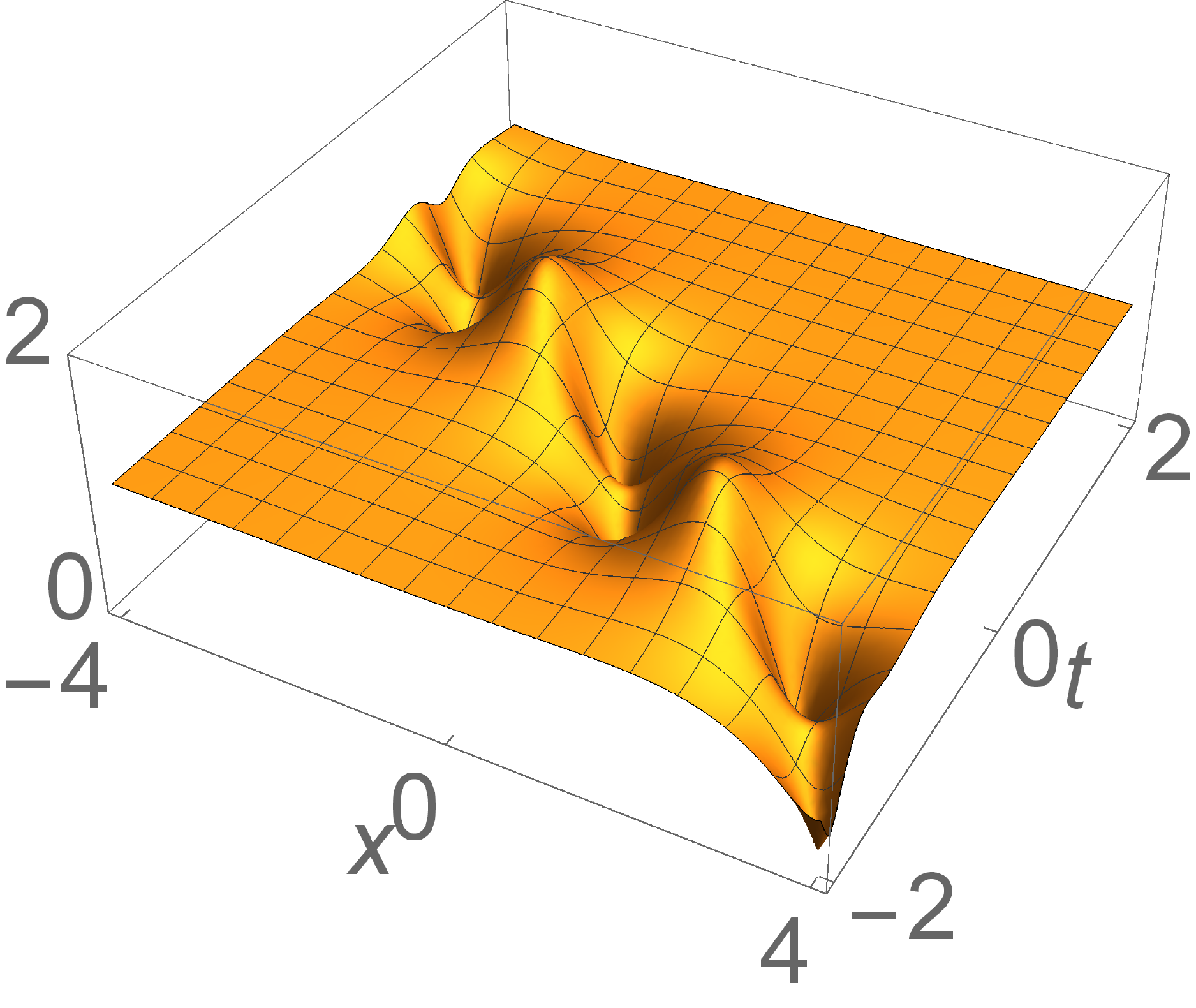}\,
    \includegraphics[scale=0.208]{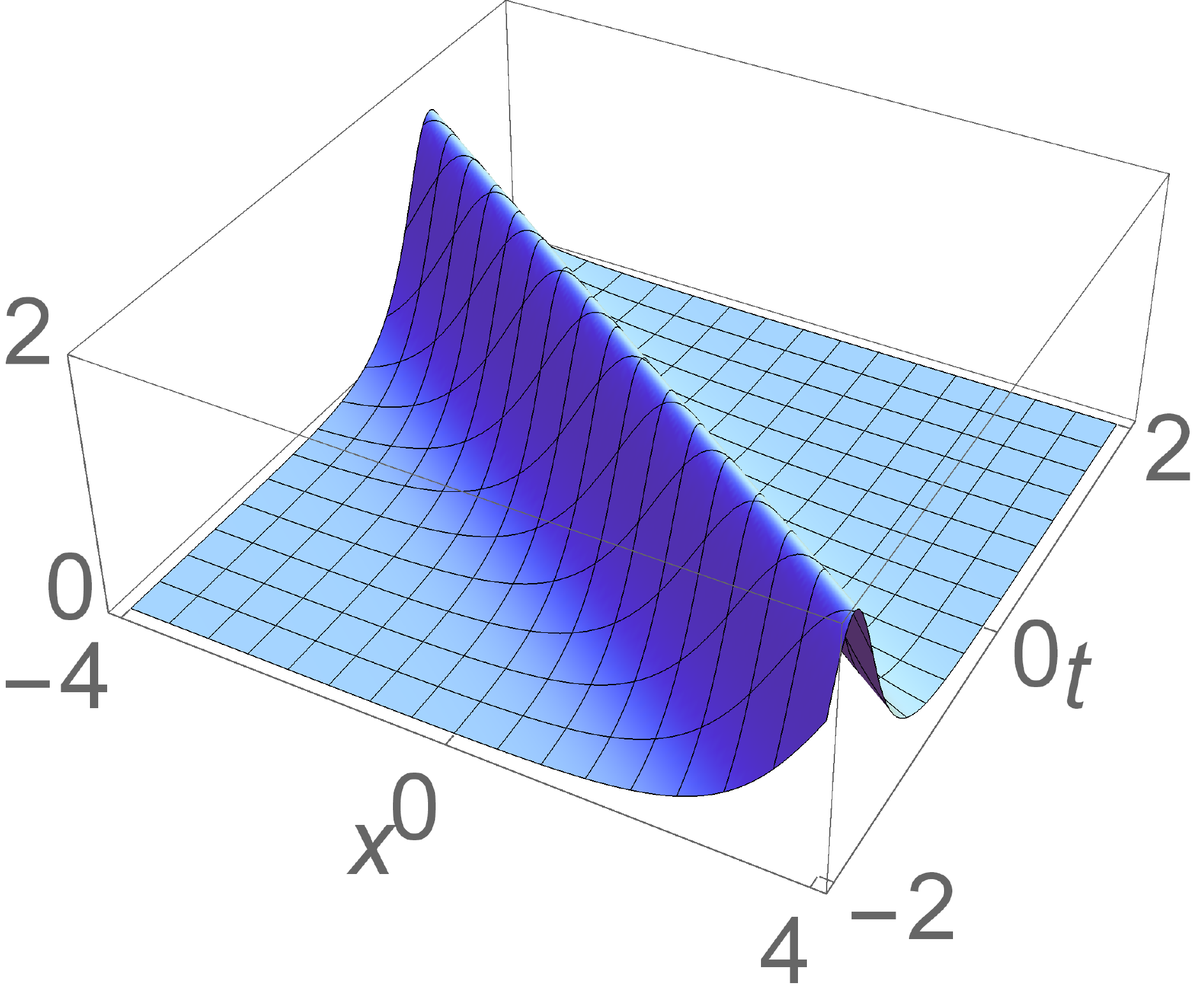}\,
    \caption{
        Amplitudes of a one-soliton solution of the spinor BEC model with NZBC
        with $k_o = 1$, discrete eigenvalue $\zeta = 2i\e^{-i\pi/6}$ and $\gamma = 1$ in class~C.
        Left: the solution $\Phi(x,t)$ with $U_C$ defined in Eq.~\eref{e:KUC}
        and parameters $\beta_1 = -\pi/2$, $\beta_2 = \pi$ and $n = 0$.
        From top to bottom: $|\phi_1(x,t)$, $|\phi_0(x,t)|$ and $|\phi_{-1}(x,t)|$.
        Right: the corresponding core soliton component $Q(x,t)$ from Eq.~\eref{e:QC}, with
        $|q_{1,1}(x,t)| = |q_{2,2}(x,t)|$ (top),
        $|q_{1,2}(x,t)|$ (center) and $|q_{2,1}(x,t)|$ (bottom).
%
    }
    \label{f:solitonC}
\end{figure}

Combining $\Gamma_C$ in Eq.~\eqref{e:Schurparametrization} with Eq.~\eref{e:Q_NZBC},
after some calculations we obtain the core soliton component as
\[
\label{e:Qmatrix}
Q(x,t) = (q_{j,k}(x,t))_{j,k=1,2}\,,
\]
where
\bse
\label{e:QC}
\begin{gather}
q_{1,1}(x,t) = i k_o \e^{-i\alpha} (\sin\alpha - i \cos\alpha \tanh\chi(x,t)) \,,\\
q_{1,2}(x,t) = -i \frac{k_o}{Z} \e^{-i (2 \alpha + s(x,t))} \cos\alpha \,\sech\chi(x,t)\,,\\
q_{2,1}(x,t) = -i k_o Z \e^{i s(x,t)} \cos\alpha \, \sech\chi(x,t)\,,\\
q_{2,2}(x,t) = q_{1,1}(x,t)\,,
\end{gather}
\ese
with
\begin{gather*}
\chi(x,t) = c_{-,1} k_o x \cos\alpha - k_o^2 c_{+,2} t\sin(2\alpha)
\kern8em\nonumber\\
\kern14em {} + \log[(2Zk_o\cos\alpha)/\xi]\,,\\
s(x,t) = c_{+,1} k_o x\sin\alpha + c_{-,2} k_o^2 t \cos2\alpha - \phi\,,
\end{gather*}
and $\gamma = \xi \,\e^{i\phi}$.
In this case $Q(x,t)$ has the form of a dark-bright soliton, similar to those obtained
for the vector focusing NLS equation (i.e., the so-called Manakov system) with NZBC in \cite{KBK15}.
The general one-soliton solution $\Phi(x,t)$ of the spinor model in this case
is given by Eq.~\eref{e:PhiQ} with $Q(x,t)$ given by Eq.~\eqref{e:QC} and $U_C$ given by Eq.~\eref{e:KUC}.
The resulting solution is a superposition of dark and bright solitons, coupled by $U_C$,
which generically produces a breather-type solution
due to out-of-phase oscillations resulting from the off-diagonal entries of $Q(x,t)$.
For brevity, we omit the explicit expressions for the entries of~$\Phi(x,t)$.

A soliton solution $\Phi(x,t)$ in class~C and the corresponding core soliton component $Q(x,t)$
are shown in Fig.~\ref{f:solitonC}.
We reiterate that, unlike $\Phi(x,t)$, $Q(x,t)$ is not symmetric in this case, hence it is not itself a solution of the spinor BEC model.

\paragraph{\bf Class~D.}

\begin{figure}[t!]
    \centering
    \includegraphics[scale=0.22]{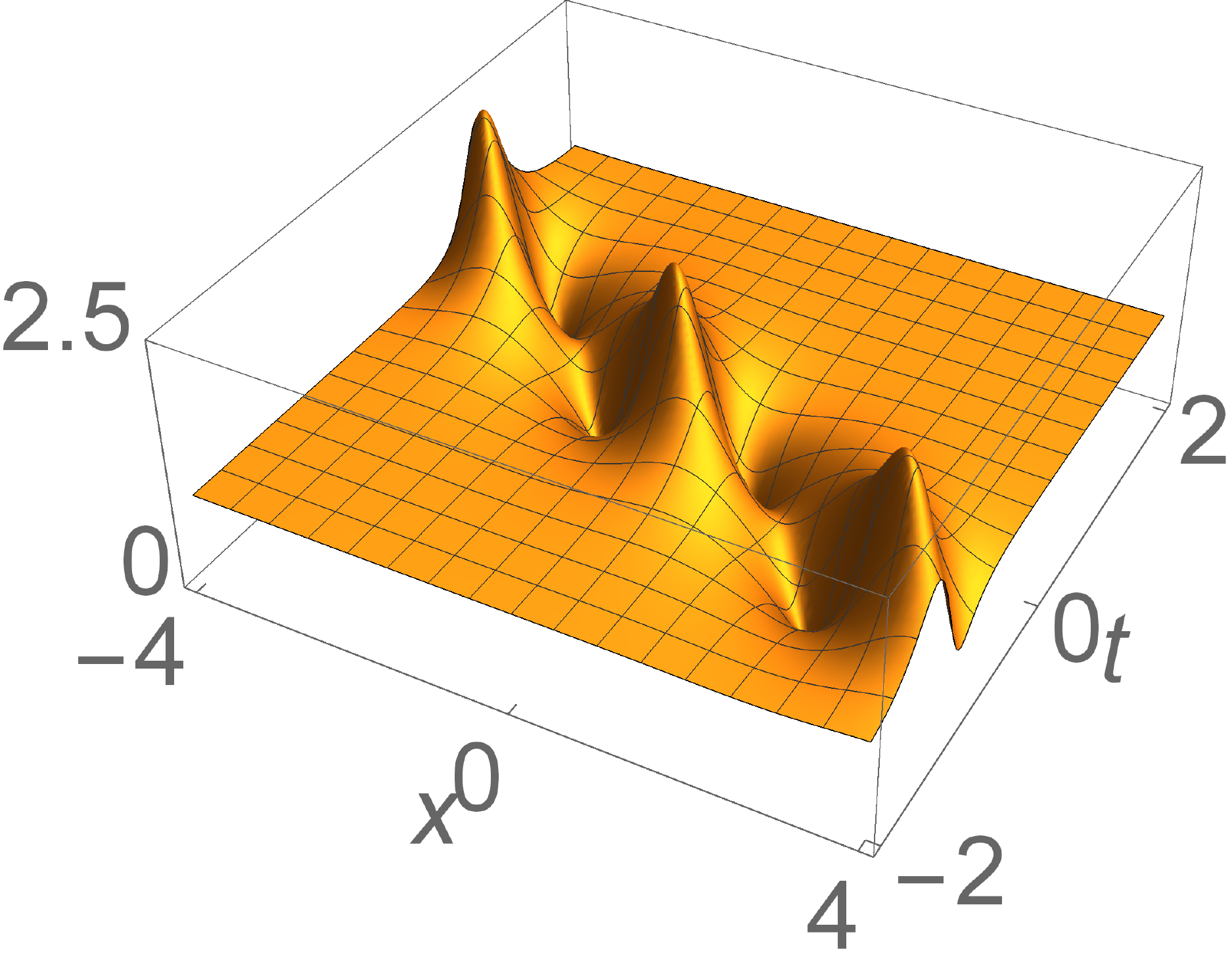}\,
    \includegraphics[scale=0.22]{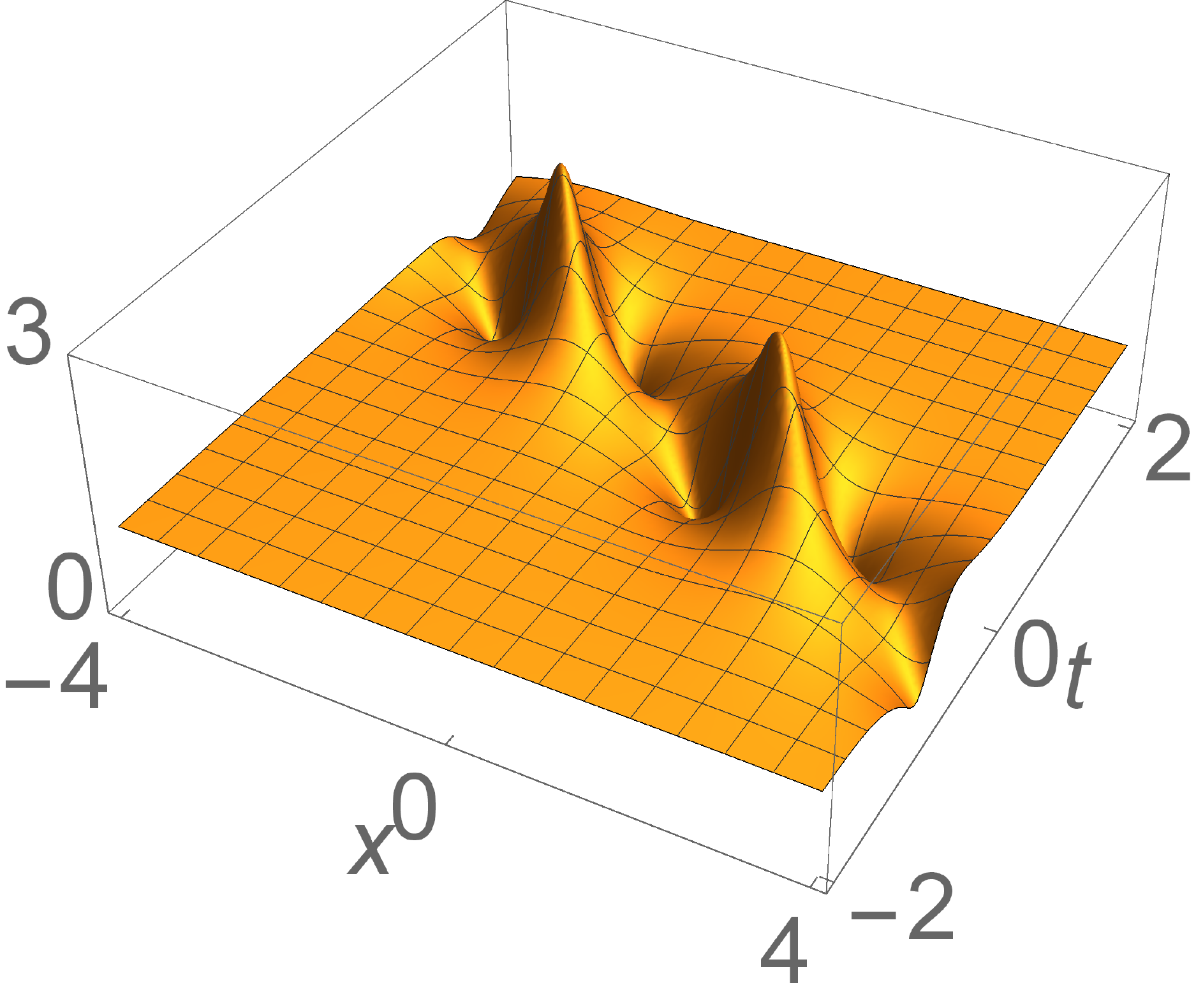}\\
    \includegraphics[scale=0.22]{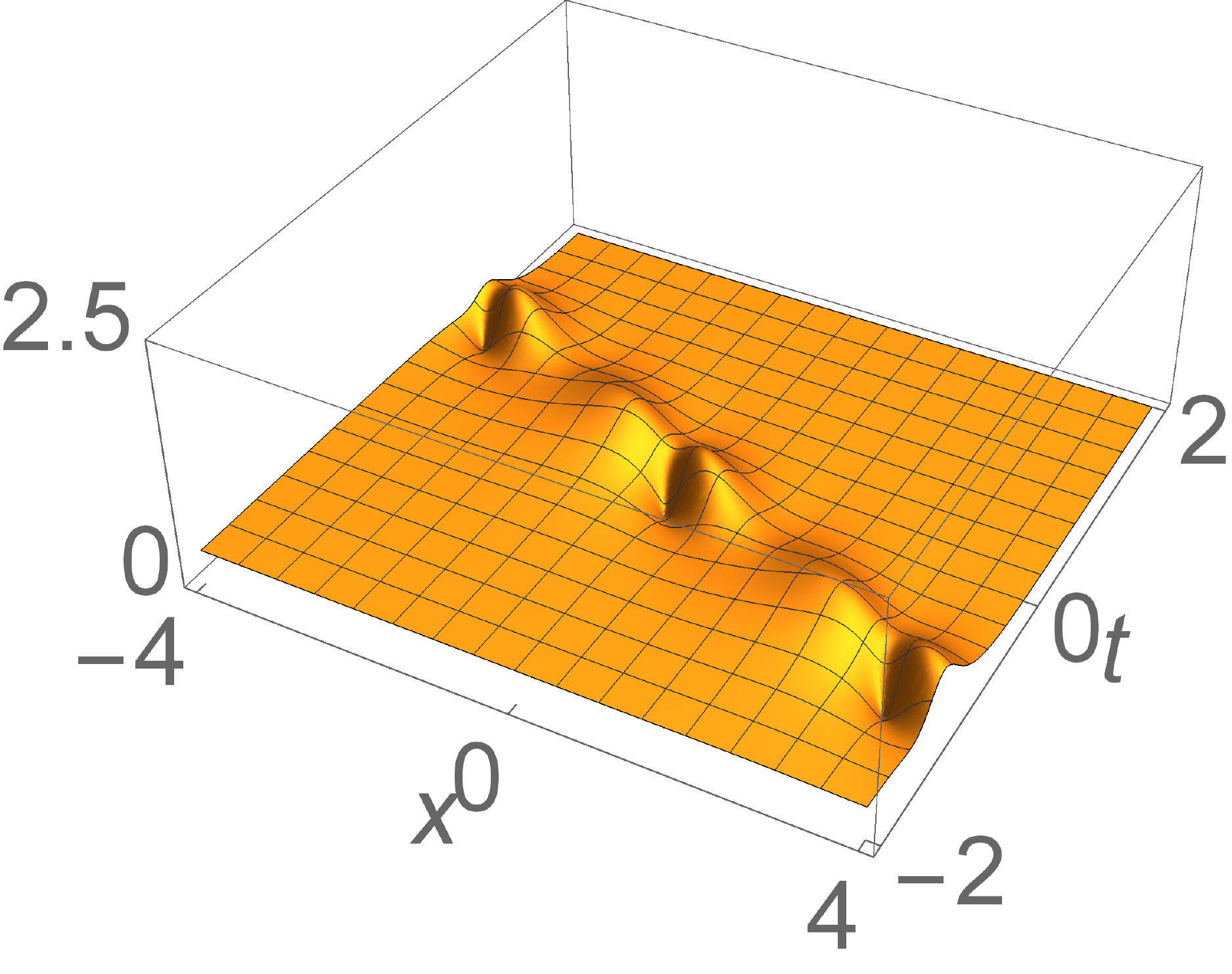}\,
    \includegraphics[scale=0.22]{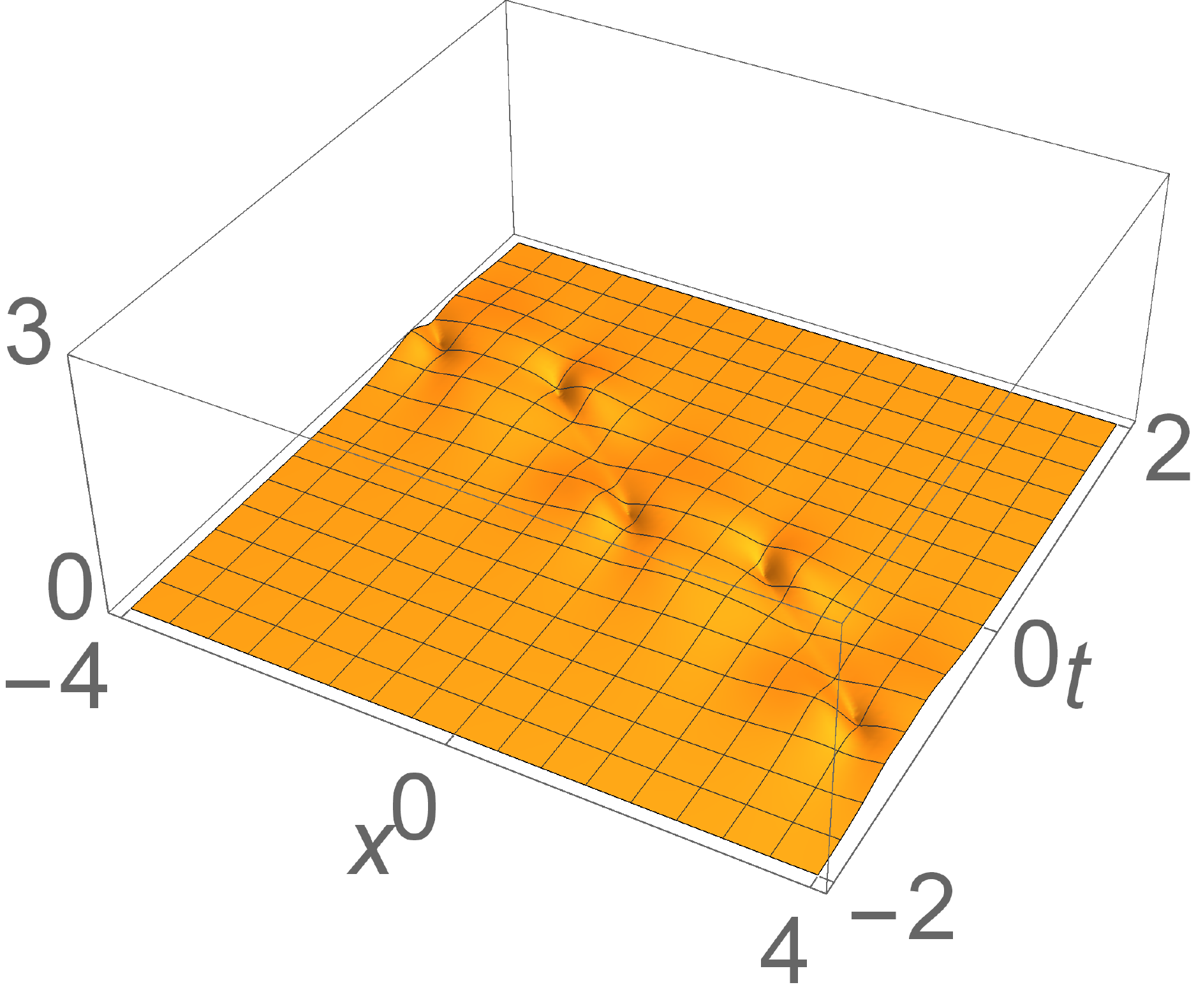}\\
    \includegraphics[scale=0.22]{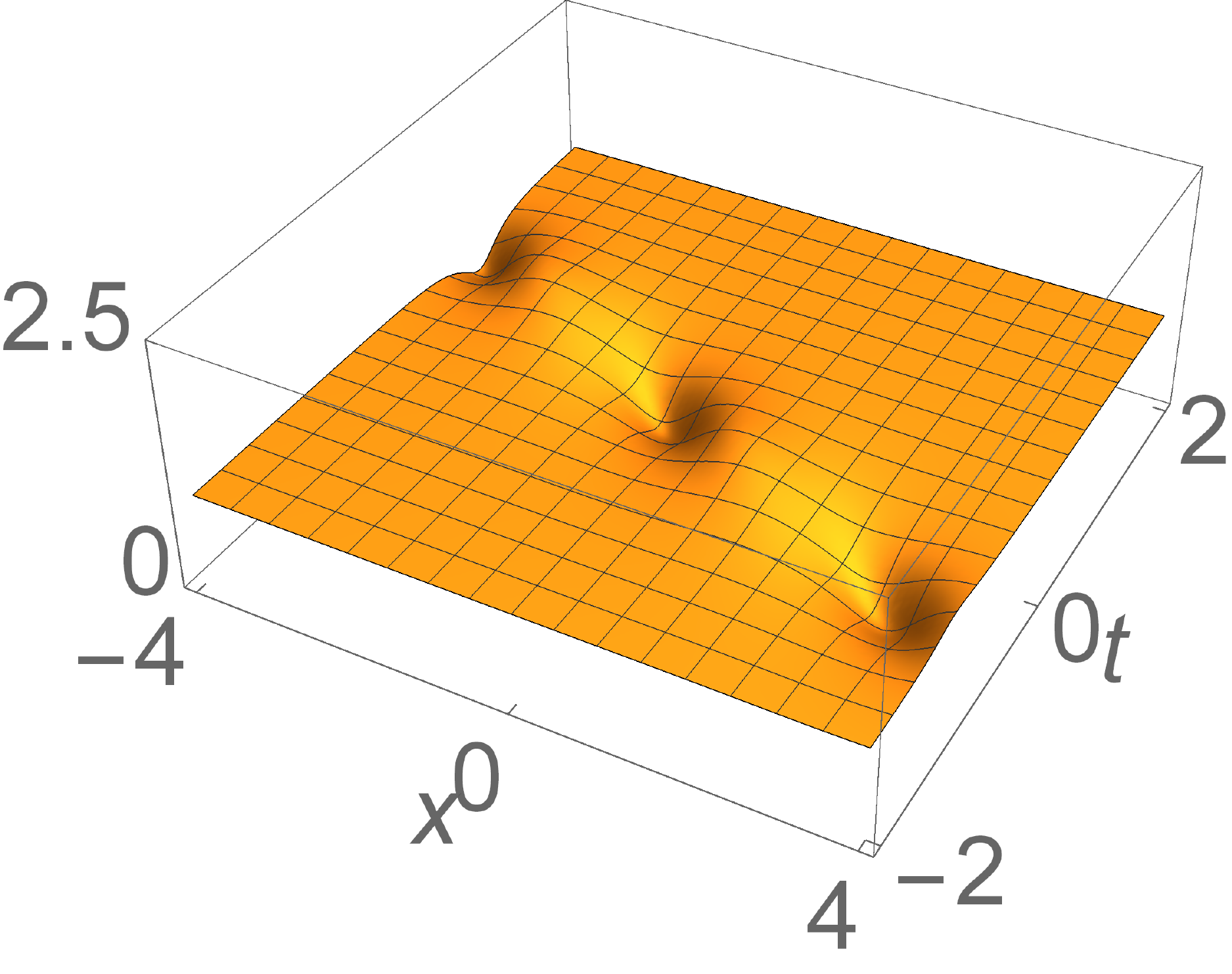}\,
    \includegraphics[scale=0.22]{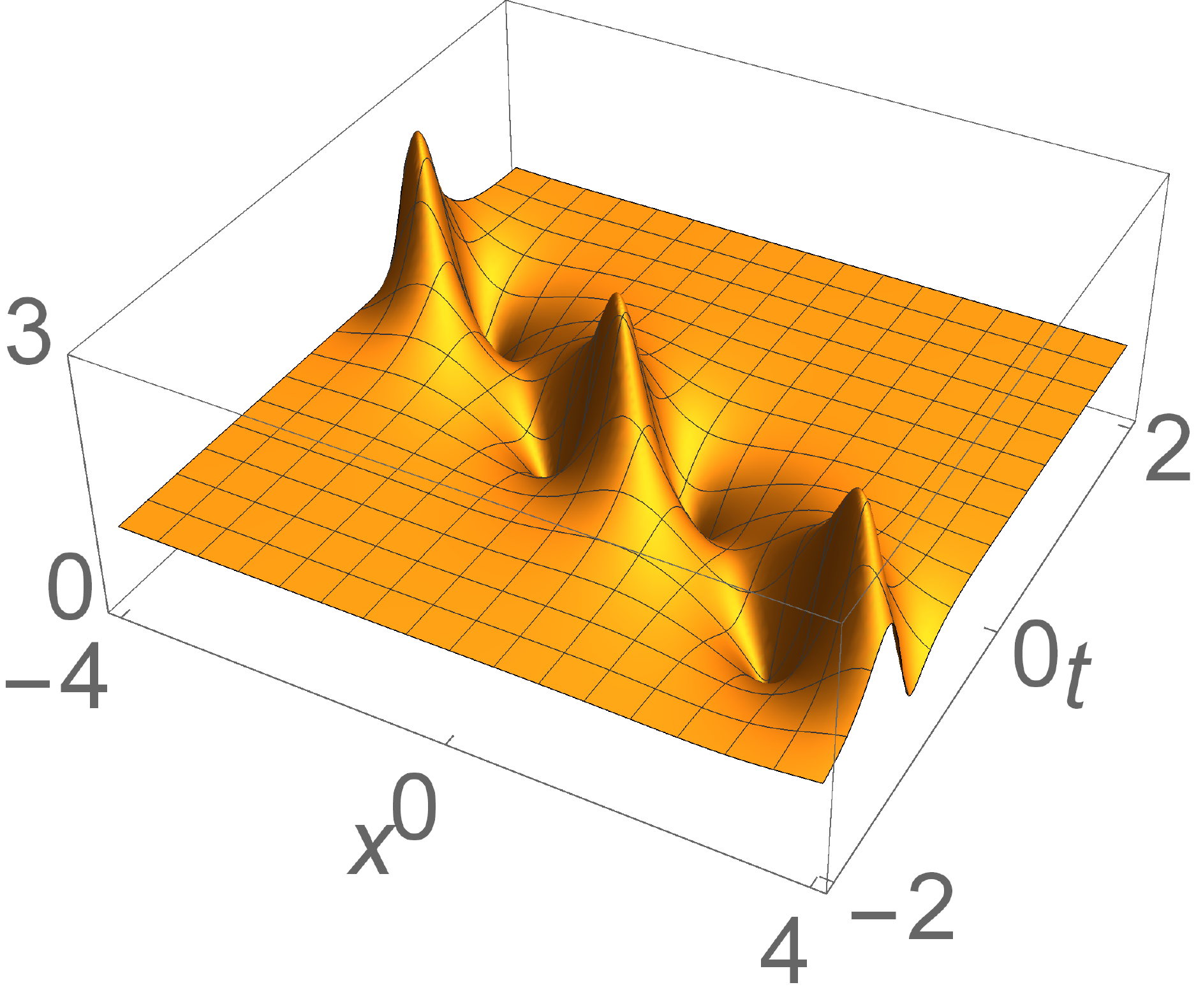}
    \caption{
        Similarly to Fig.~\ref{f:solitonC}(left),
        but for two one-soliton solutions in classes~D (left) and~E (right) with $\zeta = 2i\e^{-i\pi/4}$.
        The unitary matrix $U_D$ is given by Eq.~\eref{e:KUD} with
        $\gamma = 1$, $\eta = \pi/8$ and $\beta_2 = \pi/2$.
        The unitary matrix $U_E$ is given by Eq.~\eref{e:KUE} with
        $\gamma_{-1} = 1$, $\gamma_0=2i$, $\beta = \pi/2$ and $\eta = \pi/32$.
    }
\label{f:solitonDE}
\end{figure}

\begin{figure}[t!]
    \centering
    \includegraphics[scale=0.22]{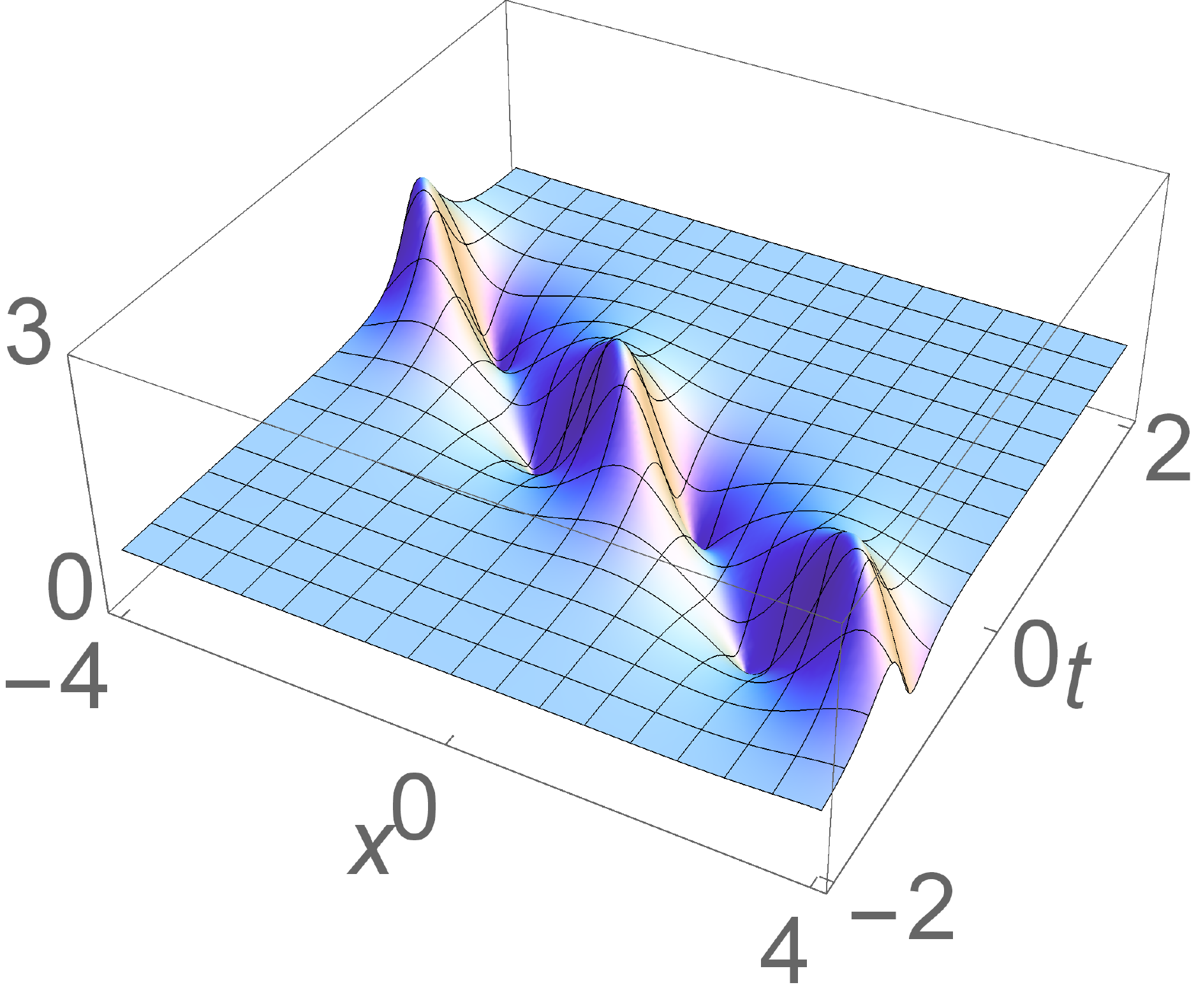}\,
    \includegraphics[scale=0.22]{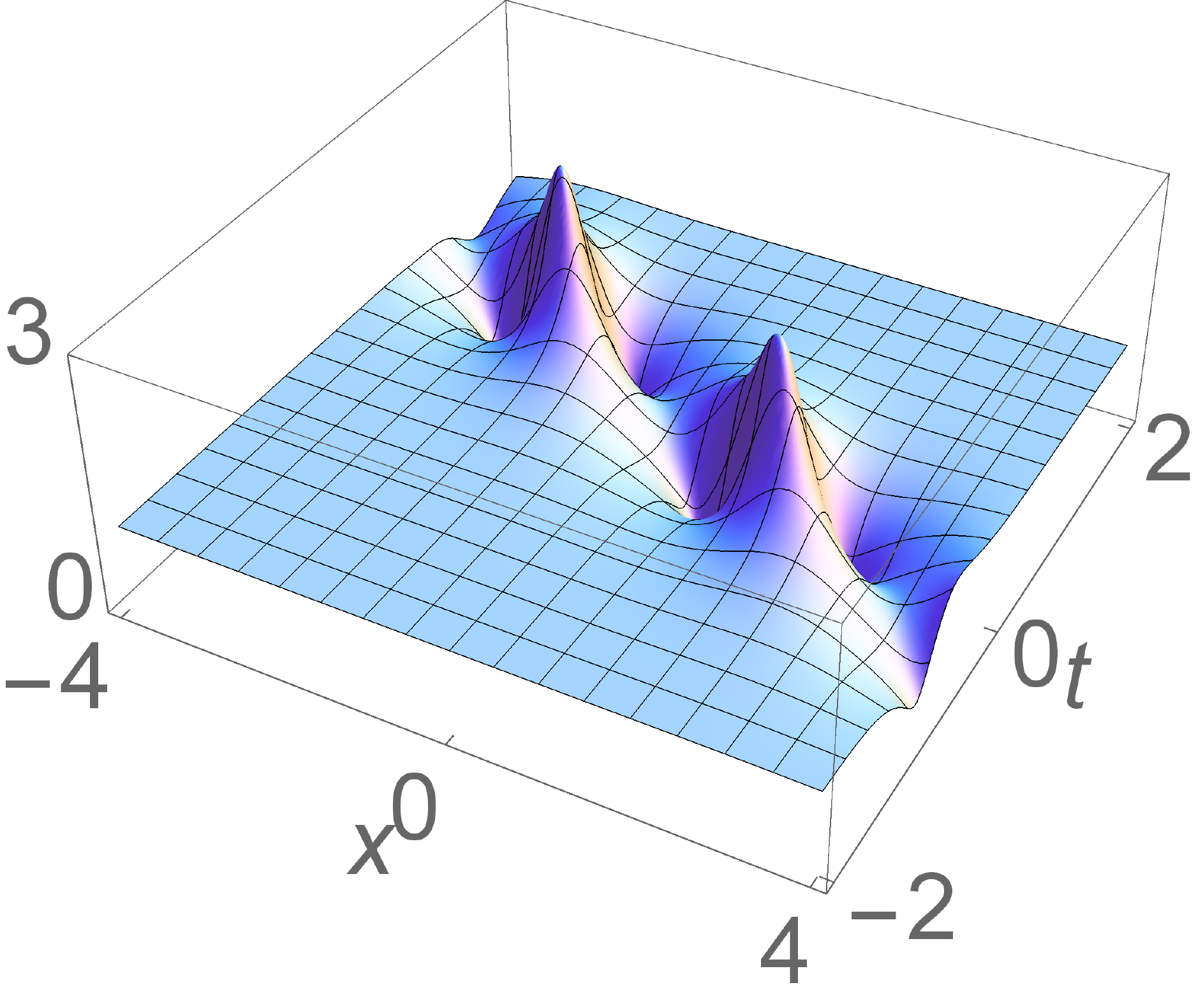}\\
    \includegraphics[scale=0.22]{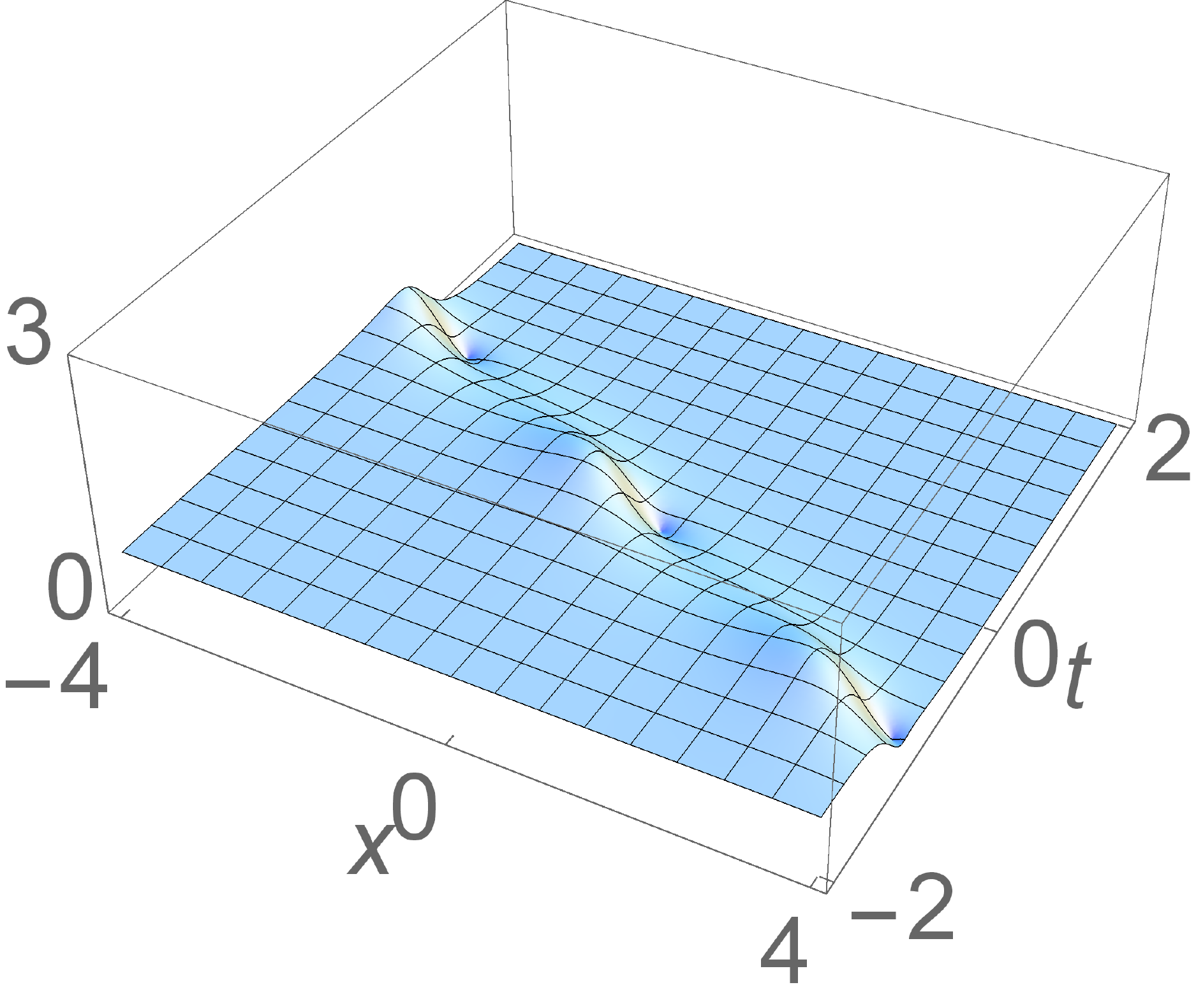}\,
    \includegraphics[scale=0.22]{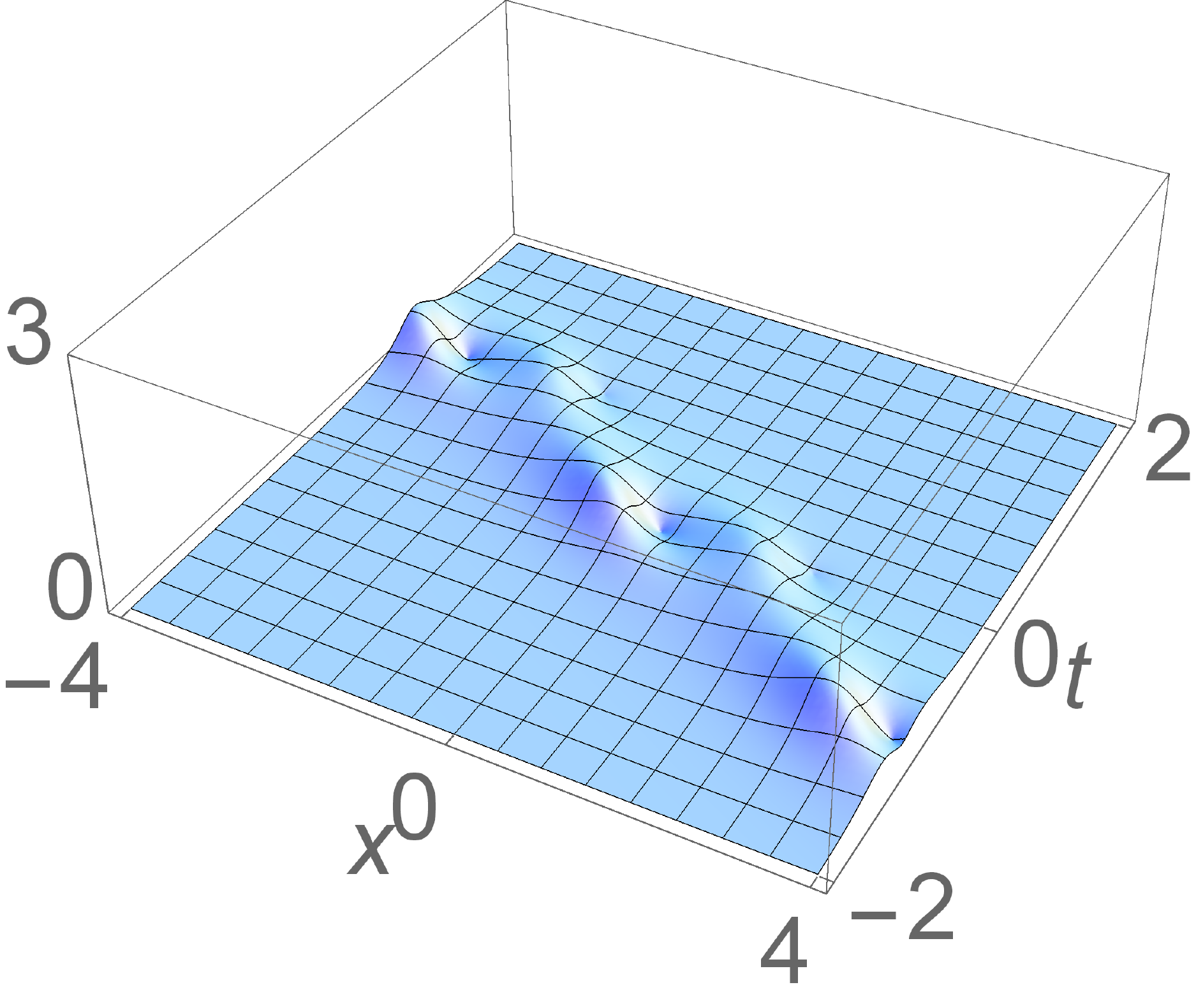}\\
    \includegraphics[scale=0.22]{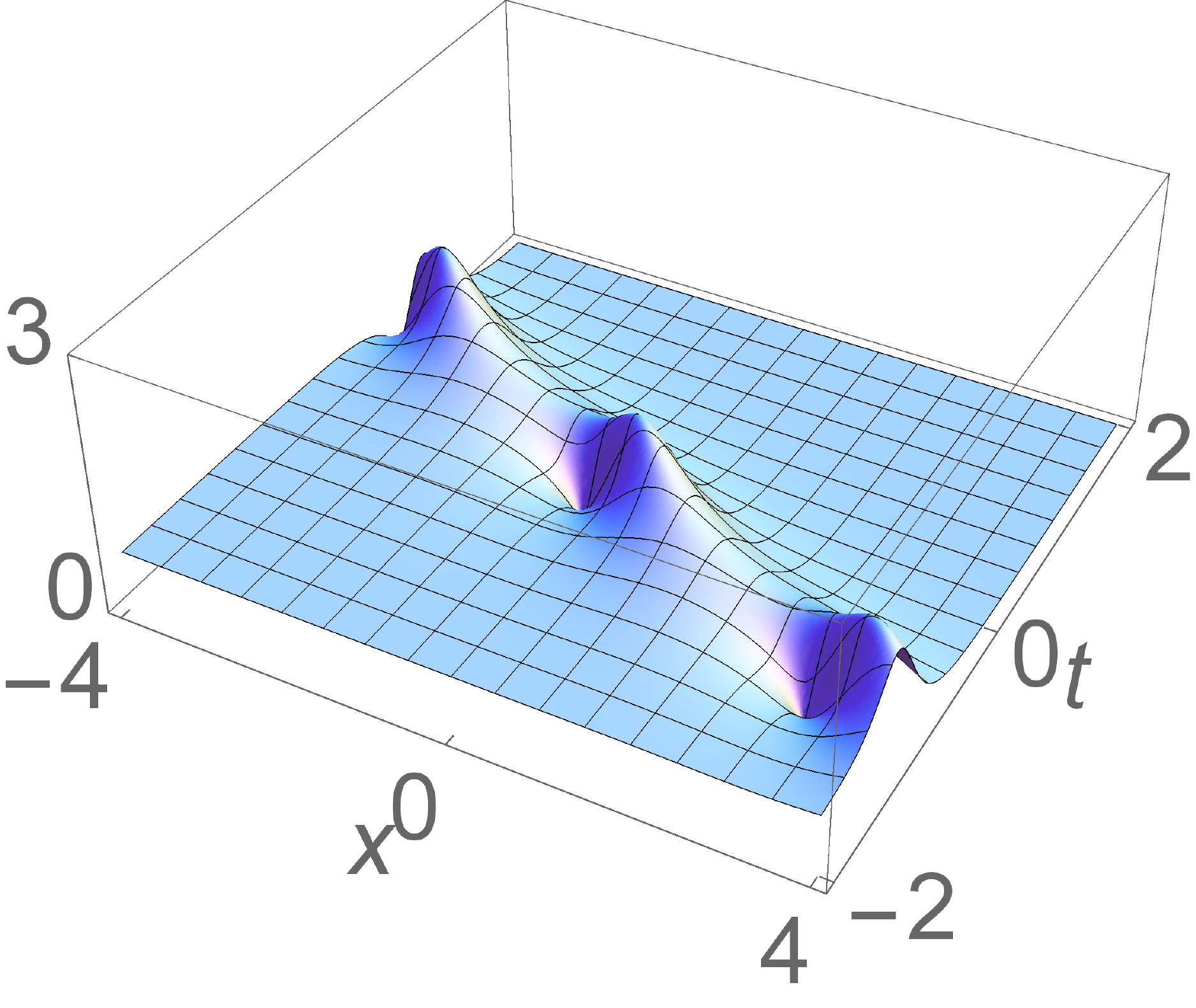}\,
    \includegraphics[scale=0.22]{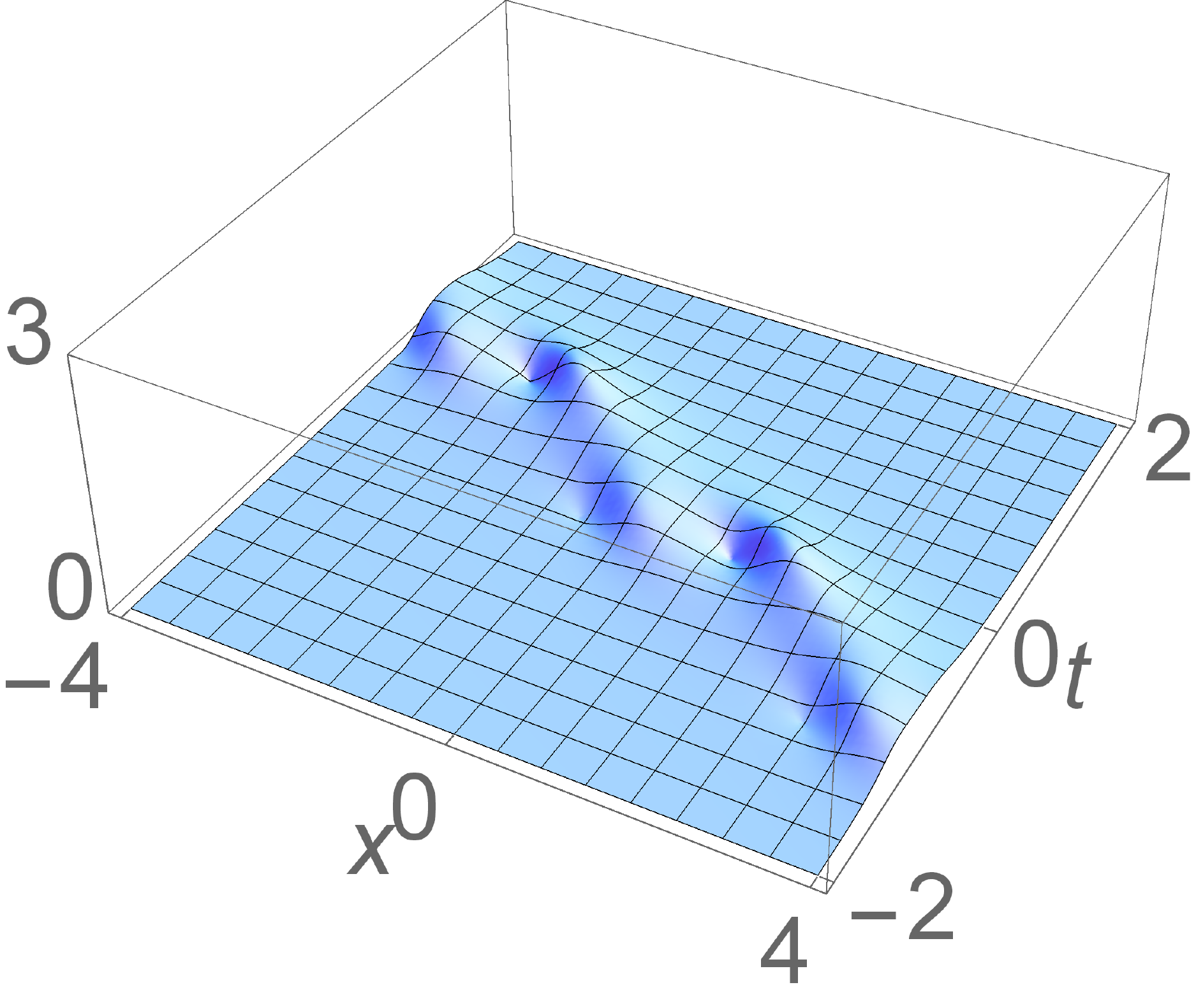}\\
    \includegraphics[scale=0.22]{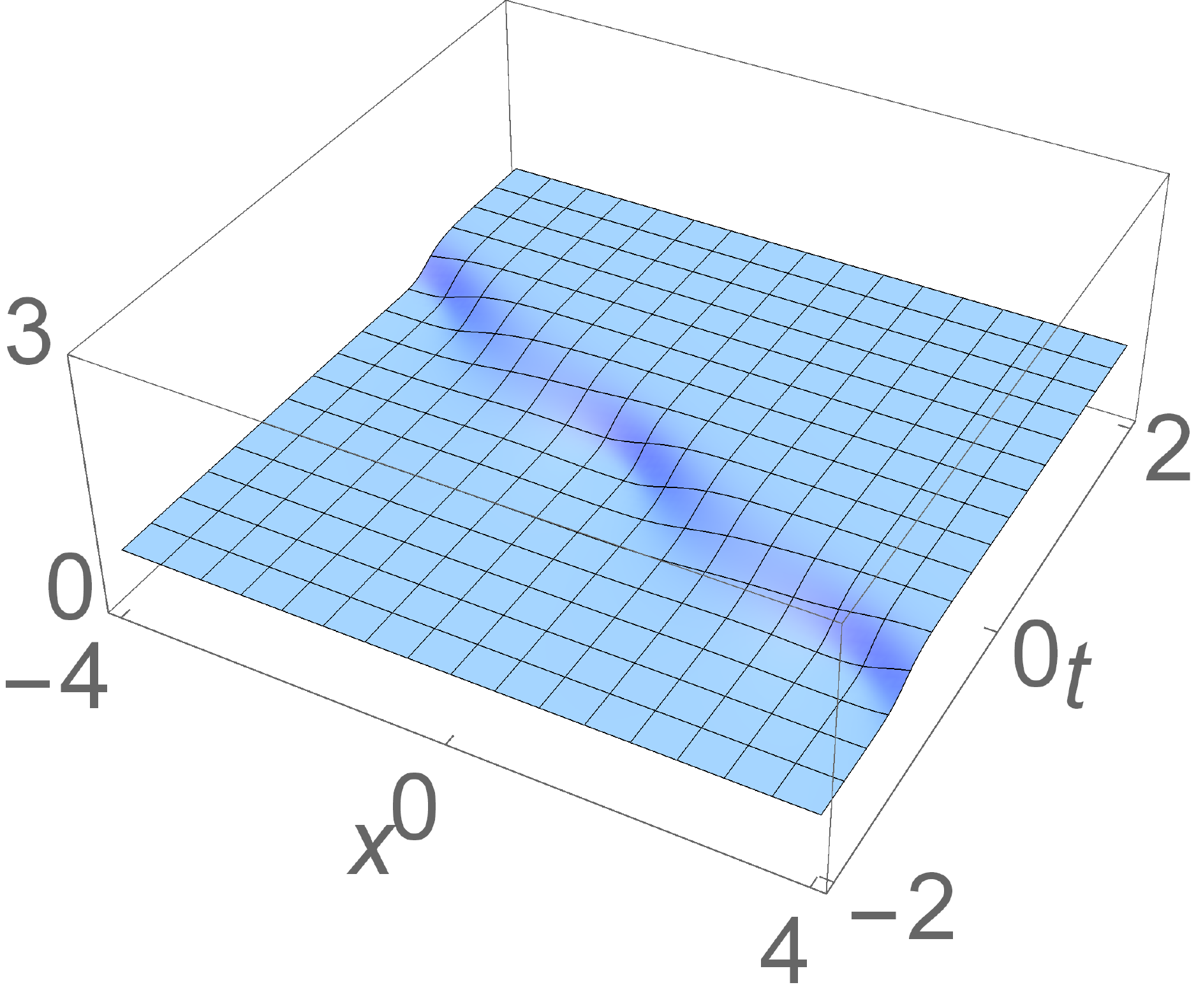}\,
    \includegraphics[scale=0.22]{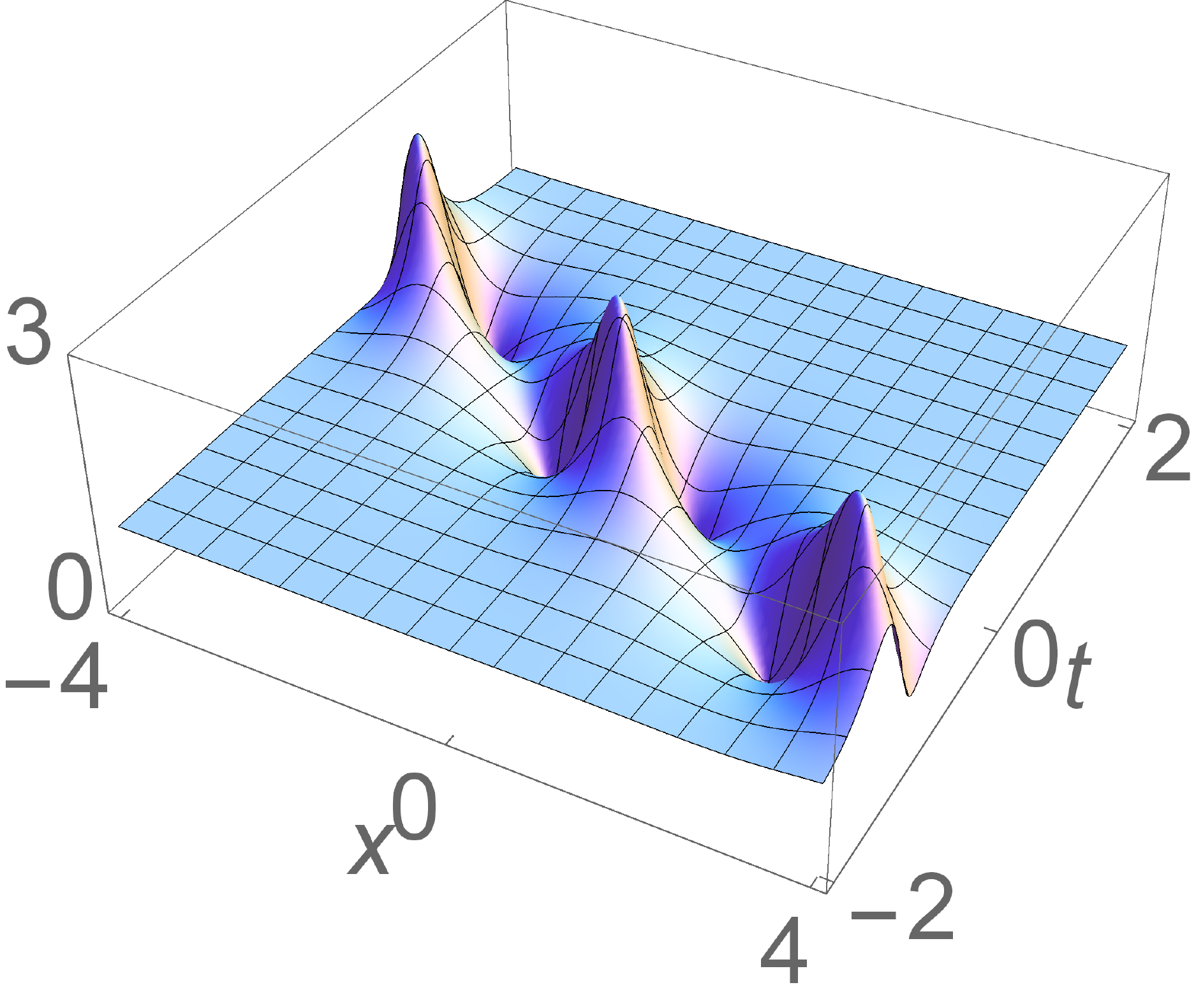}
    \caption{
        Entries of the core soliton component $Q(x,t)$ for the solutions in class~D (left) and class~E (right) shown in Fig.~\ref{f:solitonDE}.
        From top to bottom: $|q_{1,1}(x,t)$, $|q_{1,2}(x,t)|$, $|q_{2,1}(x,t)|$ and $|q_{2,2}(x,t)|$.
    }
    \label{f:QsolitonDE}
\end{figure}

Combining $\Gamma_D$ in Eq.~\eqref{e:Schurparametrization} with Eq.~\eref{e:Q_NZBC},
the core soliton component is still given by a full $2\times2$ matrix $Q(x,t)$ as in Eq.~\eqref{e:Qmatrix}, but
where now the individual entries are given explicitly in Eq.~\eref{e:QDcomponent} in Appendix~II for brevity.
As for class~C, the core soliton component $Q(x,t)$ is not a solution of the spinor BEC model~\eref{e:spinor_nzbc},
because it is not symmetric.
The general one-soliton solution $\Phi(x,t)$ is given by Eq.~\eref{e:PhiQ} as usual.
A three-parameter family of unitary matrices $U_D$ that convert the norming constant into its Schur form, and
hence the core soliton component into the general one-soliton soliton,
is given by Eq.~\eref{e:KUD}.
One example of a soliton solution with norming constant in class~D is shown in Fig.~\ref{f:solitonDE}(left).
The entries of the corresponding core soliton component are shown in Fig.~\ref{f:QsolitonDE}(left).

The solution in class~D also exhibits a domain wall behavior.
One can still see a small asymptotic amplitude difference in $\phi_0$ as $x\to\pm\infty$ in Fig.~\ref{f:solitonDE}(left).
Analytically, this is consistent with Ref.~\cite{pdlhf},
where the asymptotics of one soliton solutions as $x\to\pm\infty$ shows that $\Phi(x,t)$ has different asymptotic amplitudes in each spin state as $x\to\pm\infty$ when $\det K = 0$.


\paragraph{\bf Class~E.}

By combining $\Gamma_E$ in Eq.~\eqref{e:Schurparametrization} with the general solution formula~\eref{e:Q_NZBC},
one obtains the core soliton component $Q(x,t)$ in class~E.
One example of a soliton solution with norming constant in class~E is shown in Fig.~\ref{f:solitonDE}(right).
The entries of the corresponding core soliton component are shown in Fig.~\ref{f:QsolitonDE}(right).
Unlike the previous classes, in this case we were unable to write $Q(x,t)$ in compact form:
the entries of the core soliton component do not seem to be reducible to simple superposition of scalar solutions,
and are not simpler than the solution $\Phi(x,t)$ itself, as one can see by comparing Fig.~\ref{f:solitonDE}(right) and
Fig.~\ref{f:QsolitonDE}(right).
Similarly to class~A,
the locations of peaks in $\phi_{\pm1}$ and holes in $\phi_0$ are clearly in one-to-one correspondence.
Thus this solution also exhibits a dark-bright behavior.

\paragraph{\bf Additional remarks.}
The results of this section can be summarized in Table~\ref{t:solutiontypes}
(which labels the different types of solution) and Table~\ref{t:summary}
(identifying which solution type is obtained for various classes of norming constants and kinds of discrete eigenvalues).
At this stage it is also worth highlighting some of the common features of the soliton solutions
derived above.

(i) We reiterate that traveling solitons, stationary solitons and periodic solutions in each equivalence class
share the same expressions for the core soliton components, and the only difference between them is the
different kind of discrete eigenvalue.

(ii)
Solitons obtained from eigenvalues of the first three kinds (cf.\ Fig.~\ref{f:zeta})
are localized along a line
\[
c_{-,1} k_o x \cos\alpha - k_o^2 c_{+,2} t\sin2\alpha = \mathrm{const}\,.
\nonumber
\]
The left-hand side of the above expression is determined solely by the discrete eigenvalue $\zeta$,
yielding the soliton velocity as
\vspace*{-1ex}
\[
\label{e:velocity}
V = 2k_o c_{+,2}\sin\alpha/c_{-,1}\,.
\]
This velocity also characterizes the domain walls in classes~B and~D.
For a discrete eigenvalue in a generic position in the complex plane (i.e., an eigenvalue of the first kind, cf.\ Fig.~\ref{f:zeta}),
the soliton travels with $V\ne0$,
thus it is a genuine traveling soliton.
For a purely imaginary discrete eigenvalue $\zeta = i k_o Z$, (i.e., $\alpha = 0$ corresponding to an eigenvalue of the second kind)
the soliton velocity is zero,
i.e., the soliton is stationary.
Moreover, if $Z = 1$ with $\alpha\ne0$ (i.e., for an eigenvalue of the third kind), the soliton velocity becomes infinite.
The soliton is localized in time and the solution becomes a periodic in space.
Finally, as we show in the next section,
the limit $Z\to1$ with $\alpha = 0$, corresponding to
an eigenvalue of the fourth kind in Fig.~\ref{f:zeta},
may give rise to rational solutions.

\begin{table}[t!]
\begin{tabular}{|c|l|}
\hline
Type~I & ~Reducible, two (shifted) scalar soliton solutions~ \\
Type~II & ~Reducible, one scalar soliton solution \\
Type~III & ~Irreducible, dark-bright soliton solution \\
Type~IV & ~General irreducible solution \\
Type~V & ~Constant solution \\
\hline
\end{tabular}
\caption{The various types of core soliton components.}
\label{t:solutiontypes}
\bigskip
\begin{tabular}{|c|c|c|}
    \hline
    Schur class & Eigenvalues 1--3 & Eigenvalue 4\\
    \hline
    $\Gamma_A$ & Type I, Eq.~\eref{e:QA} & Type I, Eq.~\eref{e:PhirationalA} \\
    \hline
    $\Gamma_B$ & Type II, Eq.~\eref{e:QB} & Type II, Eq.~\eref{e:PhirationalB} \\
    \hline
    $\Gamma_C$ & Type III, Eq.~\eref{e:QC} & Type V, Eq.~\eref{e:QrationalC} \\
    \hline
    $\Gamma_D$ & Type IV, Eq.~\eref{e:QDcomponent} & Type IV, Eq.~\eref{e:QrationalD} \\
    \hline
    $\Gamma_E$ & Type IV, Eq.~\eref{e:Q_NZBC} & Type IV, Eq.~\eref{e:QrationalD} \\
    \hline
\end{tabular}
    \caption{Summary of the various types of core soliton components (cf.\ Table~\ref{t:solutiontypes}) obtained
        according to the Schur form of the norming constant (classes A--E, cf.\ Table~\ref{t:norming}) and
        the location of the discrete eigenvalue $\zeta$
        (eigenvalues of kinds 1--2 yield standard soliton solutions;
        eigenvalues of the third kind yield periodic solutions and an eigenvalue of the fourth kind yields rational solutions,
        cf.\ Fig.~\ref{f:zeta}).}
\label{t:summary}
\end{table}

(iii) 
Unlike the soliton velocities, the specific location and phase of an individual soliton are determined by the Schur form of the norming constant and by the unitary transformation that reduces the norming constant to its Schur form.
The explicit expressions for these quantities in general are different for the five classes.

(iv) Finally, we reiterate that solutions in equivalence classes A and E represent polar states,
for which the total spin of the condensate is zero,
and the asymptotic state $\Phi_-$ is also equals $k_o$ times the identity matrix up to a phase.
Conversely, solutions in equivalence classes B, C and D describe ferromagnetic states, for which the total spin of the condensate
is non-zero and the corresponding asymptotic state $\Phi_-$ is not diagonal in general.
Since the asymptotics $\Phi_\pm$ in a polar state are diagonal with only an overall phase difference~\cite{pdlhf},
each $\phi_j$ has the same asymptotic amplitudes as $x\to\pm\infty$.
Therefore, domain walls cannot form in a polar state.
On the other hand, in a ferromagnetic state $\Phi_-$ is not diagonal while $\Phi_+$ is,
implying that each $\phi_j$ has different asymptotic amplitudes as $x\to\pm\infty$.
One then expects kink-like behavior in some solutions corresponding to domain walls.
This can clearly be seen from Fig.~\ref{f:solitonB}.
Thus, polar states and ferromagnetic states in general have different topological properties.

\section{IV.~ ROGUE-WAVE SOLUTIONS}

Next we derive rogue-wave (i.e., rational) solutions of the spinor BEC model
by taking suitable limits of the stationary soliton solutions for each equivalence class of norming constants.
For simplicity, and without loss of generality, in this section we take $k_o = 1$,
using the scaling invariance of Eq.~\eref{e:spinor_zbc}:
if $\Phi(x,t)$ is a solution of Eq.~\eref{e:spinor_zbc},
$c\Phi(cx,c^2t)$ is also a solution for any real constant~$c$.

Recall that the scalar NLS equation~\eref{e:nls_nzbc} admits a rational solution
known as the Peregrine soliton:
\bse
\label{e:Peregrine}
\begin{gather}
q_\P(x,t) = 1 - 4f(x,t)\,,\\
f(x,t) = \frac{4 i t + 1}{4 x^2+16 t^2+1}\,,
\end{gather}
\ese
which, for instance, can be derived by taking the limit of
a stationary (Kuznetsov-Ma) soliton solution~\cite{k1977,m1979},
i.e., the TW solution~\eref{e:qtw} with $\zeta = i Z$,
as the discrete eigenvalue approaches the branch point $i$,
i.e., as $Z\to1$,
with a suitable rescaling of the norming constant, namely letting $\gamma = \gamma_\P$, with
\[
\label{e:rationallambda}
\gamma_\P = 2i Z ({Z^2-1})/({Z^2+1})\,.
\]
The existence of branch points in the spectral plane is a result of the NZBC in the formulation of the IST.
The solution~\eref{e:Peregrine} is centered at the origin.
However, one can easily derive a Peregrine soliton centered at an arbitrary point $(x_o,t_o)$
using a different rescaling for the norming constant.
Indeed, taking
\[
\label{e:rationallambda1}
\gamma = \gamma_\P\,\e^{c_{-,1}x_o + i c_{-,2}t_o}\,,
\]
where $c_{-,j}$ is defined in Eq.~\eref{e:qtw},
the limit $Z\to1$ leads to a displaced Peregrine soliton $q_\P(x-x_o,t-t_o)$ centered at $(x_o, t_o)$.

Following a similar procedure,
we next consider the limit of the solutions obtained from a purely imaginary discrete eigenvalue $\zeta = i Z$ with $Z>1$
for all five equivalence classes discussed in Section~III.B,
and we derive corresponding families of rational solutions with suitable rescalings of the norming constants.

\begin{figure}[t!]
    \centering
    \includegraphics[scale=0.22]{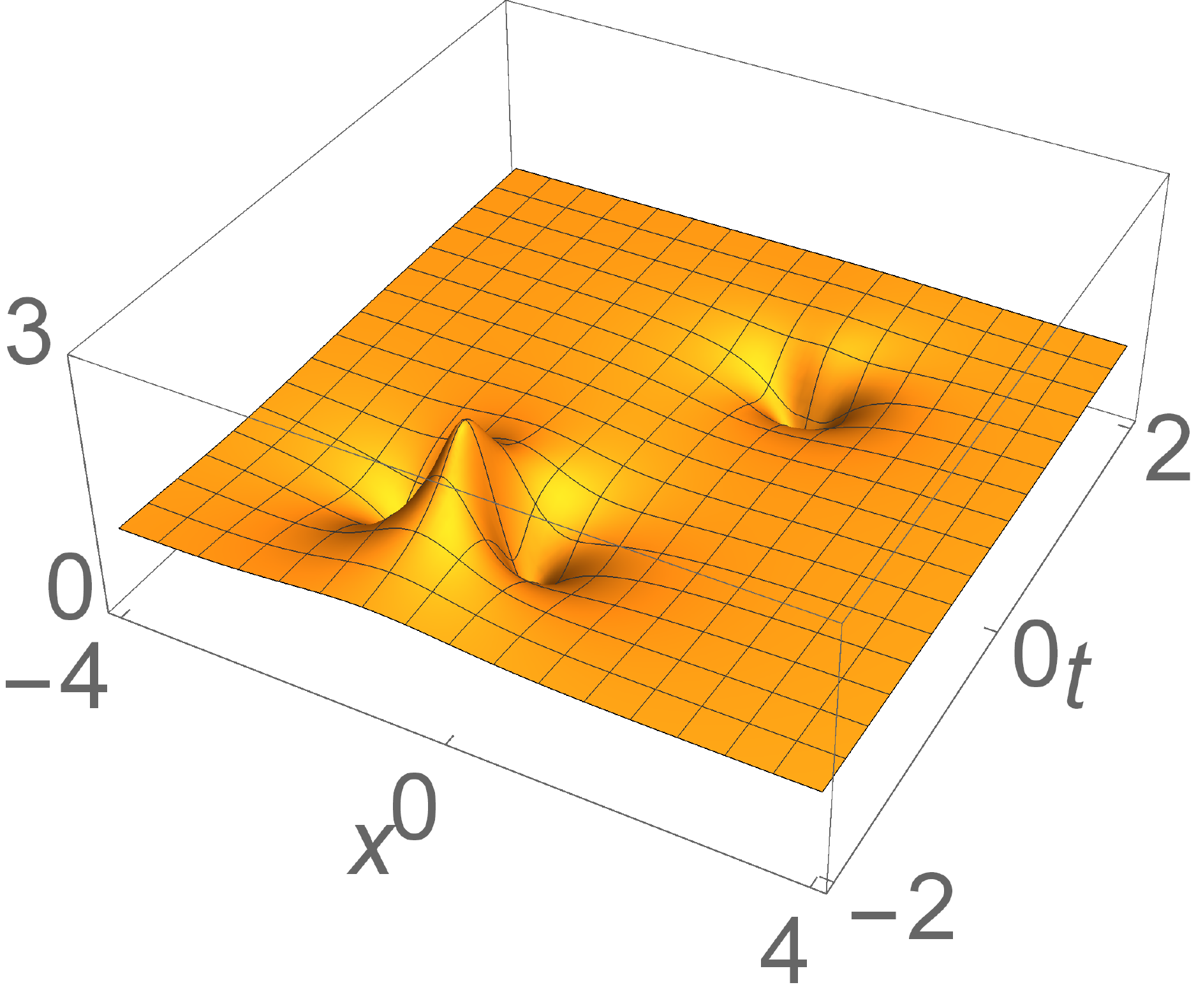}\,
    \includegraphics[scale=0.22]{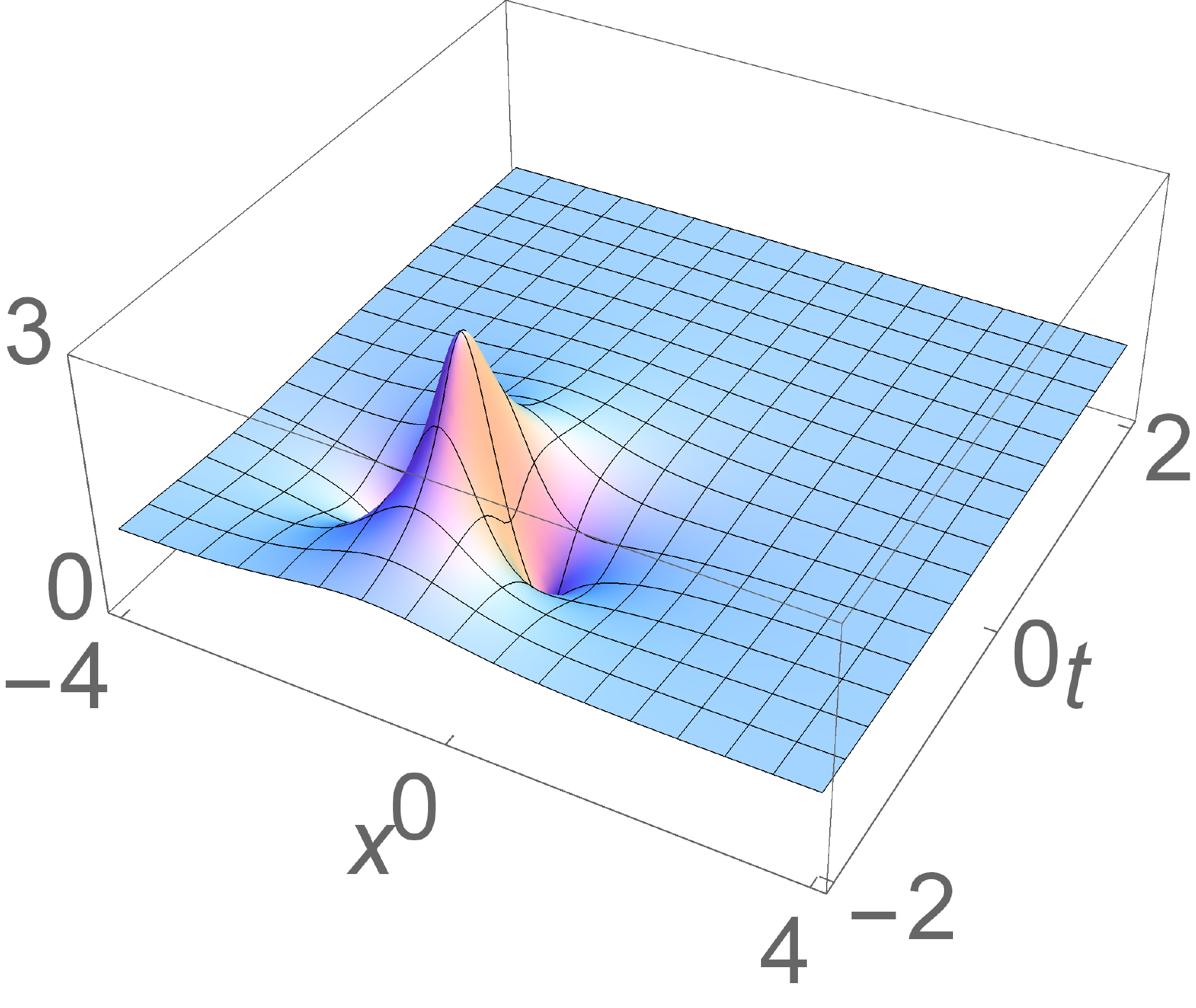}\\
    \includegraphics[scale=0.22]{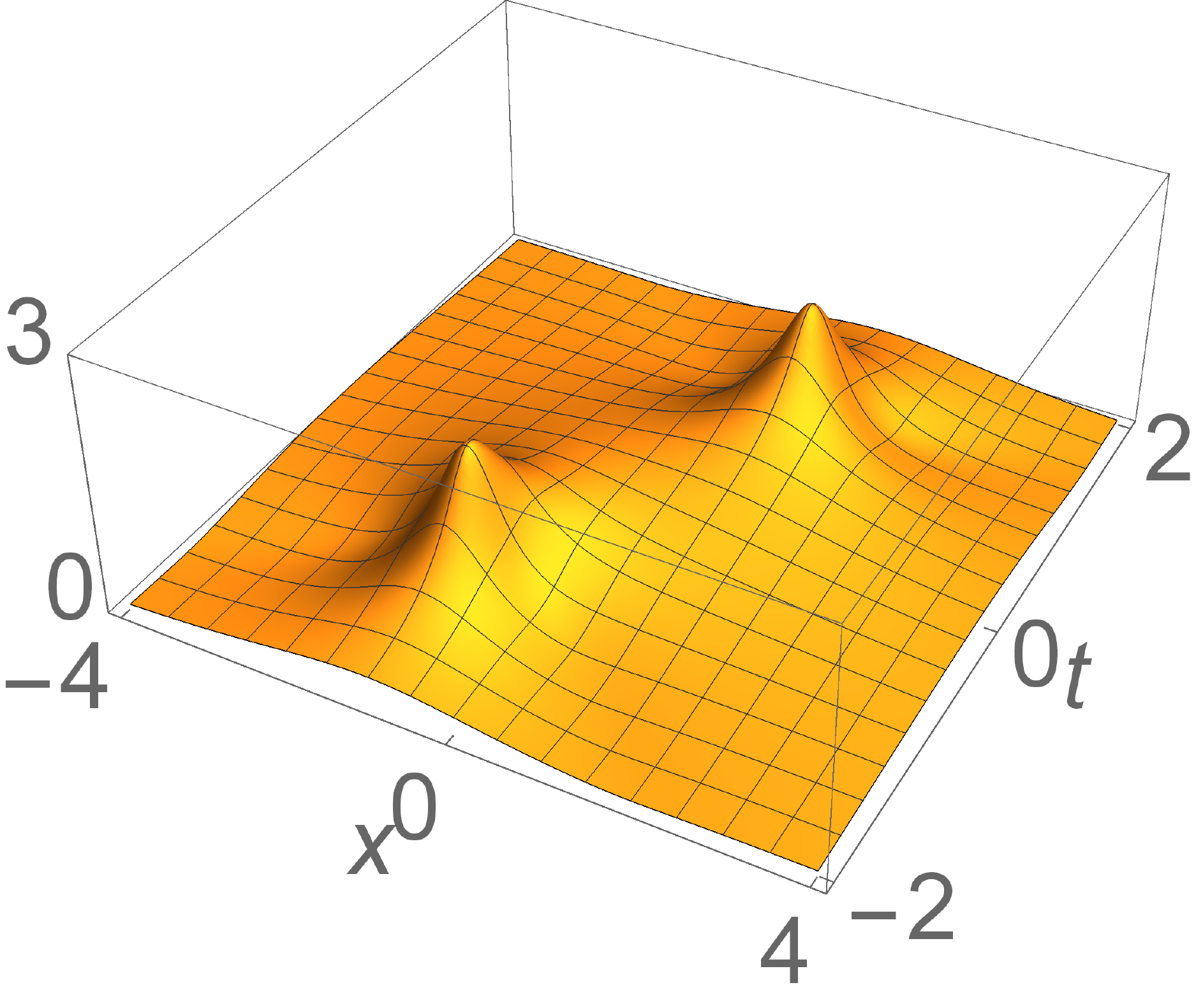}\,
    \includegraphics[scale=0.22]{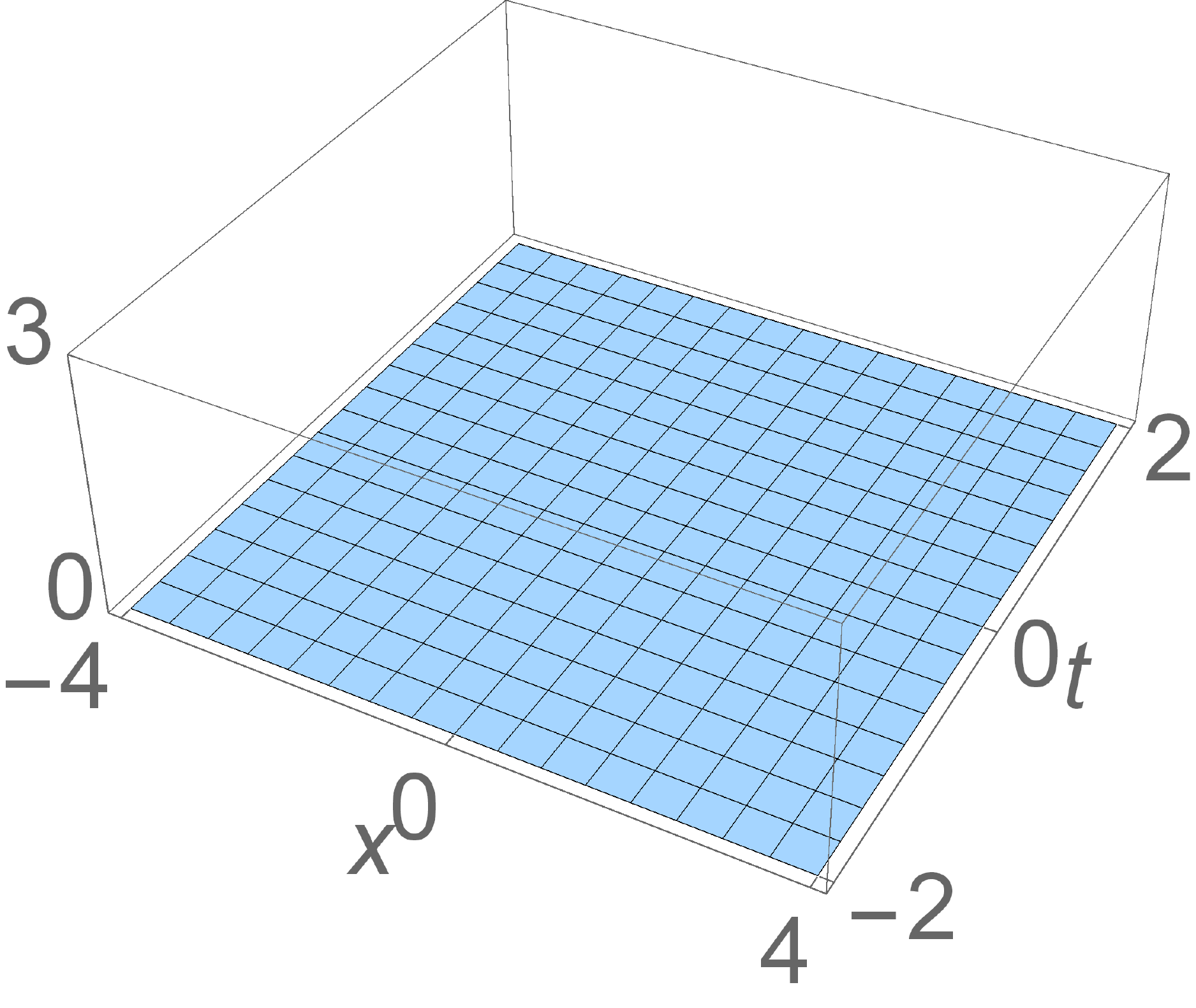}\\
    \includegraphics[scale=0.22]{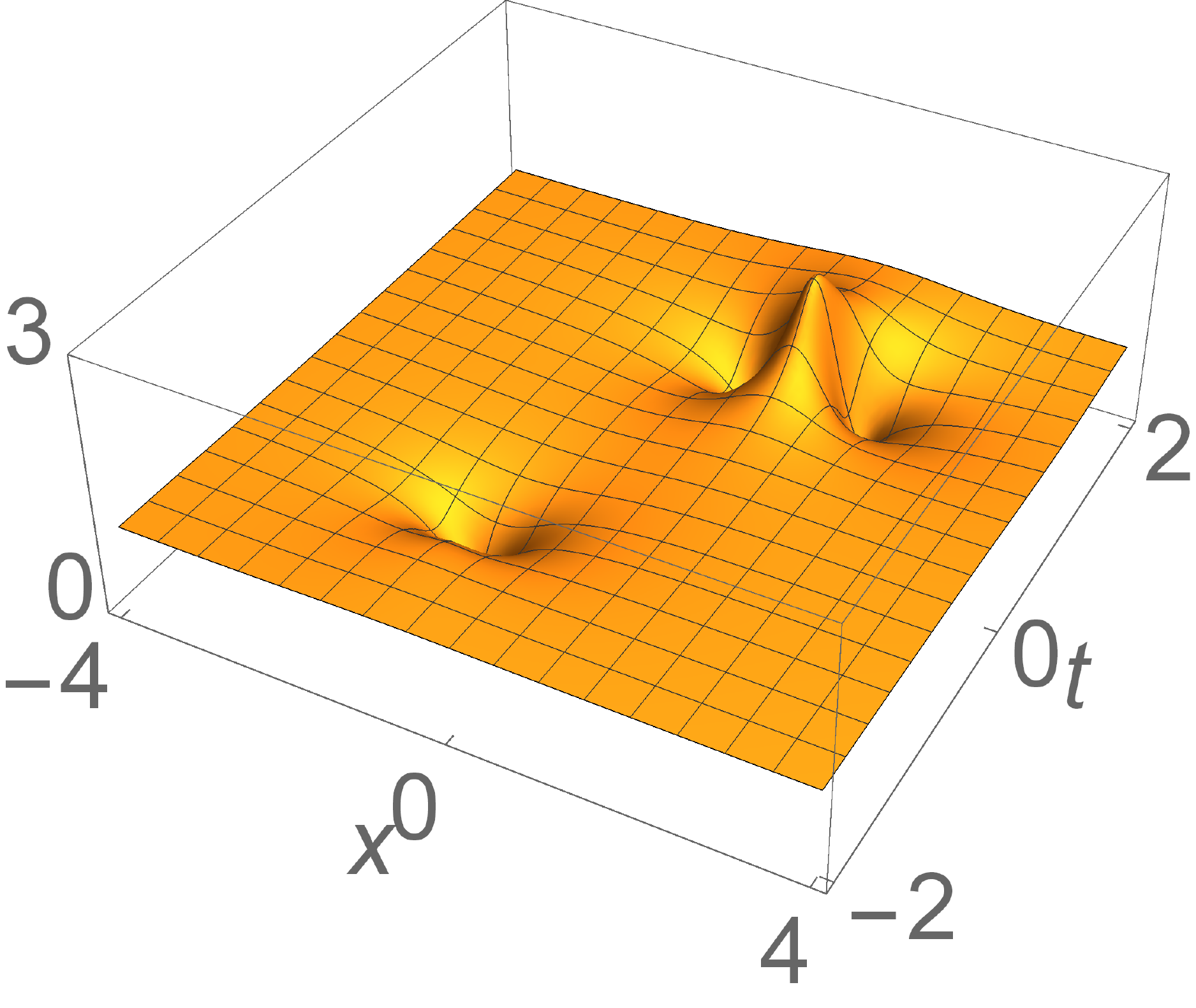}\,
    \includegraphics[scale=0.22]{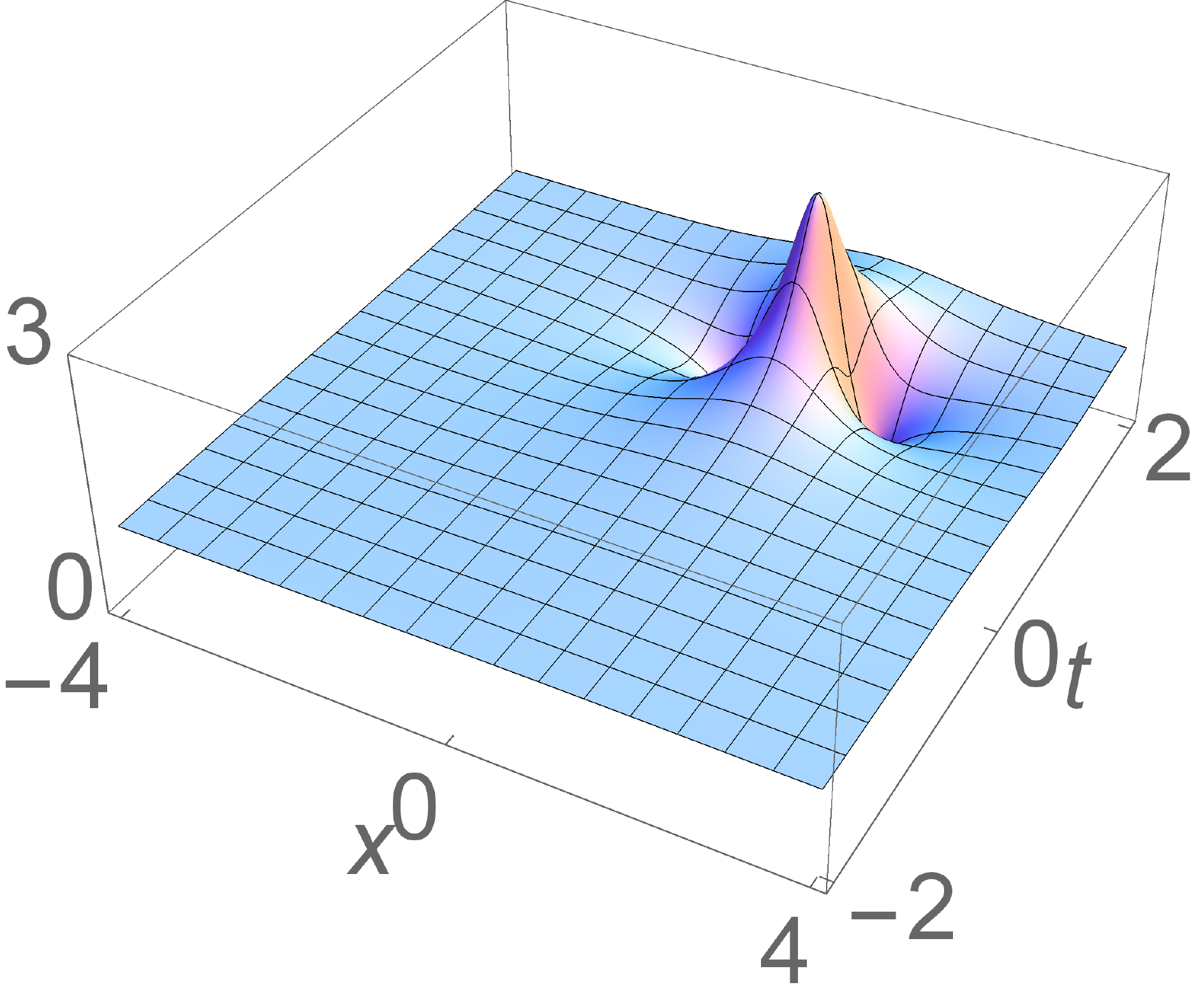}
    \caption{
        Amplitudes of a rational solution of the spinor BEC model in class~A with $k_o = 1$.
        Left: entries of $\Phi(x,t)$ from Eq.~\eref{e:PhirationalA} with  $\eta = \pi/3$,  $(x_1,t_1) = (-1,-1)$ and $(x_{-1},t_{-1}) = (1,1)$.
        From top to bottom: $|\phi_1(x,t)$, $|\phi_0(x,t)|$ and $|\phi_{-1}(x,t)|$.
        Right: entries of the corresponding core soliton component $Q(x,t)$ from Eq.~\eref{e:QrationalA}.
        From top to bottom: $|q_{1,1}(x,t)|$, $|q_{1,2}(x,t)| = |q_{2,1}(x,t)|$ and $|q_{2,2}(x,t)|$.
    }
    \label{f:rationalA}
\end{figure}

\paragraph{\bf Class~A.}
Recall first that in this case the core component~\eref{e:QA} has two independent scalar TW solitons.
Similarly to the rescaling~\eref{e:rationallambda1},
we rewrite the Schur form of the norming constant as
\vspace*{-0.6ex}
\[
\nonumber
\Gamma_A = \gamma_\P\,\diag(\e^{c_{-,1}x_1 + i c_{-,2}t_1},\e^{c_{-,1}x_{-1} + i c_{-,2}t_{-1}})\,,
\]
where $x_{\pm1}$ and $t_{\pm1}$ are four real constants determined by the eigenvalues $\gamma_{\pm1}$ of the norming constant~$K$.
By changing~$\gamma_{\pm1}$, one can change the values of $x_{\pm1}$ and $t_{\pm1}$.
In the limit $Z\to1$, the core soliton component~\eref{e:QA} with discrete eigenvalue $\zeta = iZ$ becomes
\vspace*{-0.6ex}
\[
\label{e:QrationalA}
Q_\P(x,t) = \diag\big(q_{\P,1}(x,t),q_{\P,-1}(x,t)\big)\,,
\]
where $q_{\P,\pm1}(x,t) = q_\P(x-x_{\pm1},t-t_{\pm1})$.
If $x_1 = x_{-1}$ and $t_1 = t_{-1}$ (corresponding to a norming constant with $\gamma_1=\gamma_{-1}$),
the centers of the two Peregrine solitons coincide.
Using the general unitary matrix $U_A$ from Eq.~\eref{e:UA},
we then obtain the general family of rational solutions in class~A as
\bse
\label{e:PhirationalA}
\begin{gather}
\phi_1(x,t) = q_{\P,1}(x,t) \sin^2\eta + q_{\P,-1}(x,t) \cos^2\eta\,,\\
\phi_0(x,t) = \frac{1}{2}(q_{\P,1}(x,t) - q_{\P,-1}(x,t)) \sin2\eta\,,\\
\phi_{-1}(x,t) = q_{\P,1}(x,t) \cos^2\eta + q_{\P,-1}(x,t) \sin^2\eta\,,
\end{gather}
\ese
where $-\pi/2 <\eta < \pi/2$.
An example of a rational solution $\Phi(x,t)$ in class~A and the corresponding core component $Q(x,t)$
are shown in Fig.~\ref{f:rationalA}.

The spinor BEC model~\eref{e:spinor_nzbc} possesses a translation invariance,
namely, if $\Phi(x,t)$ is a solution, then $\Phi(x-x_o,t-t_o)$ is also a solution.
Using this invariance, one can eliminate two of the parameters without loss of generality,
for example $x_{-1}$ and $t_{-1}$.
However, the relative position of the two peaks in the $(x,t)$-plane cannot be changed using this invariance.
Hence, Eq.~\eref{e:PhirationalA} defines a three-parameter family of rational solutions.
The parameter $\eta$ characterizes the spin rotation of this solution.
The other two parameters describe the relative position of the two peaks.

Clearly from Fig.~\ref{f:rationalA}(left),
the potential traps in $\phi_{\pm1}$ pair with peaks in $\phi_0$.
Again, the mixed dark-bright behavior is observed in this rogue-wave solution.
However, unlike the soliton solutions in class~A,
this dark-bright behavior does not oscillate.

\paragraph{\bf Class~B.}

\begin{figure}[b!]
    \centering
    \includegraphics[scale=0.22]{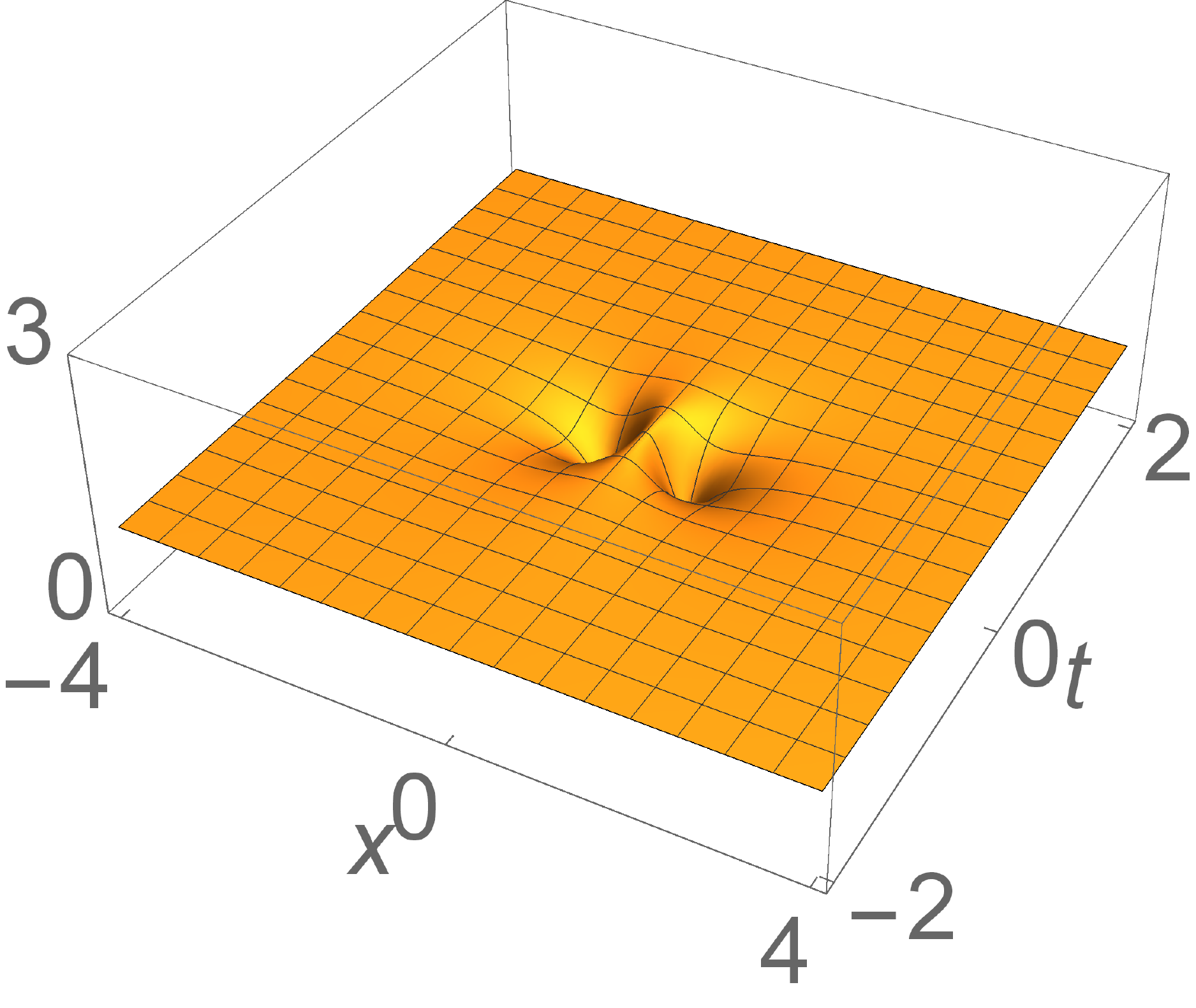}\,
    \includegraphics[scale=0.22]{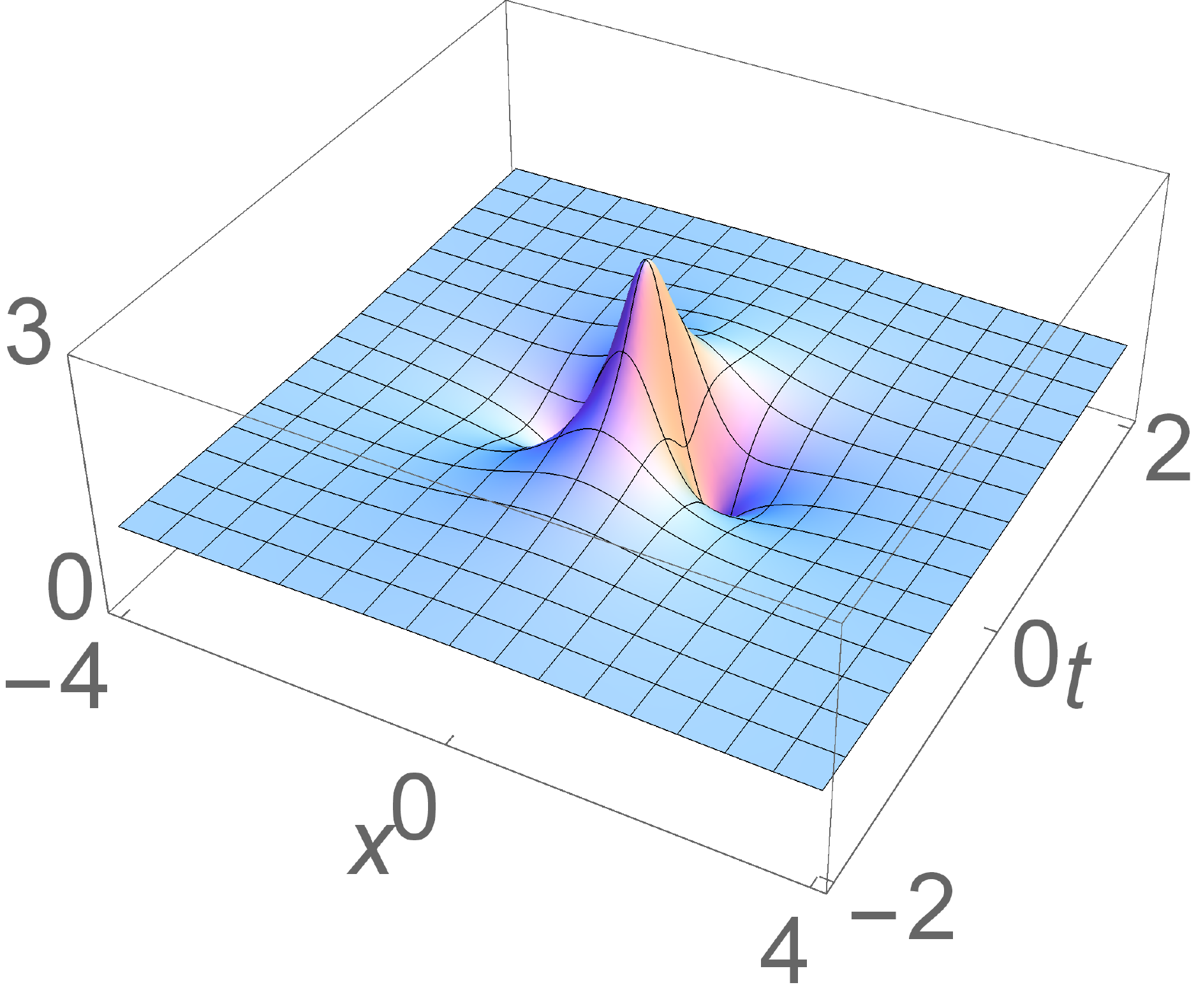}\\
    \includegraphics[scale=0.22]{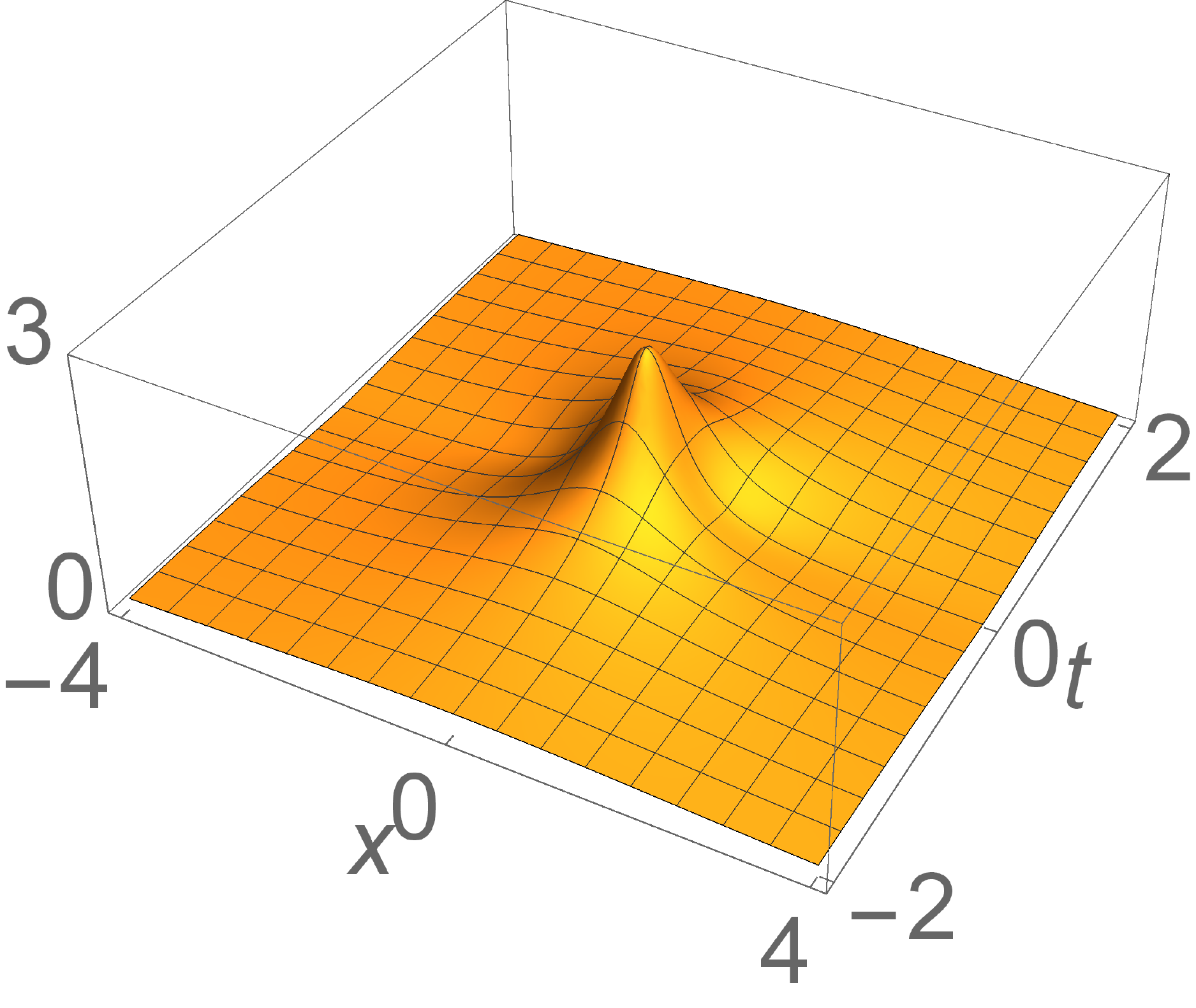}\,
    \includegraphics[scale=0.22]{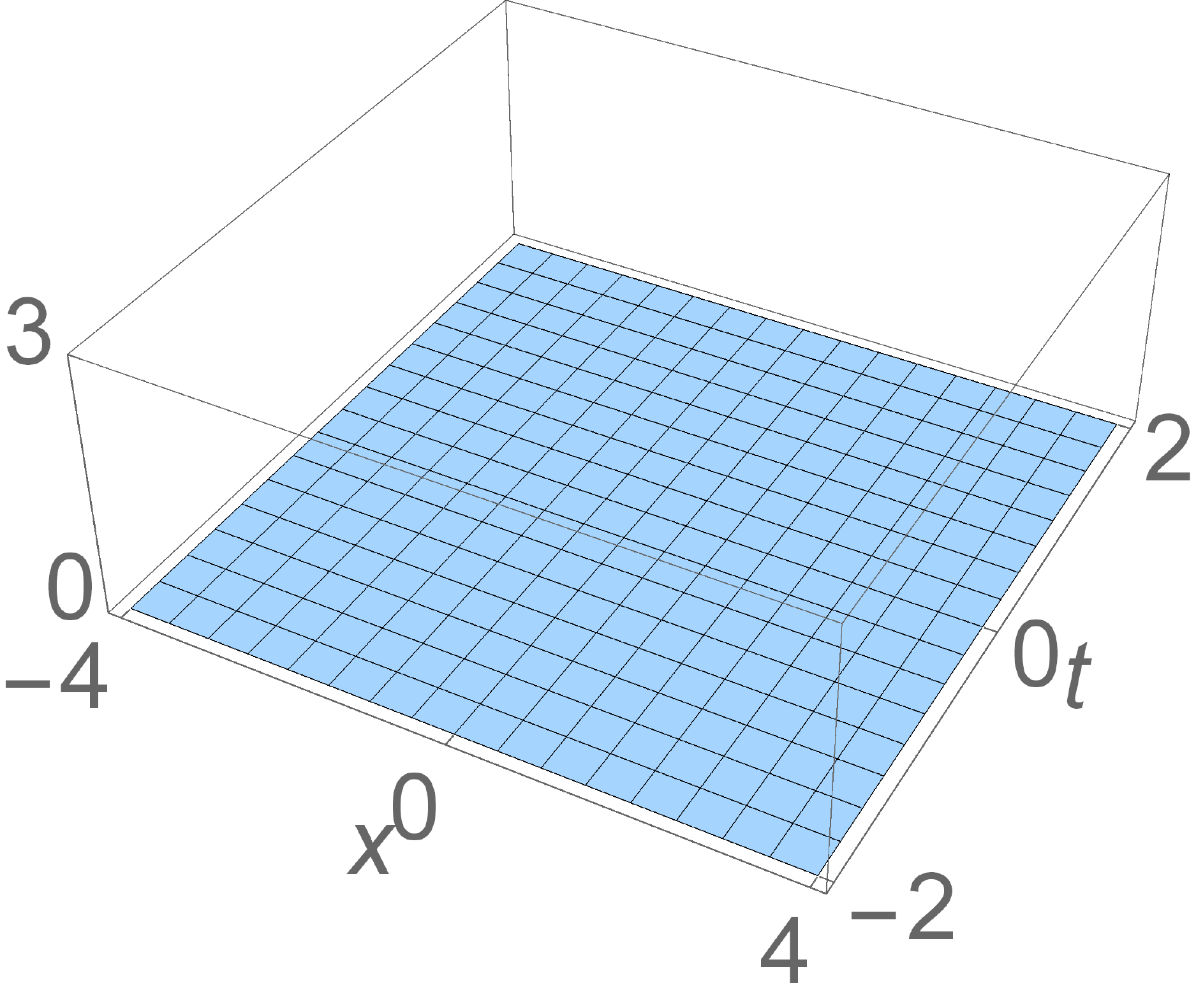}\\
    \includegraphics[scale=0.22]{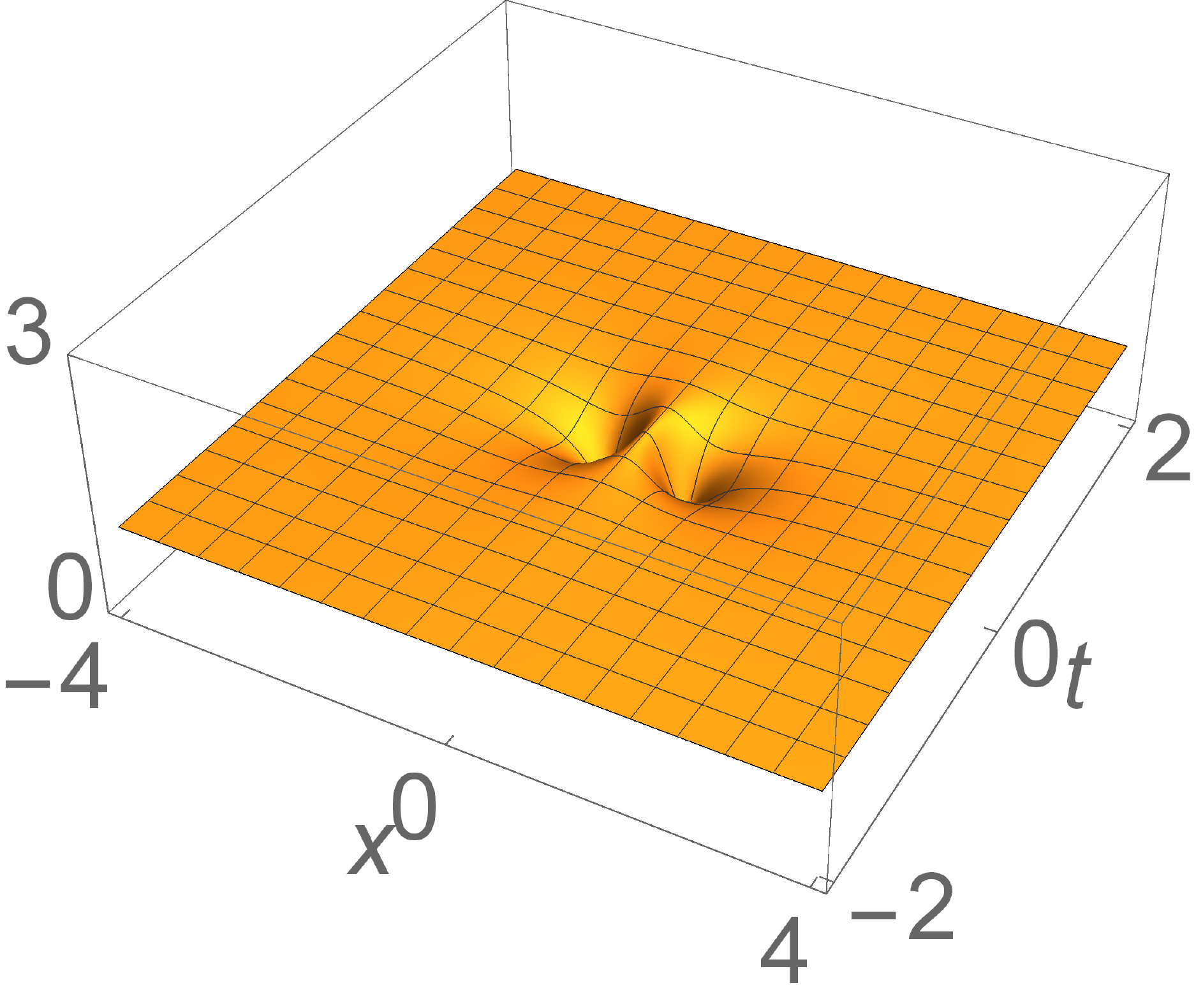}
    \includegraphics[scale=0.22]{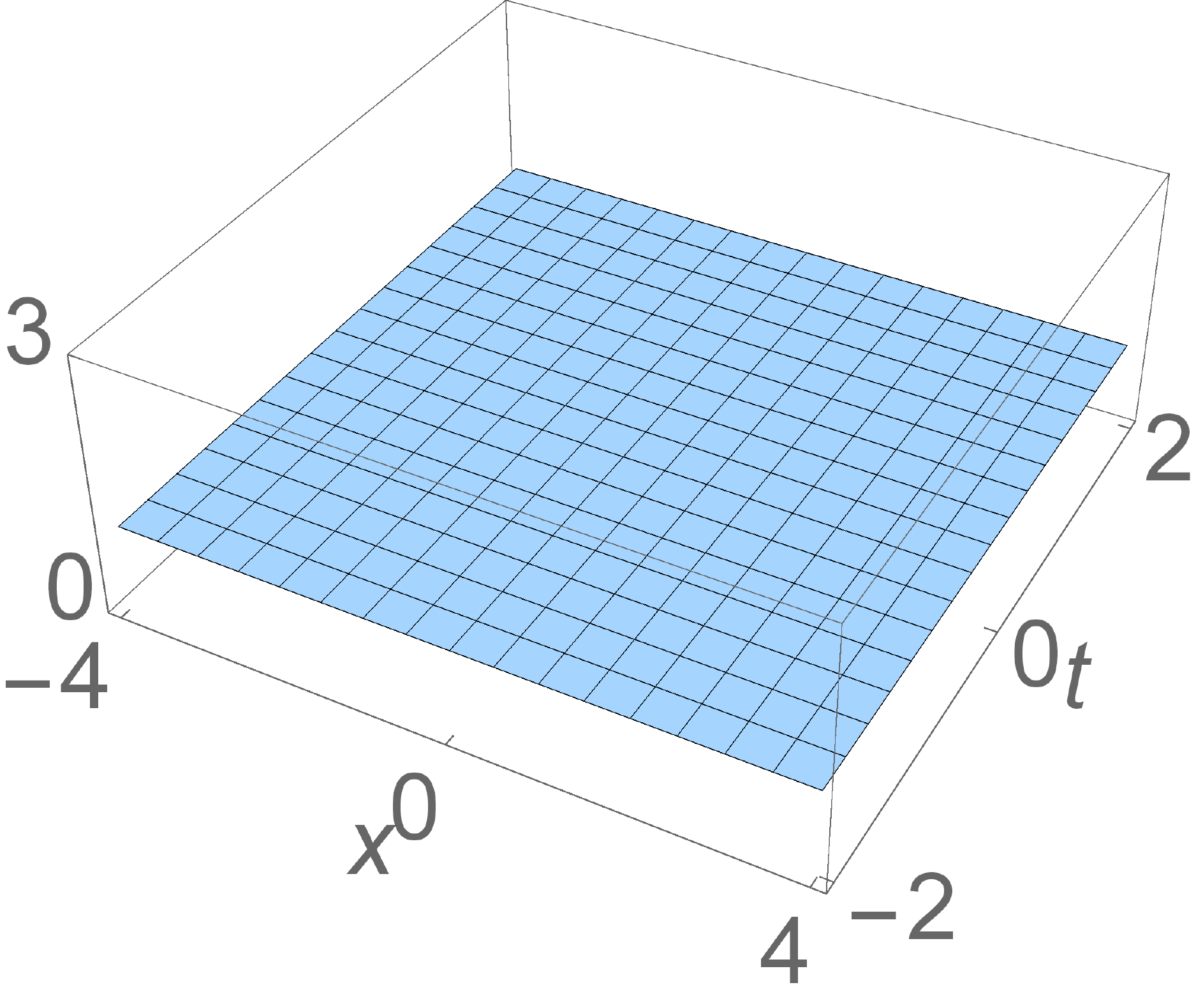}
    \caption{
        Same as Fig.~\ref{f:rationalA},
        but for rational solutions in class~B with $(x_o,t_o) = (0,0)$.
        Left: the rational solution from Eq.~\eref{e:PhirationalB} with $\eta = \pi/4$.
        Right: the corresponding core component from Eq.~\eref{e:QrationalB}.
    }
    \label{f:rationalB}
\end{figure}

In this case $Q(x,t)$ is given by Eq.~\eref{e:QB}.
As $Z\to1$ with $\zeta = iZ$ and $\gamma$ as in Eq.~\eref{e:rationallambda1},
one obtains
\vspace*{-0.6ex}
\[
\label{e:QrationalB}
Q_\P(x,t) = \diag(q_\P(x-x_o,t-t_o),1)\,,
\]
which has a peak at $(x_o,t_o)$.
Of course, the general rational solution is obtained from Eq.~\eref{e:PhiQ},
with the general unitary matrix $U = U_B = U_A$ in Eq.~\eref{e:UA}, i.e.
\vspace*{-0.6ex}
\bse
\label{e:PhirationalB}
\begin{gather}
\phi_1(x,t) = q_\P(x-x_o,t-t_o) \sin^2\eta + \cos^2\eta\,,\\
\phi_0(x,t) = [q_\P(x-x_o,t-t_o) - 1]\sin(2\eta)/2\,,\\
\phi_{-1}(x,t) = q_\P(x-x_o,t-t_o) \cos^2\eta + \sin^2\eta\,,
\end{gather}
\ese
where again $-\pi/2 <\eta < \pi/2$.
Since one can shift the position of the peak by using the translation invariance of the spinor model,
there is only one genuine parameter $\eta$ in Eq.~\eref{e:PhirationalB},
which characterizes the spin rotation.
We point out that, when $\eta = \pi/4$, the solution~\eref{e:PhirationalB}
coincides with the one obtained by direct methods in Ref.~\cite{qm2012}.
An example of a rational solution $\Phi(x,t)$ in class~B and the corresponding core component $Q(x,t)$ are shown in Fig.~\ref{f:rationalB}.
With a generic parameter $\eta$,
the peak in $\phi_0$ corresponds to the potential traps in both $\phi_{\pm1}$.
The essential difference between rational solutions in classes~A and~B is obviously the number of Peregrine solitons in the core component,
which determines the number of peaks in the component $\phi_0$.

\paragraph{\bf Class~C.}
Substituting $\zeta = i Z$ with $\alpha = 0$ into the expression for $Q(x,t)$ from Eq.~\eref{e:QC},
we have
\vspace*{-0.4ex}
\begin{gather*}
q_{1,1}(x,t) = q_{2,2}(x,t) = \tanh\chi(x) \,,\\
q_{1,2}(x,t) = -i \frac{1}{Z} \e^{-i s(t)} \sech\chi(x)\,,\\
q_{2,1}(x,t) = -i Z \, \e^{i s(t)} \sech\chi(x)\,,
\end{gather*}
where now
\vspace*{-1ex}
\[
\chi(x) = c_{-,1} x + \log(2Z/\xi)\,,\quad s(t) = c_{-,2} t - \phi\,,
\nonumber
\]
and $\gamma = \xi \e^{i\phi}$ as before.
Then it is easy to show that $Q(x,t)$ tends to a trivial constant matrix
regardless of the norming constant as $Z\to1$, namely
\vspace*{-1ex}
\[
Q_\P(x,t) = \frac{1}{\xi^2 + 4}\begin{pmatrix}
4 - \xi^2 & -4 i \xi \e^{i \phi } \\
-4 i \xi \e^{-i \phi } & 4 - \xi^2
\end{pmatrix}\,.
\label{e:QrationalC}
\]
Thus, no nontrivial rational solutions are obtained as limits of soliton solutions in class~C.

\begin{figure}[t!]
        \centering
        \includegraphics[scale=0.22]{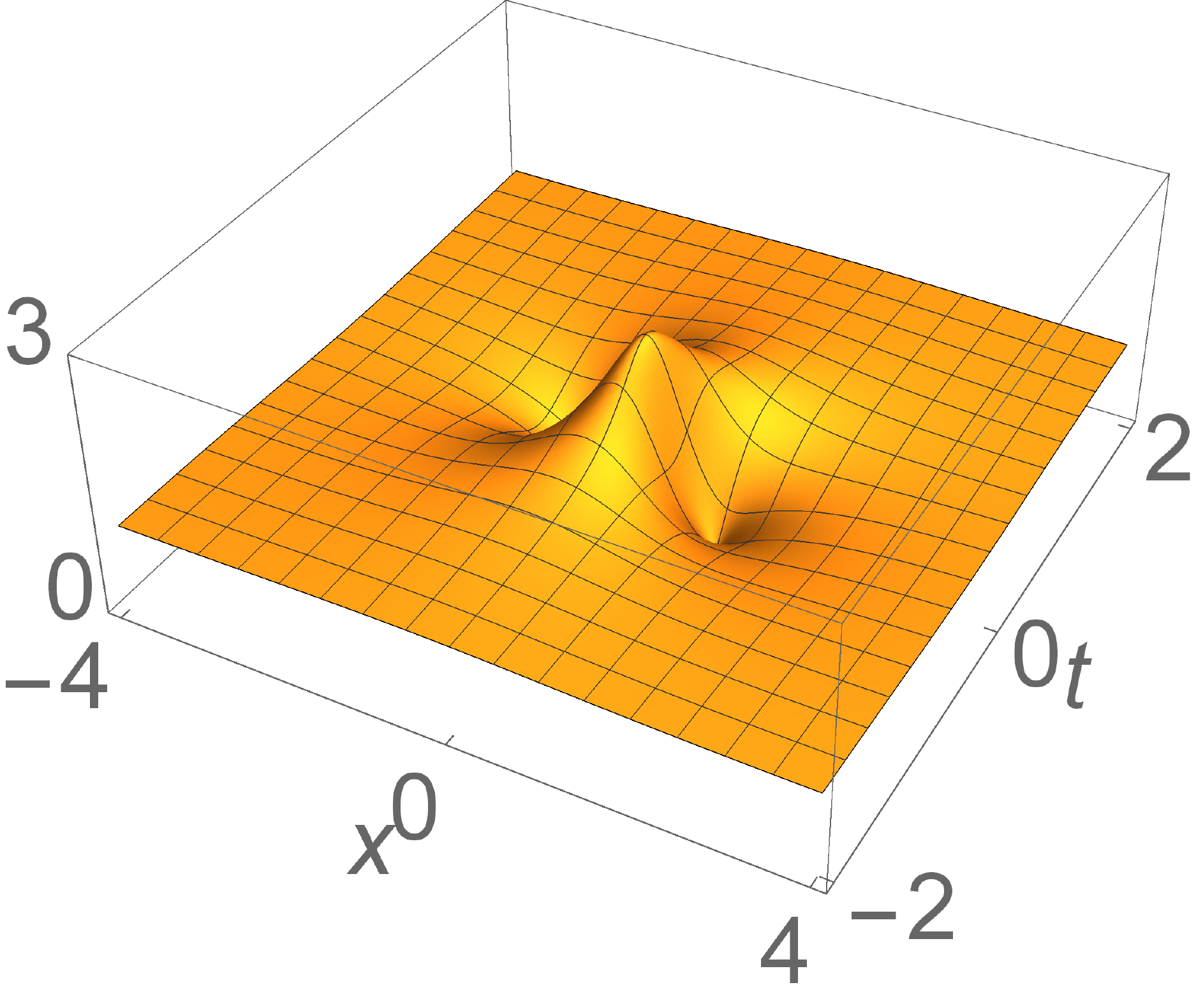}\,
        \includegraphics[scale=0.22]{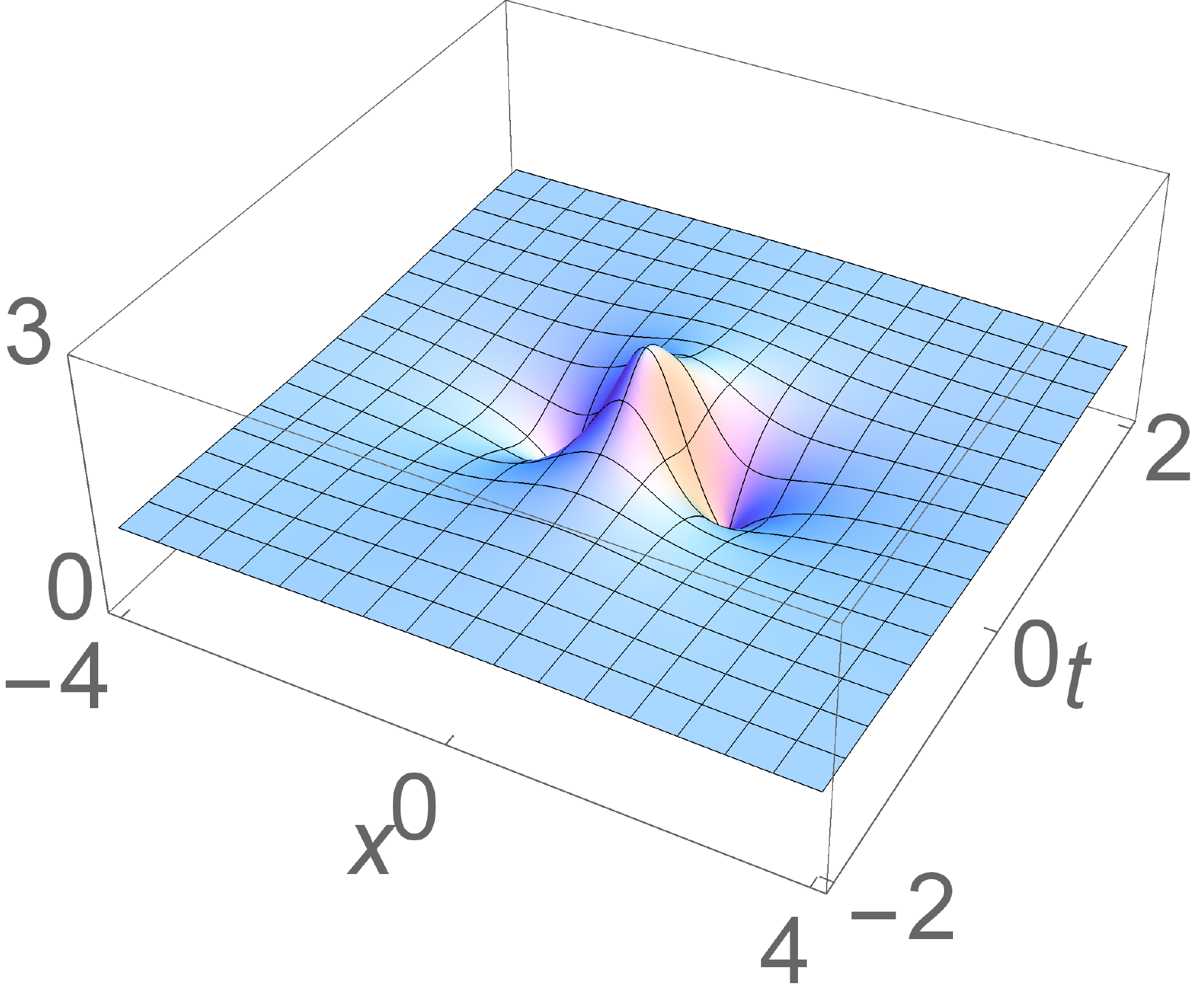}\\
        \includegraphics[scale=0.22]{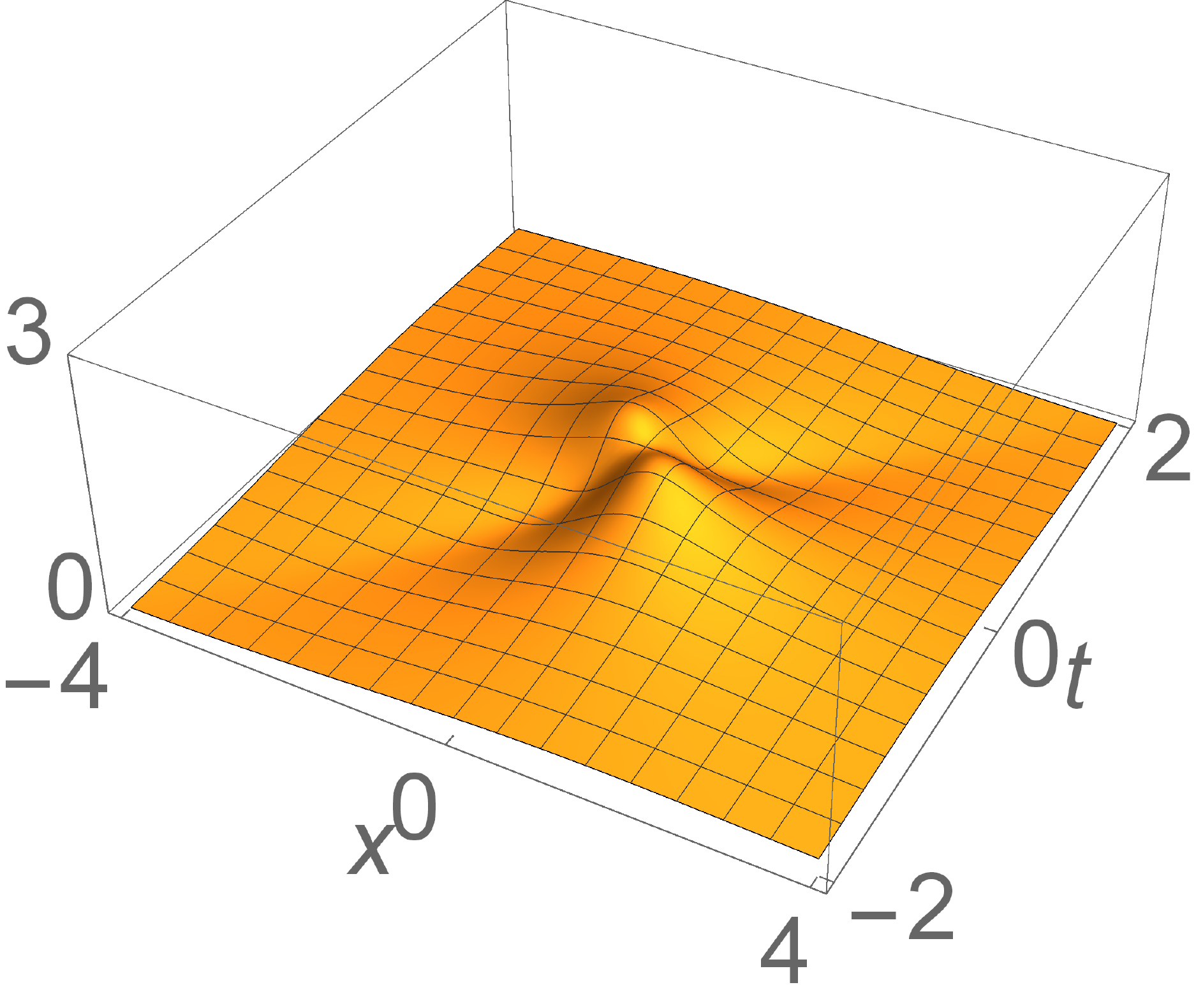}\,
        \includegraphics[scale=0.22]{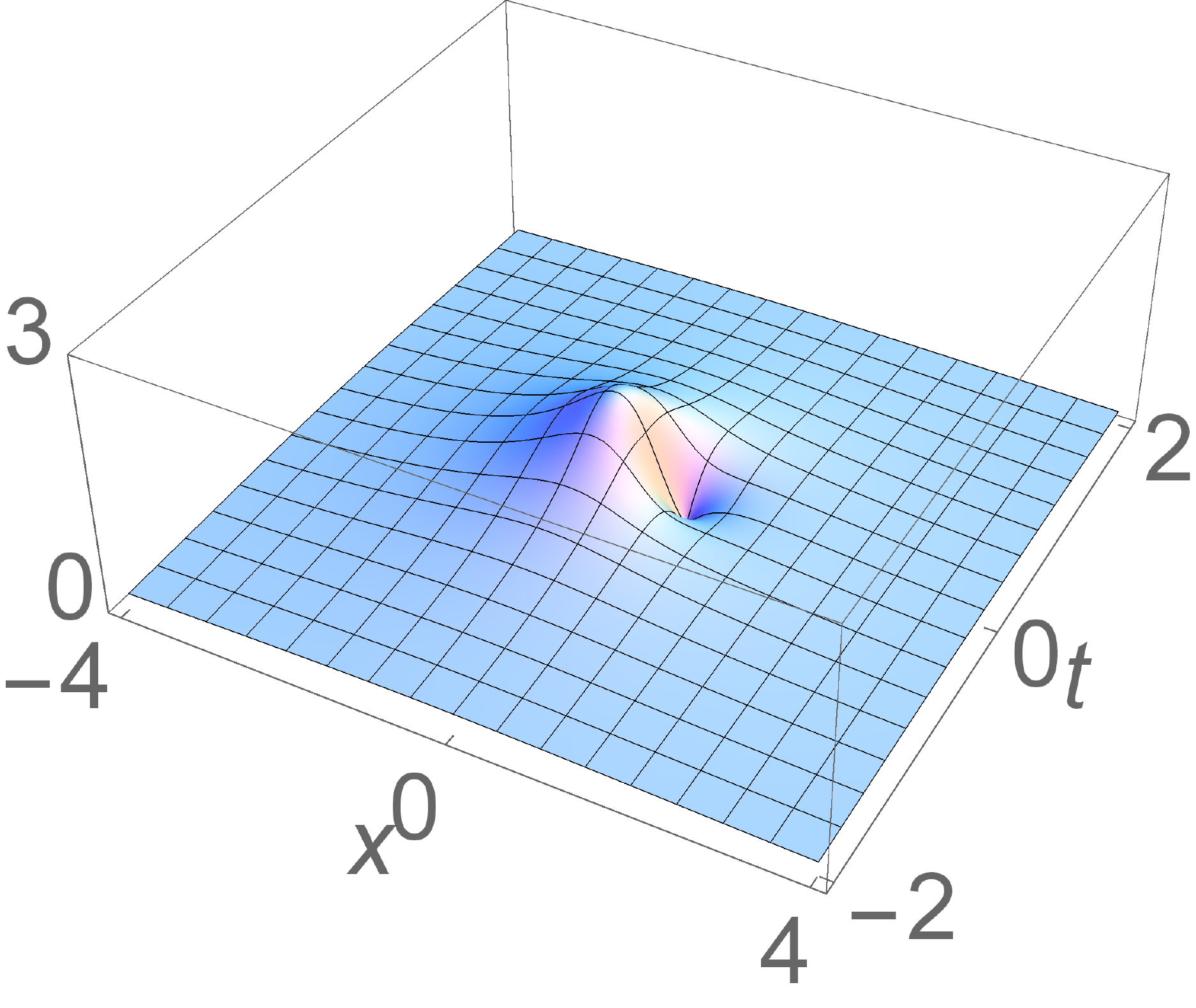}\\
        \includegraphics[scale=0.22]{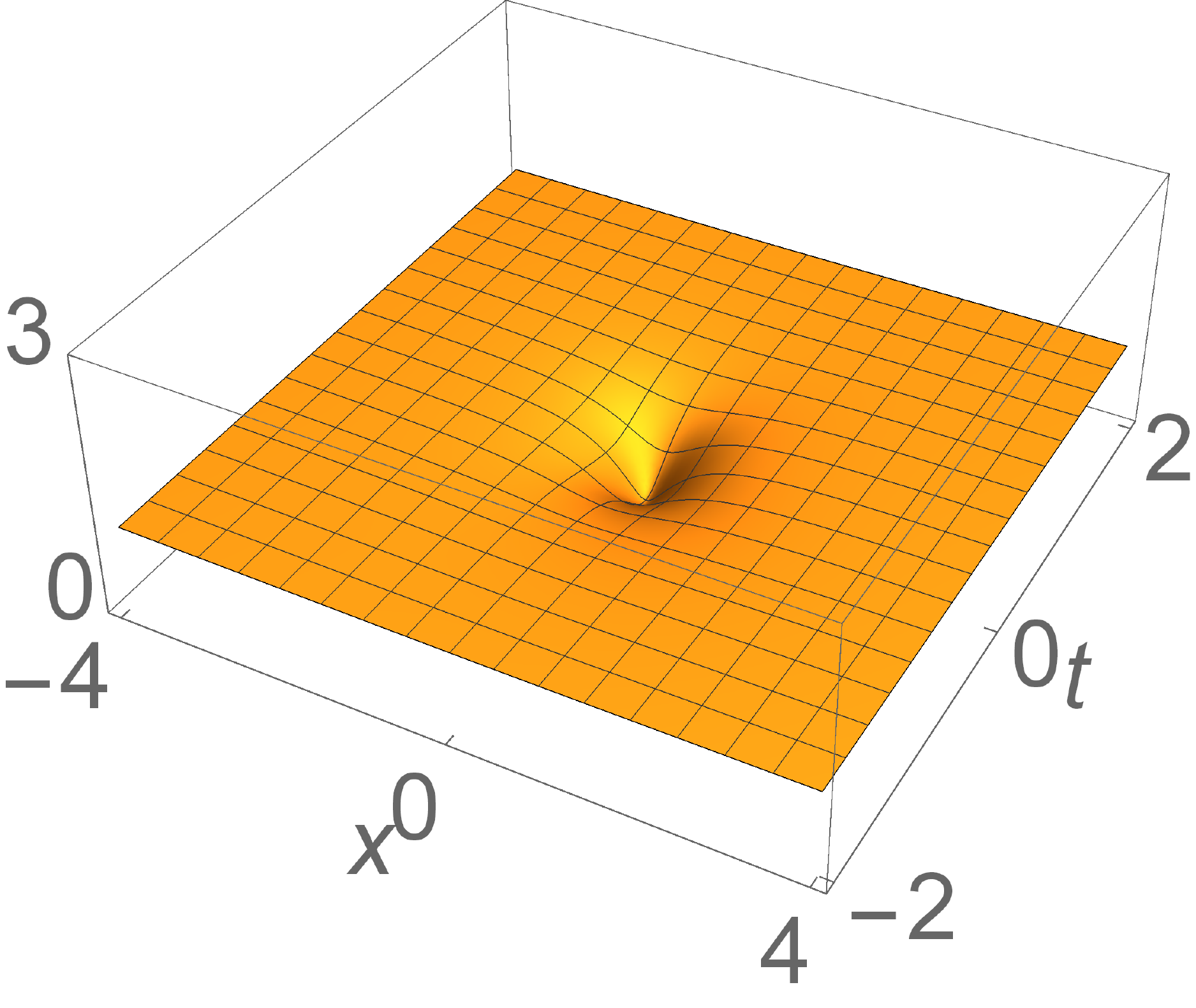}\,
        \includegraphics[scale=0.22]{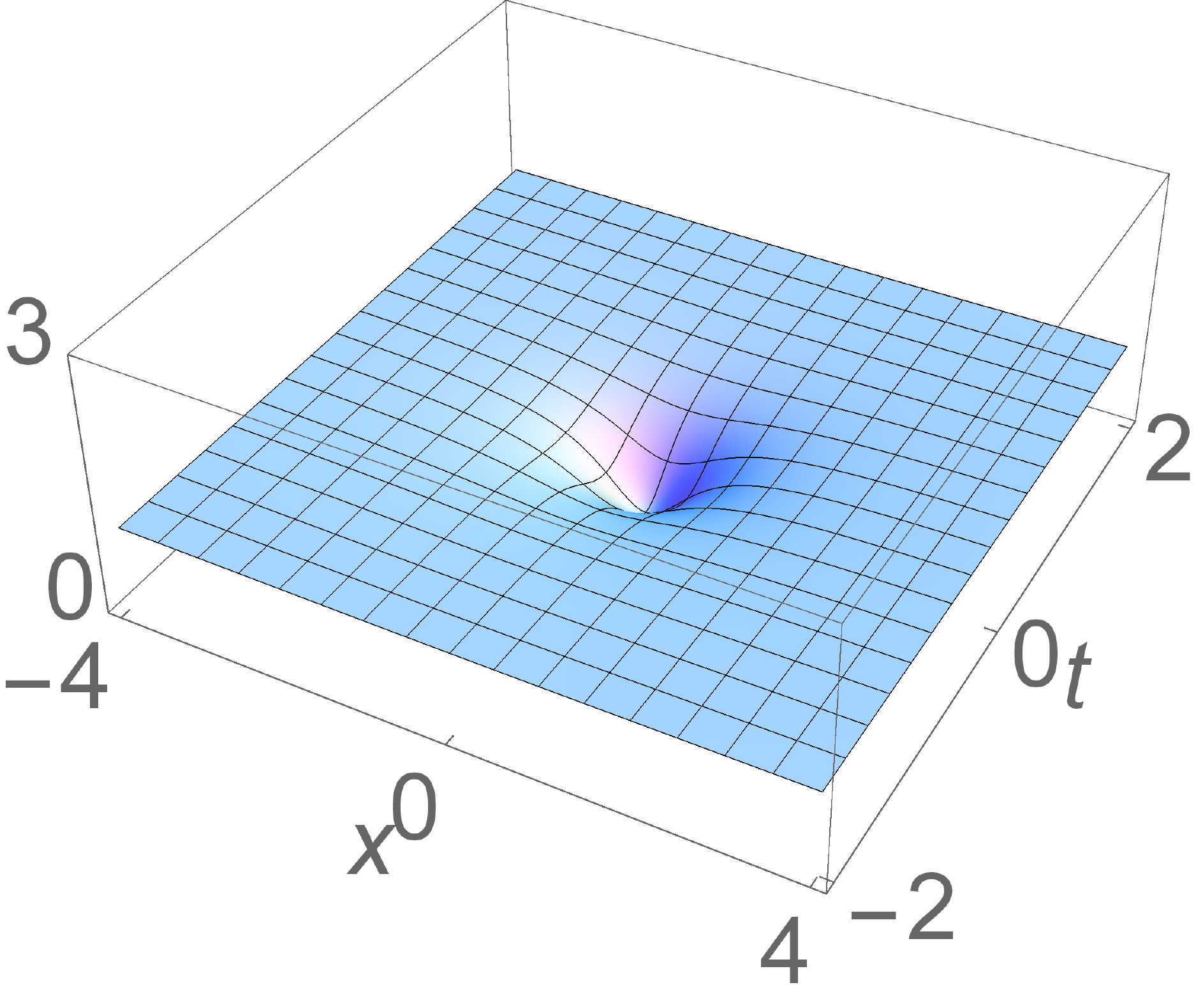}
    \caption{
        Similarly to Fig.~\ref{f:rationalA},
        but for rational solutions in class~D centered at the origin with $\eta = \pi/8$.
        Left: the rational solution from Eq.~\eref{e:PhiQ} with $U_D$ defined in Eq.~\eref{e:KUD} and $\beta_2 = \pi/3$ .
        Right: the core component from Eq.~\eref{e:QrationalD} with $|q_{1,1}(x,t)|$ (top),
        $|q_{1,2}(x,t)| = |q_{2,1}(-x,t)|$ (center) and $|q_{2,2}(x,t)|$ (bottom).
    }
    \label{f:rationalD}
\end{figure}

\paragraph{\bf Class~D.}
Taking the limit $Z\to1$ of $Q(x,t)$ from Eq.~\eref{e:QDcomponent} with
\[
\nonumber
\gamma = \frac{2i Z(Z^2-1) |\cos(2\eta)|}{\sqrt{Z^4 + 2Z^2\cos(4\eta) +1}}\e^{c_{-,1}x_o + i c_{-,2}t_o}\,,
\]
we have
\vspace*{-1ex}
\bse
\label{e:QrationalD}
\begin{gather}
q_{1,1}(x,t) = 1 -2 \frac{8i (t - t_o) + \sec ^2(2 \eta )  + 1}{\Delta(x-x_o,t-t_o)}\,,\\
q_{1,2}(x,t) = -2 \tan (2 \eta ) \frac{4 i (t -  t_o) - 2 (x - x_o) + 1}{\Delta(x-x_o,t-t_o)}\,,\\
q_{2,1}(x,t) = -2 \tan (2 \eta ) \frac{4 i (t - t_o) + 2 (x - x_o) + 1}{\Delta(x-x_o,t-t_o)}\,,\\
q_{2,2}(x,t) = 1-\frac{2 \tan ^2(2 \eta )}{\Delta(x-x_o,t-t_o)}\,,
\end{gather}
\ese
where
\vspace*{-1ex}
\[
\label{e:Delta}
\Delta(x,t) = 16 t^2 + 4 x^2 + \sec ^2(2 \eta )\,.
\]
The corresponding solution $\Phi(x,t)$ is obtained by
composing Eq.~\eqref{e:QrationalD} with the unitary matrix in Eq.~\eref{e:KUD}.
This yields a two-parameter family of rational solutions, whose explicit expression is omitted for brevity.
An example of a rational solution $\Phi(x,t)$ in class~D and the corresponding core component $Q(x,t)$ from Eq.~\eref{e:QrationalD}
are shown in Fig.~\ref{f:rationalD}.
The one-to-one correspondence between potential traps and peaks is observed again among the spin states.

\paragraph{\bf Class~E.}
In this case there are two possible approaches to obtain a rational solution.
One can show that in order to obtain a nontrivial limit $Z\to1$ with $\zeta = i Z$,
one needs $\det \Gamma_E \to 0$, but the two diagonal entries of $\Gamma_E$ cannot both tend to zero.
Thus, we consider two cases separately.
First, we let the $(1,1)$-component of $\Gamma_E$ tend to zero,
i.e., we parametrize the Schur form $\Gamma_E$ as
\vspace*{-0.6ex}
\[
\nonumber
\Gamma_{E} = \begin{pmatrix}
\gamma & \gamma_0  \\
0 & \gamma - \e^{i \beta } \gamma_0  \cot (2 \eta )
\end{pmatrix}\,,
\]
with $\gamma = \gamma_P$ and $\gamma_P$ given by Eq.~\eqref{e:rationallambda} as before.
Then the entries of the core component as $Z\to1$ are
\begin{gather*}
q_{1,1}(x,t) = 1 -2 [8 i t +1 +\sec^2 (2 \eta)]/\Delta(x,t)\,,\\
q_{1,2}(x,t) = -2 \e^{-i \beta }\tan(2\eta) [4 i t-2 x+1]/\Delta(x,t)\,,\\
q_{2,1}(x,t) = -2 \e^{i \beta }\tan(2\eta) [4i t+2 x + 1]/\Delta(x,t)\,,\\
q_{2,2}(x,t) = 1 - 2\tan^2(2\eta)/\Delta(x,t)\,,
\end{gather*}
with $\Delta(x,t)$ given by Eq.~\eref{e:Delta}.
Using the same matrix $U_E$ from Eq.~\eref{e:KUE},
we have
\begin{gather*}
\phi_1(x,t) = 1-4\cos^2\eta\sec(2\eta)[4it+\sec(2\eta)]/\Delta(x,t)\,,\\
\phi_0(x,t) = 4i\tan(2\eta) x/\Delta(x,t)\,,\\
\phi_{-1}(x,t) = 1+4\sin^2\eta\sec(2\eta)[4it-\sec(2\eta)]/\Delta(x,t)\,.
\end{gather*}
One can show that the above solution $\Phi(x,t)$ coincides with the one from class~D centered at the origin,
obtained by using $Q(x,t)$ from Eq.~\eref{e:QrationalD} and the unitary matrix $U_D$ from Eq.~\eref{e:KUD} with $\beta_2 = \pi/2$.
Thus, this kind of rational solutions obtained from class~E are equivalent to those from class~D.

Alternatively, one can let the $(2,2)$-component of $\Gamma_E$ tend to zero.
i.e., take $\Gamma_{E}$ as
\vspace*{-0.6ex}
\begin{gather*}
\gamma_{E} = \begin{pmatrix} \gamma - \e^{i \beta } \gamma_0  \cot(2 \eta) & \gamma_0 \\
0 & \gamma
\end{pmatrix},
\end{gather*}
with $\gamma$ as above.
Note however that the resulting solution is equivalent to the one obtained above.
The reason why this is the case is that the two parametrizations above for the norming constant can be obtained
from each other by simply switching the diagonal entries, which can be done by a unitary similarity transformation.
Therefore the corresponding solutions are in the same equivalence class.
Thus, all rational solutions obtained from class~E are equivalent to those from class~D.

\paragraph{\bf Inequivalence of rational solution families.}

Even though soliton solutions in different Schur classes are inequivalent, this might not be the case for the corresponding
rational solutions.
In other words,
since the limit $Z\to 1$ is a singular limit for the norming constants, and inequivalent Schur forms might reduce to the same ones in the limit,
it is not obvious a priori that the rational solutions obtained from different Schur classes would be inequivalent.
As a matter of fact, we have already seen that the rational solutions obtained from classes~D and~E are equivalent.
On the other hand, we next show that the rational solutions obtained from classes~A,~B and~D are indeed inequivalent.

Recall that the trace and the determinant of a matrix are invariant under similarity transformations.
Thus, the equality of their trace and their determinant
is a necessary condition for two solutions $\Phi_1(x,t)$ and $\Phi_2(x,t)$
to be equivalent.
That is, if two solutions have different trace or different determinant, they are inequivalent.
In light of this observation, we compute the traces of rational solutions in classes A, B and D, obtaining:
\vspace*{-0.6ex}
\begin{align*}
&\tr\,\Phi_A(x,t)
\nonumber\\
&\kern2em { }
  = 2 - 4 [ f(x-x_1,t-t_1) + f(x-x_{-1},t-t_{-1})]\,,\\
&\tr\,\Phi_B(x,t) = 2 - 4 f (x-x_o,t-t_o)\,,
\\
&\tr\,\Phi_D(x,t)
\nonumber\\
&\kern2em{ }
  = 2 - 4[4i(t - t_o) - \sec^2(2\eta)]/\Delta(x-x_o,t-t_o)\,,
\end{align*}
where $f(x,t)$ and $\Delta(x,t)$ are given by Eq.~\eref{e:Peregrine} and Eq.~\eref{e:Delta}, respectively.
If $\eta = 0$, one has $\tr\,\Phi_D(x,t) = \tr\,\Phi_B(x,t)$ up to a spatial and temporal shift.
Note however that, in the classification of the norming constants, $\eta \ne0$ in class~D,
because when $\eta = 0$, class~D reduces to class~B.
So, all three traces are distinct.
Consequently, the three classes of rational solutions are all inequivalent,
and therefore represent three distinct families of rational solutions of the spinor BEC model.

As mentioned before, the rational solutions in class~B are equivalent, up to unitary transformations,
to one of the solutions derived in Ref.~\cite{qm2012}.
The other two families of solutions however are new to the best of our knowledge.
Moreover, while the rational solutions in classes~A and~B are reducible,
in the sense that they are equivalent to scalar rational solutions (i.e., Peregrine solitons of the scalar NLS equation),
the rational solutions in class~D cannot be reduced to scalar rational solutions.

Finally, we would like to comment on the singular nature of the limit $Z\to1$.
We have shown that this limit can give rise to rational solutions.
On the other hand,
it is evident from Table~\ref{t:norming}
that some of the Schur forms are special reductions of others.
(For example, $\Gamma_A$ with $\gamma_{-1}=0$ reduces to $\Gamma_B$.)
The soliton solutions with eigenvalues of kinds~1--3 inherit the reductions of the Schur forms.
(For example, one can easily check that the soliton solution~\eref{e:PhiA} in class~A reduces to the one~\eref{e:PhiB} in class~B,
when $\gamma_{-1}=0$,
in which case the TW soliton $q_{\tw,-1}(x,t)$ reduces to the trivial non-zero background $k_o$.)
However, in order to obtain the rational solution~\eref{e:PhirationalB} in class~B
from the solution~\eref{e:PhirationalA} in class~A
one must consider either of the singular limit $x_{-1}\to\infty$ or $t_{-1}\to\infty$.


\paragraph{\bf Spin state.}

Direct calculations show that all three classes of rational solutions have a zero spin $\@F=(0,0,0)$
(cf. Appendix~I for definition),
regardless of all other parameters.
Thus all three classes of spinor BECs rogue waves correspond to the polar states.
One should note that the rational solutions in classes~B and~D are derived from a ferromagnetic state,
which confirms the singular nature of the limit.
Moreover, because the spin of these rogue waves is zero,
when they interact with other waves,
they will not affect their spin state.

\section{VI.~ Concluding remarks}

We have presented a classification of one-soliton solutions of spinor BECs with ZBC and NZBC,
and we have derived novel families of rogue-wave solutions of spinor BECs.
We have shown that one-soliton solutions with ZBC are always reducible,
in the sense that there always exists a unitary transformation that relates them to solutions of single-component BECs.
On the other hand, we have also shown that
solutions with NZBC are divided into reducible and irreducible classes.
Moreover, we showed that there exist two inequivalent classes of one-soliton solutions with ZBC
(corresponding to ferromagnetic versus polar states),
five inequivalent classes of one-soliton solutions with NZBC,
and three inequivalent classes of rational solutions.
The classification of all inequivalent solitons and rational solutions is of course important in order to single out the fundamental properties of the solutions,
and peel off the complications introduced by simple rotations of the quantization axes.
In particular, in this work we also used the classification to prove the inequivalence of the three families of rational solutions presented,
and hence the novelty of the two families obtained as limits of soliton solutions in classes~B and~D.
%

We also discussed the physical properties of solitons and rational solutions.
We showed that some solutions exhibit oscillating pairs of potential traps and peaks,
that resemble the behavior of dark-bright soliton complexes in the focusing vector NLS.
Other solutions, on the other hand, are topological solitons and form domain walls.
The domain walls can be analyzed through the core solutions.
In particular, the velocity in Eq.~\eref{e:velocity} also corresponds to the velocity of the wall.
From a physical point of view,
all soliton solutions with NZBC can be categorized into either polar or ferromagnetic state depending on their total spin being zero or not,
and this corresponds to having topological or non-topological solitons.

We emphasize that
the classification introduced in this work also applies to the full matrix NLS equation, either with ZBC or NZBC,
namely Eq.~\eqref{e:matrixNLS}, but where now $\Phi(x,t)$ is not necessarily a symmetric matrix,
i.e., when the constraint~\eqref{e:symmetryconstraint} does not apply.
The only differences from the analysis of this work are that, for the full matrix NLS equation,
the core soliton components are themselves always solutions of the matrix system,
and different solutions can be combined via arbitrary (i.e., unconstrained) unitary matrices.

Another open question is whether even more general soliton solutions can be obtained which do not satisfy the
constraint~\eqref{e:constraint}.
It should be noted that this is also an open problem in the case of the vector NLS equation.

While matter-wave solitons in one- and two-component systems have been extensively studied and observed experimentally,
an extension to three components (and hence to spinor systems) had not been attempted in experiments until most recently:
in Ref.~\cite{2017arXiv170508130B} the existence of robust dark-bright-bright and dark-dark-bright solitons in a defocusing spinor $F=1$ condensate has been reported.
Although in general, the systems considered in the experiments are non-integrable
(see, for instance Refs.~\cite{ku2012,sku2013,kf2016,2017arXiv170508130B,ppw2008}),
one can get useful insight into their behavior using perturbation techniques of related integrable systems.
For instance, the model equation in Ref.~\cite{2017arXiv170508130B} can be considered a small perturbation of a 3-component vector NLS equation.
In this respect, the theoretical predictions for the soliton solutions in the integrable case can be an extremely valuable tool for the investigation of the non-integrable solitary waves in regimes that are not too far from the integrable one.
While to date there is not yet an experimental realization of the exact focusing system on a non-zero background that we considered in our work,
we note that the polar and ferromagnetic solitons analyzed in Ref.~\cite{llmml2005} for the $F=1$ spinor system were found to be structurally stable,
i.e., robust under random changes of the relevant nonlinear coefficients in time.
This suggests that the solitons,
and possibly the rogue waves derived in our work,
could be soon observed experimentally,
in models that may be at least perturbatively close to ours.
\section{Acknowledgments}

This work was partially supported by the National Science Foundation under grant numbers
DMS-1615524, DMS-1614623 and DMS-1614601.

\addcontentsline{toc}{section}{Appendix}

\setcounter{section}1
\setcounter{subsection}0
\setcounter{equation}0
\def\thesection{\Alph{section}}
\def\theequation{\Alph{section}.\arabic{equation}}

\section*{Appendix~I: Total spin and spin states}

As shown in Ref.~\cite{h1998},
the spin-1 BECs are either in a polar state or in a ferromagnetic state,
depending on a conserved quantity---total spin---in the ground state.
In particular, a polar state corresponds to a zero total spin,
whereas a ferromagnetic state corresponds to a nonzero total spin.
We next investigate the total spin of the one-soliton solutions~\eref{e:1soliton_nzbc} explicitly.

The spin density $\@f = (f_{-1},f_0,f_1)$ is defined by
\[
\nonumber
\@f = \tr(\Phi^\dagger\bsigma\Phi)\,,\qquad
\bsigma = (\sigma_1,\sigma_2,\sigma_3)\,,
\]
where $\sigma_j$ for $j=1,2,3$ are the Pauli matrices,
$\Phi = \Phi(x,t)$ is given by Eq.~\eref{e:1soliton_nzbc},
so the spin is $\@F = \int\@f \,\d x$,
and the total spin is $\|\@F\|$~\cite{imw2004}.
We first rewrite the spin $\@F$ as
\[
\nonumber
\@F = \tr(\bsigma\int\nolimits_\Real\Phi\Phi^\dagger\d x) = \tr(\bsigma\int\nolimits_\Real(\Phi\Phi^\dagger - k_o^2I_2)\,\d x)\,.
\]
It is then evident that one only needs to compute the following integral
\[
\label{e:int}
I = \int\nolimits_\Real (\Phi\Phi^\dagger - k_o^2I_2)\,\d x\,,
\]
in order to determine the spin $\@F$ corresponding to the solution $\Phi$.
In other words, the spin $\@F$ is the projection of $I$ onto the Pauli matrices.
Note that the term $k_o^2 I_2$ must be added so that this integral is convergent on the line.
It is worth mentioning that in this work the total spin is used to
characterize the polar state, whereas in \cite{h1998} the spin
state is defined by the local spin, i.e., the polar state
satisfies $\@f=0$, instead of $\@F=0$.

The integral in Eq.~~\eref{e:int} is difficult to compute directly from the solution~\eref{e:1soliton_nzbc}.
However, it is well known that the IST provides an easier way to get conserved quantities in terms of asymptotics of eigenfunctions and scattering data.
In particular, one can derive for the asymptotic behavior of one of the eigenfunctions as the following expression:
\[
\label{e:Nup}
Y = I_2 - \frac{i}{2}\int_x^\infty (\Phi\Phi^\dagger - k_o^2 I_2)\,\d x' + O(1/z^2)\,,\quad
z\to\infty\,,
\]
where $z$ is the spectral parameter~\cite{pdlhf}.
(Note that $Y(x,t,z)=\bar{N}^{up}(x,t,z)$ in the notation of Ref. [48])
Relating the integral $I$ in Eq.~\eref{e:int} to the asymptotics~\eref{e:Nup},
we have
\[
\nonumber
I = -i\lim_{x\to-\infty}\lim_{z\to\infty} z(I_2 - Y)\,.
\]
The eigenfunction $Y$ can be reconstructed from the inverse problem of the IST.
Specifically,, $Y$ is given by a linear system and can be computed explicitly in the case of pure soliton solutions.
Omitting the details,
the following reconstruction formula for $I$ holds,
\[
\label{e:Ireal}
I = \lim_{x\to-\infty} \bigg( \frac{k_o}{\zeta}\e^{2i\theta}X_2 K + \frac{k_o}{\zeta^*}\e^{-2i\theta^*}X_1 K^\dagger\bigg)\,,
\]
where $\zeta$ is the discrete eigenvalue,
$K$ is the norming constant,
$\theta$ is given by Eq.~\eref{e:theta}
and $X_j$ with $j = 1,2$ are given by the same linear system~\eref{e:X1X2}.
We refer the reader to the recent IST formulation presented in Ref.~\cite{pdlhf} for details.

Solving the linear system~\eref{e:X1X2},
we have
\begin{gather*}
X_1 = [I_2 - i k_o (D^\dagger)^{-1}c/\zeta][D+c^\dagger (D^\dagger)^{-1}c]^{-1}\,,\\
X_2 = (I_2 - i \zeta D^{-1} c^\dagger/k_o)(D^\dagger + c D^{-1}c^\dagger)^{-1}\,.
\end{gather*}
In order to further simplify the expression for $X_j$,
we need to consider the two cases $\det K=0$ and $\det K\ne0$ separately.
Also, one should notice that for all $t$ as $x\to-\infty$,
\[
\nonumber
O(\e^{2i\theta}) = O(\e^{-2i\theta^*})\to\infty\,,\quad
O(\e^{-2i\theta}) = O(\e^{2i\theta^*})\to0\,.
\]
Below we use the above formulas to compute the total spin of the soliton solutions.

\paragraph{\bf Full rank norming constant $K$.}

If the norming constant $K$ is such that $\det K \ne 0$,
after some tedious calculations, we have the following asymptotics as $x\to-\infty$,
\begin{gather*}
X_1 = -i\frac{k_o(\zeta-\zeta^*)((\zeta^*)^2 _+ k_o^2)}{\zeta(k_o^2 + |\zeta|^2)}\e^{2i\theta^*}(K^\dagger)^{-1} + O(\e^{-4i\theta})\,,\\
X_2 = i \frac{\zeta(\zeta-\zeta^*)(\zeta^2+k_o^2)}{k_o(k_o^2 + |\zeta|^2)}\e^{-2i\theta} K^{-1} + O(\e^{-4i\theta})\,.
\end{gather*}
Substituting the above asymptotics into Eq.~\eref{e:Ireal}, we have
\[
\nonumber
I = \frac{\zeta-\zeta^*}{k_o^2 + |\zeta|^2}\bigg\{\frac{k_o^2}{|\zeta|^2}[(\zeta^*)^2 + k_o^2] - \zeta^2 - k_o^2\bigg\}I_2\,.
\]
Notice that as we expected,
$I$ is independent of $t$.
Moreover, $I$ is diagonal implying that
\[
\nonumber
f_j = \tr(\sigma_{j+2} I) =0\,,\qquad j = -1,0,1\,.
\]
Thus, we conclude that the total spin $\|\@F\| = 0$ and hence the BEC is in a polar state.

\paragraph{\bf Rank one norming constant $K$.}

We next consider the case where $\det K =0$ and $K\ne0$,
which implies that $K$ cannot be diagonal.
Similarly to the previous case, we can calculate asymptotics of all needed quantities.
However, because $\det K = 0$,
these asymptotics are more complicated than before.
Without showing the details, we give the final result
\begin{gather*}
I = \frac{2i\Re[d_o(\tr K)^* K/\zeta] + c_o^* K^\dagger K + k_o^2 c_o K K^\dagger/|\zeta|^2}
{|d_o|^2 \tr K \,\tr K^\dagger + |c_o|^2 \tr(K^\dagger K)}\,,
\end{gather*}
where
\begin{gather*}
c_o = (\zeta^*-\zeta)^{-1}\,,\qquad
d_o = i k_o/[(\zeta^*)^2 + k_o^2]\,.
\end{gather*}
Notice that $I$ contains three matrices $K$, $K^\dagger K$ and $K K^\dagger$.
Recall that $K$ is not diagonal in this situation,
so $K^\dagger K$, $K K^\dagger$ and $I$ cannot be diagonal.
Moreover, the two nondiagonal matrices $K^\dagger K$ and $K K^\dagger$ are Hermitian.
Their projections on the Pauli matrices cannot be identically zero.
This implies that $\@F\ne(0,0,0)$.
In other words, the total spin is nonzero and the BEC is in a ferromagnetic state.

One can then obtain the corresponding results for one-soliton solutions~\eref{e:1soliton_zbc} with ZBCs,
by simply taking the limit $k_o\to0$.
One shows that the BECs are in a polar state when $\det K \ne 0$,
and are in a ferromagnetic state when $\det K = 0$.

\section*{Appendix~II: Equivalence classes, norming constants and unitary transformations}

Recall the Schur decomposition~\eref{e:Schur}.
We next discuss the equivalence classes of norming constants $K$ according to their Schur form $\Gamma$,
and we identify the corresponding families of unitary similarity transformations allowed in each class.
Recall that the diagonal entries of $\Gamma$ are simply the eigenvalues of $K$.
As discussed earlier, there are five different equivalence classes.
Note that any unitary matrix can be parametrized by four independent real parameters in general,
but at least one parameter is fixed by the requirement that $K = U\Gamma U^\dagger$ be symmetric.
Consequently, unitary matrices defining similarity transformations of solutions of the spinor BECs
contain at most three independent real parameters.

\textbf{A.}
If $K$ is diagonalizable by a unitary matrix and has two nonzero eigenvalues,
its Schur form is given by $\Gamma_A$ in Eq.~\eqref{e:Schurparametrization},
where $\gamma_{\pm1}$ are the nonzero complex eigenvalues of $K$.
The most general parameterization of the unitary matrix in this case is $U = U_A$, with
\vspace*{-1ex}
\[
\label{e:UA}
U_A = \e^{i\beta_1}\begin{pmatrix}
\sin\eta & \e^{i \beta_2} \cos\eta \\
\cos\eta & -\e^{i \beta_2} \sin\eta \\
\end{pmatrix}\,,
\]
with $0\le\beta_1<2\pi$, $0\le\beta_2<\pi$ and $0\le\eta<2\pi$.
As a result, the most general form for the norming constant in class~A is given by
\vspace*{-1ex}
\[
\nonumber
K_A = \frac{1}{2} \begin{pmatrix}
K_{1,1} & K_{1,2} \\ K_{1,2} & K_{2,2} \end{pmatrix}\,.
\]
with $K_{1,1} = 2\gamma_1\sin^2\eta + 2\gamma_{-1}\cos^2\eta$,
$K_{1,2} = (\gamma_1-\gamma_{-1}) \sin(2\eta)$
and $K_{2,2} = 2\gamma_1\cos^2\eta + 2\gamma_{-1}\sin^2\eta$.

\textbf{B.}
If $K$ is diagonalizable by a unitary matrix and has one zero eigenvalue,
its Schur form is given by $\Gamma_B$ in Eq.~\eqref{e:Schurparametrization}.
Note that one could interchange the two diagonal entries by letting $\tilde\Gamma_B = \sigma_1\Gamma_B\sigma_1$, with
\vspace*{-0.6ex}
\[
\nonumber
\sigma_1 = \begin{pmatrix}0 & 1 \\ 1 & 0\end{pmatrix}\,.
\]
Thus, one can use the Schur form $\Gamma_B$ without loss of generality.
We have $U_B = U_A$ from Eq.~\eref{e:UA}.
As a result, any norming constant in this class has the form
\vspace*{-1ex}
\[
\nonumber
K_B = \frac{\gamma}{2}\begin{pmatrix}
1-\cos(2\eta) & \sin(2\eta) \\
\sin(2\eta) & \cos(2\eta)+1
\end{pmatrix}.
\]

\textbf{C.}
If $K$ is non-diagonalizable by a unitary matrix and has two zero eigenvalues,
its Schur form is given by $\Gamma_C$ in Eq.~\eqref{e:Schurparametrization},
where $\gamma$ is an arbitrary non-zero complex number.
The most general parametrizations of norming constants and unitary matrices are
\vspace*{-1ex}
\bse
\label{e:KUC}
\begin{gather}
K_C = - i\frac{\gamma}{2}\e^{i (\beta_1 - \beta_2 )}\begin{pmatrix}
i  & (-1)^n  \\
(-1)^n & -i
\end{pmatrix}\,,\\
U_C =
\frac{\sqrt{2}}{2}\begin{pmatrix}
i (-1)^n \e^{i \beta_1} & -i (-1)^n \e^{i \beta_2} \\
\e^{i \beta_1} & \e^{i \beta_2} \\
\end{pmatrix}\,,
\end{gather}
\ese
where $n=0,1$, and $0\le\beta_j<2\pi$ with $j = 1,2$.
Note that, unlike classes~A and~B, $U_C$ contains only two independent real parameters.

\textbf{D.}
If $K$ is non-diagonalizable by a unitary matrix and has one zero eigenvalue,
its Schur form is given by $\Gamma_D$ in Eq.~\eqref{e:Schurparametrization},
where $\gamma$ is the complex non-zero eigenvalue and $\eta\in(-\pi/4,0)\cup(0,\pi/4)$.
As before, one could interchange the two diagonal entries by a unitary transformation, and thus choose
the $(1,1)$-component of $\Gamma$ to be the non-zero eigenvalue without loss of generality.
Also, without loss of generality
we can take the entries in the first row to have the same complex phase.
The most general unitary matrix in this case is given by
\vspace*{-1ex}
\bse
\label{e:KUD}
\begin{gather}
U_D = \e^{i\beta_1}\begin{pmatrix}
U_{1,1} & U_{1,2} \\
U_{2,1} & U_{2,2}
\end{pmatrix}\,,
\end{gather}
where $0\le\beta_1<2\pi$, $\beta_2 \in [-\pi/2,-2|\eta|)\cup(2|\eta|,\pi/2]$,
$\beta_3 = \arccos(1-2 \csc^2\beta_2 \sin ^2(2 \eta ))/4$,
and
\vspace*{-1ex}
\begin{gather*}
U_{1,1} = \cos\beta_3\,,\qquad
U_{2,1} = \e^{i \beta_2} \sin\beta_3\,,\\
U_{1,2} = \frac{1}{2} [\cos(2\beta_3) + i \cot\beta_2] \sec\beta_3 \tan (2 \eta )\,,\\
U_{2,2} = -\frac{i}{2}\,\e^{i\beta_2} [\cot\beta_2 - i \cos(2\beta_3)] \csc\beta_3 \tan(2\eta)\,.
\end{gather*}
Thus, the most general norming constant in this class is
\vspace*{-1ex}
\[
K_D = \gamma\begin{pmatrix}
K_{1,1} & K_{1,2} \\ K_{1,2} & K_{2,2}
\end{pmatrix}\,,
\]
where
\vspace*{-1ex}
\begin{gather*}
K_{1,1} = (\e^{2 i \beta_2} \tan ^2\beta_3+1)^{-1}\,,\\
K_{1,2} = \frac{1}{2}(\cos\beta_2 \csc(2\beta_3) - i \sin\beta_2 \cot(2\beta_3))^{-1}\,,\\
K_{2,2} =  1-(\e^{2 i \beta_2} \tan ^2\beta_3+1)^{-1}\,.
\end{gather*}
\ese

\textbf{E.}
If $K$ is non-diagonalizable by a unitary matrix and has two non-zero eigenvalues,
its Schur form is given by $\Gamma_E$ in Eq.~\eqref{e:Schurparametrization},
where $\gamma$ and $\gamma_0$ are two complex numbers,
$0\le\beta<2\pi$ and $0<\eta\le\pi/4$.
As a special case, if $\eta = \pi/4$ the two discrete eigenvalues coincide.
The most general norming constant and unitary matrix for this class are given by
\vspace*{-1ex}
\bse
\label{e:KUE}
\begin{gather}
K_E = \gamma\, I_2 + \frac{\gamma_0}{2}\e^{i\beta}
\begin{pmatrix}
\cot\eta & i  \\
i & -\tan\eta
\end{pmatrix}\,,\\
U_E = \e^{i\beta_1}\begin{pmatrix}
\cos\eta & \e^{-i \beta } \sin\eta \\
i \sin\eta & -i \e^{-i \beta } \cos\eta \\
\end{pmatrix}\,,
\end{gather}
\ese
where $0\le\beta_1<2\pi$.


\textbf{Core soliton component in class~D.}
Finally, for completeness, here we give explicitly the entries of the core soliton component $Q(x,t)$ in class~D:
\vspace*{-1ex}
\bse
\label{e:QDcomponent}
\begin{widetext}
    \begin{gather}
    q_{1,1}(x,t) = k_o \e^{-2i \alpha}
    [|c_o| c_1^2 Z \sec\alpha\cosh (\chi + 2i \alpha) - |c_o|^3\e^{ - \chi-i\alpha} Z \tan^2 (2\eta) - i c_1 \e^{- i s} (c_o - c_o^*\e^{2 i s} Z^4)]/
    (Z\Delta_C)\,,\\
    q_{1,2}(x,t) = c_o^* k_o\e^{ -3 i \alpha}\tan(2\eta)
    (|c_o| c_{+,1} \e^{-\chi} Z^2-i c_1 c_o \e^{i \alpha-is})/
    (Z\Delta_C)\,,\\
    q_{2,1}(x,t) = -k_o Z |c_o|\tan2\eta
    (i |c_o| c_{1} \e^{i s}+c_o c_{+,1} \e^{-\chi-i \alpha })/\Delta_C\,,\quad
    q_{2,2}(x,t) = k_o - [|c_o|^3 k_o \e^{-\chi-i \alpha } \tan^2 (2\eta)]/\Delta_C\,,
    \end{gather}
\end{widetext}
\ese
where $\gamma = \xi\e^{i\phi}$,
for brevity we suppressed the $x$ and $t$ dependence of all quantities in the right-hand side,
and
\vspace*{-1ex}
\begin{gather*}
\chi(x,t) = k_o c_{-,1} x \cos\alpha - k_o^2 c_{+,2} t \sin2\alpha + \log\frac{2Zk_o |c_o|\cos\alpha}{c_1\xi}\,,\\
s(x,t) = c_{+,1} k_o x\sin\alpha + c_{-,2} k_o^2 t \cos2\alpha - \phi\,,\\
\Delta_C(x,t) = c_1(|c_o| c_{1} \sec\alpha \cosh\chi + 2 Z K_s)\,,\\
K_s(x,t) = Z^2 \sin (s+2 \alpha)-\sin s\,,\\
c_o = 1-\e^{- 2i\alpha}Z^2\,,\quad
c_1 = \sqrt{c_{+,1}^2 Z^2 + |c_o|^2 \tan^2(2\eta)}\,.
\end{gather*}



\providecommand\reftitle[1]{``#1''}%

\end{document}